\documentclass[10pt,a4paper]{article}

\usepackage[utf8]{inputenc}
\usepackage[T1]{fontenc}
\usepackage[english]{babel}

\usepackage[margin=1in]{geometry}
\usepackage{enumitem}
\usepackage{amsmath}
\usepackage{amssymb}
\usepackage{amsfonts}
\usepackage{amsthm}
\usepackage{comment}
\usepackage{rotating}
\usepackage{tabularx}
\usepackage[colorlinks=true, linkcolor=black, citecolor=black, urlcolor=black]{hyperref}
\usepackage[nameinlink, capitalise]{cleveref}
\usepackage{booktabs}\usepackage{subcaption}

\usepackage{tikz}
\usetikzlibrary{automata, positioning}
\usetikzlibrary{calc}
\usetikzlibrary{patterns}

\usepackage{algorithm}
\usepackage{algpseudocode}
\algrenewcommand\algorithmicrequire{\textbf{Input:}}
\algrenewcommand\algorithmicensure{\textbf{Output:}}

\theoremstyle{plain}
\newtheorem{thm}{Theorem}[section]
\newtheorem{pro}[thm]{Proposition}
\newtheorem{cor}[thm]{Corollary}
\newtheorem{lem}[thm]{Lemma}
\newtheorem{obs}[thm]{Observation}
\newtheorem{qst}[thm]{Question}

\newtheorem{exm}[thm]{Example}

\theoremstyle{definition} 
\newtheorem{defi}[thm]{Definition}

\DeclareMathOperator*{\cart}{\times}

\newcommand{\problemStatement}[3]{
  \begin{center}
  \begin{tabularx}{\columnwidth}{@{}lX@{}}
  \toprule
  \multicolumn{2}{@{}c@{}}{\textsc{#1}}\tabularnewline
  \midrule
  \bfseries Input:    & #2 \\
  \bfseries Question: & #3 \\
  \bottomrule
  \end{tabularx}
  \end{center}
}

\usepackage{authblk}

\author[1]{Janosch~Döcker}
\author[2]{Britta~Dorn}
\author[2]{Hannah~Van~Santvliet}
\author[2]{Katrin~Willger}

\affil[1]{Independent Researcher}
\affil[2]{Department of Computer Science, University of Tübingen, Germany}

\title{Connecting \textsc{$k$-Dimensional Matching} and \textsc{Exact Cover by $k$-Sets}: A complete complexity classification of \textsc{$k$-Partitioned 1-in-3-SAT} for planar, linear and $3$-connected instances}

\begin{document}

\maketitle
\tableofcontents

\begin{abstract}
In \textsc{$k$-Partitioned 1-in-3-Sat}, the clauses of a positive \textsc{1-in-3-Sat} formula are partitioned into~$k$ classes such that every
variable occurs exactly once in each class. This problem corresponds to \textsc{$k$-Dimensional Matching} restricted to instances in which every element lies in exactly
three tuples. For every~$k$, we provide a complete complexity classification with respect to all combinations of three structural restrictions on the incidence graph: planarity, linearity (no two clauses share more than one variable) and 3-connectivity. 
We show \textsc{NP}-completeness of \textsc{$3$-Partitioned 1-in-3-Sat} and  \#\textsc{P}-completeness of its counting version, which also serve as a starting point for the further reductions, and then enforce the desired properties by repeatedly inserting gadgets into the incidence graph. 

As a consequence, we obtain that \textsc{$k$-Dimensional Matching} and \textsc{Exact Cover by $k$-Sets} remain
\textsc{NP}-hard under the same restrictions, even if every element occurs in exactly three tuples or sets, respectively. In particular, \textsc{Planar $4$-Dimensional Matching} is \textsc{NP}-complete and we complement the \textsc{NP}-hardness result with an \textsc{FPT} algorithm with respect to the number of so-called quadrilateral components of a given plane embedding of the incidence graph.
\end{abstract}

\section{Introduction}
All \textsc{NP}-complete problems are polynomial-time reducible to one another, but some of them are particularly intertwined. An example are the three problems \textsc{$k$-Dimensional Matching} (\textsc{$k$DM}), \textsc{Exact Cover by $k$-Sets} (\textsc{X$k$C}),  and \textsc{Positive 1-in-3-Sat-E$k$}, which we briefly recall. Given a set $U$ that is the disjoint union of sets $U_1, \dots, U_k$ of equal size and a set $S \subseteq U_1 \times \dots \times U_k$, where we identify a tuple with the set of its entries. \textsc{$k$DM} asks whether there is a subset of $S$ that covers each element in $U$ exactly once. \textsc{X$k$C} is the generalization of \textsc{$k$DM}
in which $S$ is a collection of $k$-element subsets of a set $U$. Again, the goal is to decide whether there is a subcollection $S' \subseteq S$ such that every element of $U$ is contained in exactly one set of $S'$. \textsc{Positive 1-in-3-Sat} asks, given a collection of clauses containing unnegated variables only, that is, $3$-element subsets of a set of variables $X$, whether there is a truth assignment for $X$ that sets exactly one variable in each clause to $T$ and the other two to $F$. \textsc{Positive 1-in-3-Sat-E$k$} additionally requires that every variable appears in exactly~$k$ clauses. Now, let $S$ be obtained from an instance of \textsc{Positive 1-in-3-Sat-E$k$} with variables $X$ and clauses $C$ by including, for each variable $x$, a subset that contains each clause $c \in C$ with $x \in c$. Then, $U = C$ and each solution of this \textsc{X$k$C} instance corresponds to a solution of the \textsc{1-in-3-Sat} instance (cf.~Moore and Robson~\cite[Section 3.1]{Moore2001}). The problem \textsc{$k$-Partitioned 1-in-3-Sat} is the restriction of \textsc{Positive 1-in-3-Sat-E$k$} to instances whose clause set is given as the disjoint union $C = P_1 \cup \dots \cup P_k$ of $k$ partitions $P_1, \dots, P_k$ of $X$ into $3$-element sets. Via the construction above (with $U_i = P_i$), it corresponds to \textsc{$k$DM} in the same way as \textsc{Positive 1-in-3-Sat-E$k$} corresponds to \textsc{X$k$C}.

In this paper, we explore the computational landscape of \textsc{$k$-Partitioned 1-in-3-Sat} for all $k \in \mathbb{N}$ in order to highlight the connections between \textsc{Exact Cover by $k$-Sets} and \textsc{$k$-Dimensional Matching}. The partition restriction was explored by Darmann, Döcker and Dorn~\cite{Darmann2024} for \textsc{Linear Not-All-Equal-Sat}. They showed that this problem is \textsc{NP}-complete for $k \geq 4$ partitions and solvable in polynomial time for $k \leq 3$.

The computational complexity of all three problems has been studied extensively. By Schaefer's dichotomy theorem~\cite{Schaefer1978}, \textsc{Positive 1-in-3-Sat} is \textsc{NP}-complete, and Moore and Robson~\cite{Moore2001} showed that it remains \textsc{NP}-complete for instances with a cubic planar incidence graph. Mulzer and Rote~\cite[Proposition~3.3]{Mulzer2008} showed that \textsc{Planar Positive 1-in-3-Sat} remains \textsc{NP}-complete even if a planar rectilinear embedding of the incidence graph is given. Dyer and Frieze~\cite{Dyer1986} showed that \textsc{Planar 3DM} is \textsc{NP}-complete even if each element appears in only two or three triples. 
Moreover, \textsc{X3C} remains \textsc{NP}-complete if each element appears in exactly three sets (see Gonzalez~\cite{Gonzalez1985}, who calls this variant \textsc{RX3C}). We refer to it as \textsc{X3C-E3}. Moore and Robson~\cite{Moore2001} showed that \textsc{Planar X3C-E3} is \textsc{NP}-complete by establishing \textsc{NP}-completeness of \textsc{Planar Positive 1-in-3-Sat-E3}, which, by the correspondence described above, is the same problem in a different formulation.
Their question of whether the counting version \#\textsc{Planar X3C-E3} is \#\textsc{P}-complete remained open for more than two decades, until Cai and Fan~\cite{Cai2023} answered it affirmatively.  In addition, the open question on the complexity of \textsc{Planar Exact Cover by 4-Sets} posed by van Bevern, Bredereck, Chen, Froese, Niedermeier and Woeginger~\cite{bevern} was answered in a recent preprint by the MIT Hardness Group (Abel, Demaine, Diomidova, Li and Zhou)~\cite{MITHardnessGroup2026} who showed \textsc{NP}-hardness of the problem.

We also consider the complexity of the corresponding counting problems, for example, counting the perfect matchings of an instance of \textsc{Cubic Planar 3DM}. Creignou and Hermann~\cite{Creignou1996} proved a dichotomy theorem for counting problems of generalized satisfiability, showing that each such problem is either computable in polynomial time or \#\textsc{P}-complete. Moreover, they related their dichotomy to Schaefer's dichotomy theorem by showing that whenever the generalized satisfiability problem is \textsc{NP}-complete, its counting variant is \#\textsc{P}-complete (see~\cite[Corollary 4.2]{Creignou1996}). As a consequence, \#\textsc{Positive 1-in-3-Sat} is \#\textsc{P}-complete. Since these dichotomies do not capture restrictions such as planarity or bounds on the number of variable occurrences, results for such restricted classes require a different approach, for example, a parsimonious reduction from a \#\textsc{P}-complete problem. Hunt, Marathe, Radhakrishnan and Stearns~\cite{Hunt1998} studied the complexity of planar counting problems and showed, among other results, that the counting versions of \textsc{Planar X3C} and \textsc{Planar Positive 1-in-3-Sat} are \#\textsc{P}-complete.    

In this paper, for every~$k$, we classify the complexity of \textsc{$k$DM} under all combinations of the properties `planar', linear' and 3-connected', and obtain the corresponding hardness results for \textsc{X$k$C}. Figure~\ref{fig:overview} summarizes these results. Here, planarity and 3-connectivity refer to the incidence graph of the instance, and linearity means that any two tuples agree in at most one position (equivalently, in the \textsc{Sat} formulation, any two clauses share at most one variable). For further results on satisfiability problems for \textit{linear} \textsc{Sat} variants, we refer to Porschen, Speckenmeyer and Zhao~\cite{Porschen2009} and Porschen, Schmidt, Speckenmeyer and Wotzlaw~\cite{Porschen2014}. 
 
The structure of this paper follows the properties that we successively impose on \textsc{$k$-Partitioned 1-in-3-Sat}: In Section~\ref{sec:hardness3part}, we show \textsc{NP}-completeness of \textsc{3-Partitioned 1-in-3-Sat} and \#\textsc{P}-completeness of its counting variant. Sections~\ref{sec:hardnessplanarity}, \ref{sec:hardnesslinearity} and~\ref{sec:hardnessconnectivity} then add planarity, linearity and 3-connectivity, respectively, to \textsc{3-Partitioned 1-in-3-Sat}. In Section \ref{sec:4part}, we investigate \textsc{$k$-Partitioned 1-in-3-Sat} for $k>3$ and the corresponding combinations of the properties `planar', `linear' and `3-connected'. 
Besides new results such as the \textsc{NP}-completeness of \textsc{Planar 4DM} or ASP-hardness of \textsc{Planar 3DM}, our approach also yields new proofs of known results for \textsc{$k$DM} and \textsc{X$k$C}. Finally, Section~\ref{sec:matchingandexactcover} translates the results of the previous sections into the language of matchings and exact covers, therewith deduce results and compares our proofs with alternative ones from the literature. In this way, we aim at a unified view of the three problems.	


\tikzset{
  npHard/.style={
    pattern=north east lines,
    pattern color=black!40
  },
  polyTime/.style={
    pattern=north west lines,
    pattern color=black!40
  },
  trivial/.style={
    pattern=dots,
    pattern color=black!40
  },
  noInstances/.style={
    fill=gray
  }
}

\newcommand{\vennwithoutarrows}[2]{%
\begin{subfigure}[t]{0.22\textwidth}
\centering
\begin{tikzpicture}[scale=0.8, thick]

  \def\r{1.2}

  \coordinate (A) at (90:1.0);
  \coordinate (B) at (210:1.0);
  \coordinate (C) at (330:1.0);

  \path (-2.2,-1.8) rectangle (2.2,2.2);

  \fill[npHard]
    (-2.2,-1.8) rectangle (2.2,2.2);

  \begin{scope}
    \clip (A) circle (\r);
    \clip (B) circle (\r);
    \clip (C) circle (\r);
    \fill[white]
      (-2.2,-2.2) rectangle (2.2,2.2);
  \end{scope}

  #1

  \draw (A) circle (\r);
  \draw (B) circle (\r);
  \draw (C) circle (\r);

  \node[above]
    at (A) {\small planar};

  \node[below left]
    at ($(B)+(0.4,0.4)$) {\small linear};

  \node[below right]
    at ($(C)+(-0.5,0.3)$) {\small 3-con.};

\end{tikzpicture}
\caption{#2}
\end{subfigure}%
}

\begin{figure}[h]
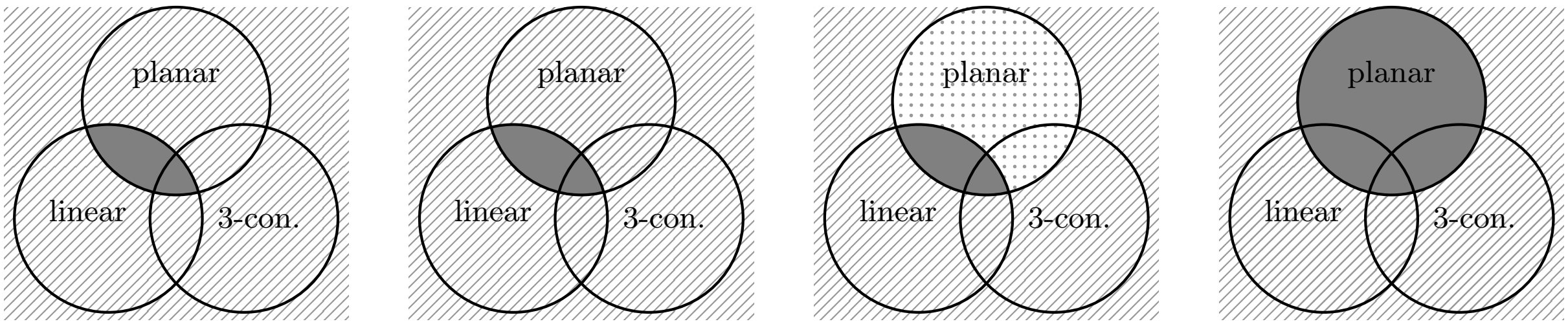

\centering

\vennwithoutarrows{

  \begin{scope}
    \clip (A) circle (\r);
    \fill[noInstances] (B) circle (\r);
  \end{scope}

}{3-partitioned}
\hfill
%
\vennwithoutarrows{

  \begin{scope}
    \clip (A) circle (\r);
    \fill[npHard] (A) circle (\r);
  \end{scope}

  \begin{scope}
    \clip (A) circle (\r);
    \fill[white] (C) circle (\r);
  \end{scope}

  \begin{scope}
    \clip (A) circle (\r);
    \fill[npHard] (C) circle (\r);
  \end{scope}

    \begin{scope}
    \clip (A) circle (\r);
    \fill[npHard] (C) circle (\r);
  \end{scope}

  \begin{scope}
    \clip (B) circle (\r);
    \fill[npHard] (C) circle (\r);
  \end{scope}

  \begin{scope}
    \clip (A) circle (\r);
    \fill[noInstances] (B) circle (\r);
  \end{scope}

}{4-partitioned}
\hfill
%
\vennwithoutarrows{

  \begin{scope}
    \clip (A) circle (\r);
    \fill[white] (A) circle (\r);
    \fill[trivial] (A) circle (\r);
  \end{scope}

  \begin{scope}
    \clip (B) circle (\r);
    \fill[npHard] (C) circle (\r);
  \end{scope}

  \begin{scope}
    \clip (A) circle (\r);
    \fill[noInstances] (B) circle (\r);
  \end{scope}

}{5-partitioned}
\hfill
%
\vennwithoutarrows{

  \fill[npHard] (B) circle (\r);

  \fill[npHard] (C) circle (\r);

  \fill[noInstances] (A) circle (\r);

}{$k$-partitioned, $k > 5$}

\caption{Complexity of \textsc{$k$-Partitioned 1-in-3-Sat} (equivalently, \textsc{$k$Dm}) for all combinations of planarity, linearity and 3-connectivity (`3-con.'). Hatched regions indicate \textsc{NP}-hardness, gray regions indicate that no instances exist, and dotted regions indicate that the problem is trivial in the sense that every instance is a no-instance.}\label{fig:overview}

\end{figure}

\section{Preliminaries}
Let $X := \{x_1, x_2, \ldots, x_n\}$ be a set of \emph{variables}. A \emph{literal} is a variable $x_i$ or its negation $\overline{x_i}$ and the set of literals corresponding to $X$ is $\mathcal{L}_X = \{x_i, \,  \overline{x_i} \mid 1 \leq i \leq n\}$. A \emph{clause} is a subset of $\mathcal{L}_X$ and is called \emph{positive} if it consists of unnegated variables only. Throughout this work, we are only dealing with \emph{$3$-clauses}, i.e., clauses containing exactly three distinct literals. A \emph{collection of clauses}, also called a~\emph{formula}, is a set $C := \{c_1, c_2, \ldots, c_m\}$ of clauses. It is sometimes convenient to index clauses with a variable to express that the variable is contained in the clause: For $k$ different clauses each containing the same variable $x$, we write $c_{x,i}$ for each $i \in \{1,2,\ldots, k\}$. By $\operatorname{var}(c)$, we denote the set of variables contained in clause~$c$. Similarly, for a gadget~$\mathcal{G}$, we use $\operatorname{var}(\mathcal{G})$ to denote the set of variables used in~$\mathcal{G}$, and  $\operatorname{cl}(\mathcal{G})$ to denote the set of clauses used in~$\mathcal{G}$. 

A \emph{truth assignment} is a mapping $\beta\colon X \rightarrow \{T, F\}$, where $T$ and $F$ represent the truth values $T$ and $F$, respectively.  
 A truth assignment $\beta$ \emph{satisfies} a literal $x_i$ if $\beta(x_i)=T$, and a literal $\overline{x_i}$ if $\beta(x_i)=F$.
We further call a clause~$c$ \emph{satisfied} under~$\beta$ if and only if at least one of its literals is set to $T$ under~$\beta$. 
If a $3$-clause~$c$ has exactly one literal set to $T$ and two literals set to $F$ under~$\beta$ (i.e., one out of the three literals belonging to the clause is set to $T$), the truth assignment~$\beta$ is said to \emph{1-in-3-satisfy} the clause~$c$. A truth assignment $\beta$ 1-in-3-satisfies a collection $C$ of clauses if and only if $\beta$ 1-in-3-satisfies each clause in $C$. Accordingly, a formula is called \emph{1-in-3-satisfiable} if there exists a truth assignment that 1-in-3-satisfies the formula.

Let $c_1, c_2, \ldots ,c_m$ be pairwise disjoint subsets of $X$. Then, $P = \bigcup_{j = 1}^{m}\{c_j\}$ is called a \emph{partition} of $X$ if the union of the elements in $P$ yields $X$, i.e., $\bigcup_{j = 1}^m c_j  = X$. A partition of~$X$ can also be seen as a collection of \emph{positive} clauses in which each variable appears in exactly one clause. Taking the union of~$k$ pairwise disjoint partitions of~$X$ yields a formula in which each variable appears exactly~$k$ times. 
\problemStatement{$k$-Partitioned 1-in-3-Sat}{
A set $X$ of variables, a collection of positive 3-clauses~$C$, and a tuple of pairwise disjoint partitions  $P = \langle P_1, P_2, \ldots , P_k\rangle$ of $X$ such that $\bigcup_{i=1}^{k} P_i = C$. 
}{
Is there a truth assignment for $X$ that 1-in-3-satisfies $C = \bigcup_{i=1}^k P_i$?
}
Throughout the paper, we always assume that for all pairs of variables $x,x' \in X$ with $x \neq x'$,  $x$ and $x'$ jointly appear in less than~$k$ clauses, i.e., 
$$|\{i \in \{1, \ldots, k\} \mid \exists c \in P_i \colon \{x,x'\} \subseteq c\}| < k.$$
\begin{exm}\label{ex:partdef} Let $X$ be a set of variables, $C$ be a set of clauses and $P=\langle P_1, P_2, P_3,P_4 \rangle$. The clauses in each $P_i$ for $i \in \{1, \dots ,4\}$ form a partition of $X$.
\begin{align*}
X &= \{x_1,x_2,\ldots, x_9\},\\
C &= \{c_1, \dots ,c_{12}\},\\
P_1 &= \{\underbrace{\{x_1, x_2, x_5\}}_{c_1}, \underbrace{\{x_3, x_4, x_6\}}_{c_2}\},\underbrace{\{x_7, x_8, x_9\}}_{c_3}\},\\
P_2 &= \{\underbrace{\{x_1, x_3, x_6\}}_{c_4}, \underbrace{\{x_2, x_4, x_8\}}_{c_5}\},\underbrace{\{x_5, x_7, x_9\}}_{c_6}\},\\
P_3 &= \{\underbrace{\{x_1, x_2, x_3\}}_{c_7}, \underbrace{\{x_4, x_6, x_7\}}_{c_8}\},\underbrace{\{x_5, x_8, x_9\}}_{c_9}\},\\
P_4 &= \{\underbrace{\{x_1, x_2, x_7\}}_{c_{10}}, \underbrace{\{x_4, x_5, x_8\}}_{c_{11}}\}, \underbrace{\{x_3, x_6, x_9\}}_{c_{12}}\}.\\
\end{align*}

The truth assignment $\beta$ with  
$\beta(x_1) = \beta(x_4) = \beta(x_9) = T 
\text{ and } 
\beta(x_i) = F \text{ for } i \in \{2,3,5,6,7,8\}$
1-in-3-satisfies all clauses. Therefore, $\mathcal{I}=(X,C,P)$ is a yes-instance of \textsc{$4$-Partitioned 1-in-3-Sat}.
\end{exm}

The \emph{\textsc{Sat} graph} or \textit{incidence graph} $G_{\textsc{Sat}}(\mathcal{I})$ of a formula with variable set $X$ and clause set~$C$ is the bipartite graph with a vertex for each clause and each variable of~$\mathcal{I}$, and an edge connecting a variable to a clause if and only if the variable is contained in the corresponding clause in~$\mathcal{I}$. Throughout this work, we assume that the \textsc{Sat} graph is connected. Otherwise, the corresponding formula essentially decomposes into two independent formulas. We represent variables as circular vertices (called variable vertices) and clauses as square vertices (called clause vertices) and, whenever relevant, label the clause vertices according to their partition: When referring to concrete partitions, we use the integers~$1, 2, \dots , k$. For gadgets that are designed to work with arbitrary partitions, we use Greek letters such as~$\lambda, \mu, \nu, \zeta \dots$ as a label. In particular, every variable vertex in an instance of \textsc{$k$-Partitioned 1-in-3-Sat} has an edge to a clause vertex labeled with $i$ for every $i \in \{1, \dots ,k\}$. Aligned with conventions in graph theory, we use the following notation: For a graph $G=(V,E)$ and a subset $V'\subseteq V$, we denote by $G\setminus V'$ the graph $G[V\setminus V']$ that is induced by the vertex set~$V\setminus V'$. Further, we say that an instance $\mathcal{I}_{\textsc{Sat}} = (X,C,P)$ of \textsc{1-in-3-Sat} is \textit{planar} if $G_{\textsc{Sat}}(X,C,P)$ is planar. We say that $\mathcal{I}_{\textsc{Sat}}$ is \textit{linear} if  no two clauses share more than one variable, i.e., for each pair of distinct clauses $c,c'\in C$, we have $|c \cap c'| \leq 1$. \\

Note that for $k=1$, \textsc{$k$-Partitioned 1-in-3-Sat} is trivial since every variable occurs only once. For $k=2$, all instances of the problem are yes-instances. This follows from Corollary 1a by Schrijver~\cite{Schrijver1998}, which implies the following Proposition. 
\begin{pro}
Every instance of \textsc{2-Partitioned 1-in-3-Sat} has at least 
\[
\left( \sqrt{\frac{4}{3}}\right)^m = \left( \sqrt[3]{\frac{4}{3}} \right)^n > \left( \frac{11}{10} \right)^n
\]
satisfying truth assignments where $m$ is the number of clauses and $n$ is the number of variables.
\end{pro}
Therefore, from here on, the paper we will focus on \textsc{$k$-Partitioned 1-in-3-Sat} for $k\geq 3$.

\section{Hardness of Deciding and Counting \textsc{3-Partitioned 1-in-3-Sat}}\label{sec:hardness3part}

We now turn to proving our first result, which constitutes the starting point for all subsequent hardness proofs. In this section, we show that \textsc{3-Partitioned 1-in-3-Sat} is \textsc{NP}-complete and that its counting variant is \#\textsc{P}-complete.
To this end, we provide a polynomial-time parsimonious reduction from \textsc{Positive 1-in-3-Sat}
to \textsc{3-Partitioned 1-in-3-Sat}. Since the reduction preserves the number of solutions, it establishes hardness of both the
decision and the counting version. 

Our strategy is similar to the method used by Moore and Robson~\cite{Moore2001}: Given the incidence graph of our input instance~$\mathcal{I}$, we use gadgets as a kind of modular plug-in system: each gadget enforces certain desired properties while preserving others, such as a
$k$-partition of the clause set.
 Formally, given a substructure~$\mathcal{S}$ of the instance~$\mathcal{I}$, a gadget is a graph (or a formula) that replaces~$\mathcal{S}$ such that the original instance~$\mathcal{I}$ and the modified instance~$\mathcal{I'}$ are equivalent with respect to satisfying assignments. When performing parsimonious reductions, we have to ensure that the gadgets do not only preserve satisfiability, but also leave the number of satisfying
assignments unchanged.\\

The gadgets have a dedicated set of vertices consisting of so-called \emph{input/output vertices}. A clause vertex that belongs to the dedicated set is an \emph{input vertex}, a variable vertex that belongs to the dedicated set is an \emph{output vertex}. We define the number of input/output vertices as the \emph{degree of a gadget}.  Assuming that each variable occurs a fixed number of times, an \emph{output vertex} in the \textsc{Sat} graph is a vertex whose degree is smaller than this number, whereas an \emph{input vertex} is a vertex whose degree is smaller than three. These vertices can be connected to the rest of the instance by adding edges, thereby fixing the position of the gadget in the graph instance. We depict input vertices by drawing edges originating from them and ending at free endpoints. If the naming of the vertices that connect to the input vertices is relevant, we draw a dotted line enclosing the vertices that belong to the gadget.

\subsection{Gadgets}
We start by introducing the gadgets used and providing short explanations of their properties. Later on, we explain how gadgets are plugged into the given instance. The first gadget was introduced by Hunt, Marathe, Radhakrishnan and Stearns~\cite[Lemma~3.6]{Hunt1998}. We call it $F$-wheel as it looks like a wheel where all variables on the outer face of the gadget have forced truth value $F$.  

\begin{figure}[h]
    \centering
\includegraphics[width=.18\textwidth]{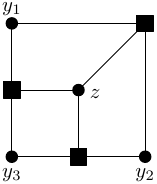}
    \caption{$F$-wheel.}
    \label{fig:F-wheel}
\end{figure}

\paragraph{$F$-wheel} The $F$-wheel consists of clauses $\{z, y_1, y_2\}$, $\{z, y_2, y_3\}$ and $\{z, y_1, y_3\}$ (see Figure~\ref{fig:F-wheel}). A truth assignment $\beta$ 1-in-3-satisfies these clauses if and only if $\beta(z) = T$ and $\beta(y_1) = \beta(y_2) = \beta(y_3) = F$.\\ 

The next gadget forces two variables to have the same truth value and was introduced by Moore and Robson~\cite[p.~581]{Moore2001}.

\paragraph{Equality verifier} The equality verifier $EQ(x,y)$ for two variables~$x$ and~$y$ consists of two clauses~$\{x,a,b\}$ and~$\{y,a,b\}$. A truth assignment $\beta$ for $\{x,y\}$ can be extended to $\{x,y,a,b\}$ if and only if $\beta(x) = \beta(y)$.\\  

Similarly to Moore and Robson, we chain equality verifiers to force multiple variables to the same truth value. The key difference is that our chains are cyclic.     

\begin{figure}[h]
    \centering
    \includegraphics[width=.95\textwidth]{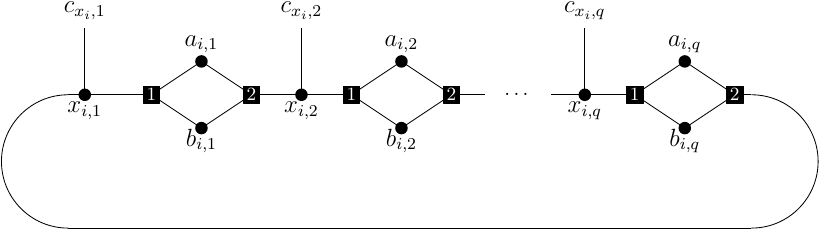}
    \caption{Equality gadget for a variable $x_i$ that appears in $q \geq 2$ clauses $c_j$ with $j \in \{1, 2, \ldots, q\}$.}
    \label{fig:equality_gadget}
\end{figure}

\paragraph{Equality gadget} Let $q \geq 2$. The equality gadget $EQ(X)$ for a set of variables $X = \{x_1, x_2, \ldots, x_q\}$ consists of a cyclic chain of $q$ equality verifiers $EQ(x_1, x_2),\, EQ(x_2, x_3),\, \ldots, EQ(x_{q-1}, x_q),\, EQ(x_q, x_1)$ as illustrated in Figure~\ref{fig:equality_gadget}.\\   

The following double clause gadget was introduced by Moore and Robson~\cite[Figure~16]{Moore2001} who called it a \textit{triangle}. 

\begin{figure}[h]
    \centering
    \includegraphics[width=.35\textwidth]{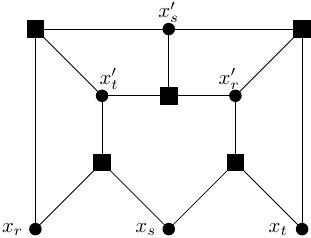}
    \caption{Double clause gadget.}
    \label{fig:double_clause_gadget}
\end{figure}

\paragraph{Double clause gadget} The double clause gadget simulates a clause $\{x_r, x_s, x_t\}$ while consuming two appearances of $x_r$, $x_s$ and $x_t$ (see Figure~\ref{fig:double_clause_gadget}). The gadget consists of the following clauses\\ 
\[
\{x_r,x_s, x_t'\}, \{x_r',x_s, x_t\}, \{x_r, x_s', x_t'\}, \{x_r',x_s', x_t\}, \{x_r',x_s', x_t'\}. 
\]

The following \emph{collection gadget} is inspired by the elegant construction by Moore and Robson~\cite[Section~3.2]{Moore2001}. As we do not need to preserve planarity (planarity will be achieved by new gadgets that we introduce in Section~\ref{sec:hardnessplanarity}), we can use a much simpler approach to increase variable appearances. The name of the gadget originates from the idea to `collect' leftover occurrences.

\begin{figure}[h]
    \centering
\includegraphics[width=.85\textwidth]{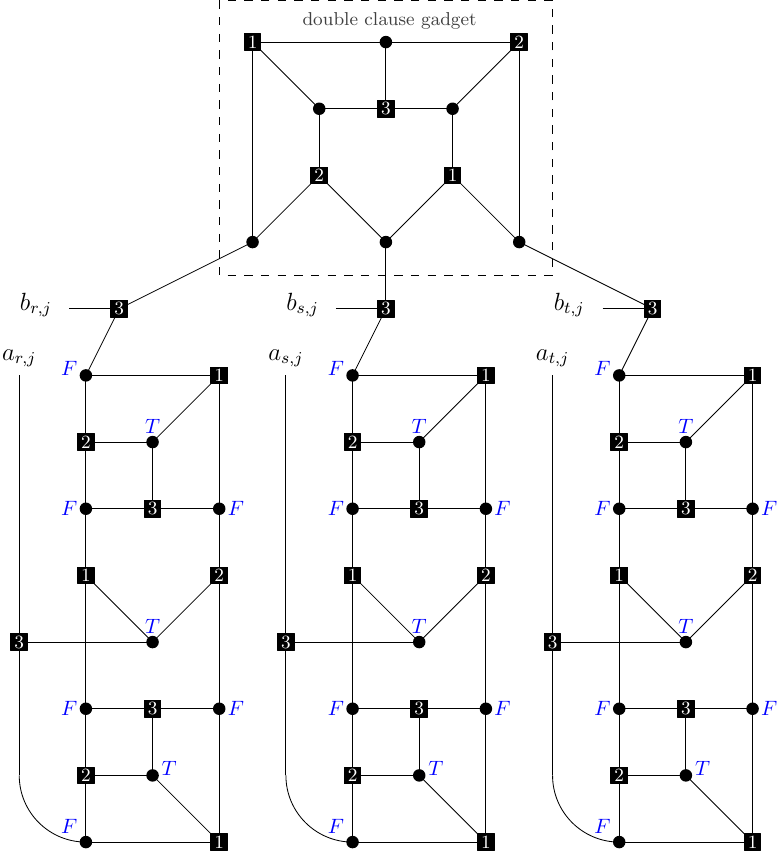}
    \caption{Collection gadget for a clause $c_j = \{x_r, x_s, x_t\}$.}
    \label{fig:collection_gadget}
\end{figure}

\paragraph{Collection gadget} Let $c_j = \{x_r, x_s, x_t\}$. The collection gadget $CO(c_j)$ consists of six $F$-wheels, a double clause gadget and nine additional clauses as illustrated in Figure~\ref{fig:collection_gadget}. It will be used to add a third appearance of the auxiliary variables $a,b$ of the equality verifiers in a way that forces $a$ to be set to $F$. 

\ \\
Note that all variables in the double clause gadget appear exactly three times.
\begin{lem}\label{lem:doubleclause_gadget}
Let $DCG(c_j=\{x_r,x_s,x_t\})$ be a double clause gadget. If $\beta$ 1-in-3-satisfies a clause $c_j$, the truth assignment can be extended to a unique 1-in-3-satisfying assignment $\beta'$ of $DCG(c_j=\{x_r,x_s,x_t\})$. In particular, we have that there are exactly three 1-in-3-satisfying assignments for $DCG(c_j=\{x_r,x_s,x_t\})$ if and only if there are exactly three 1-in-3-satisfying assignments for a clause $c_j$.
\end{lem}
\begin{proof}
Assume that $\beta$ 1-in-3-satisfies the clause $\{x_r,x_s,x_t\}$. Then, we have that exactly one variable in $\{x_r,x_s,x_t\}$ is set to $T$. Due to three equality verifiers embedded in the double clause gadget, we have for any 1-in-3-satisfying assignment $\beta':\{x_r,x'_{r},x_s,x'_{s},x_t,x'_{t}\} \to \{T,F\}$ that $\beta'(x_r) = \beta'(x'_r)$, $\beta'(x_s) = \beta'(x'_s)$ and $\beta'(x_t) = \beta'(x'_t)$. The clause $\{x'_r,x'_s, x'_t\}$ forces exactly one variable in $\{x'_r,x'_s, x'_t\}$ to $T$ and therefore, the assertion follows.
\end{proof}

Similarly, we observe that there is a unique truth assignment for the equality gadget. To highlight the fact that some variables have the same truth value in an equality gadget, we give them the same letter and first index. Further, we introduce a second index for the variables counting their appearance and enumerate the clauses correspondingly. For example, the variable $x_{1,2}$ is the second $x_1$-variable with the same truth values that appears in the clause $c_2$.

\begin{lem}\label{lem:equality_gadget}
Let $EQ(X_i)$ be an equality gadget for a set of variables $X_i = \{x_{i,1}, x_{i,2}, \ldots, x_{i,q}\}$ with $ q\geq 2$ as illustrated in Figure~\ref{fig:equality_gadget}. Then, a truth assignment $\beta$ for $X_i$ can be extended to a truth assignment $\beta'$ for $\operatorname{var}(EQ(X_i))$ that 1-in-3-satisfies each clause in $EQ(X_i)$ if and only if
$
\beta(x_{i,1}) = \beta(x_{i,2}) = \cdots = \beta(x_{i,q}). 
$
Further, if $\beta(a_{i,1}) = \beta(a_{i,2}) = \cdots = \beta(a_{i,q}) = F$, then the extensions are unique. 
\end{lem}

\begin{proof}
If a truth assignment $\beta$ for $X_i$ can be extended to a truth assignment $\beta'$ for $\operatorname{var}(EQ(X_i))$ that 1-in-3-satisfies each clause in $EQ(X_i)$, we have that $
\beta(x_{i,1}) = \beta(x_{i,2}) = \cdots = \beta(x_{i,q})
$ since the equality gadget consists of equality verifiers.

If we have $\beta(x_{i,1}) = \beta(x_{i,2}) = \cdots = \beta(x_{i,q}) = \xi$ for a truth value $\xi \in \{T,F\}$, we can extend $\beta$ to a 1-in-3-satisfying assignment $\beta'$ with $\beta'(b_{i,1}) = \beta'(b_{i,2}) = \cdots = \beta'(b_{i,q}) = \lnot \xi$. If $\beta(a_{i,1}) = \beta(a_{i,2}) = \cdots = \beta(a_{i,q}) = F$ this is a unique extension because either all $x \in X_i$ or all $b \in \{b_{i,1},b_{i,2}, \dots, b_{i,q}\}$ are set to $T$.
\end{proof}

In order to make sure that added gadgets preserve the number of 1-in-3-satisfying truth assignments, we count the 1-in-3-satisfying assignments in the gadgets. This is particularly important for the collection gadget that we insert into the \textsc{Sat} graph to compensate for additional variable occurrences.

\begin{lem}\label{lem:collection_gadget}
Let $CO(c_j=\{x_r, x_s, x_t\})$ be a collection gadget. Then, there are exactly three truth assignments $\beta_1$, $\beta_2$ and $\beta_3$ for $\operatorname{var}(CO(c_j))$ that 1-in-3-satisfy each clause in $CO(c_j)$. Further, $\beta_1$, $\beta_2$ and $\beta_3$ satisfy (P1) $\beta_1(a) =  \beta_2(a) = \beta_3(a) = F$ for each $a \in \{a_{r,j}, a_{s,j}, a_{t,j}\}$ and (P2)   

\begin{align*}
\beta_1(b_{r,j}) = T, && \beta_2(b_{r,j}) = T, && \beta_3(b_{r,j}) = F, \\
\beta_1(b_{s,j}) = T, && \beta_2(b_{s,j}) = F, && \beta_3(b_{s,j}) = T, \\
\beta_1(b_{t,j}) = F, && \beta_2(b_{t,j}) = T, && \beta_3(b_{t,j}) = T.
\end{align*}

\end{lem}
\begin{proof}
Since for each $a \in \{a_{r,j}, a_{s,j}, a_{t,j}\}$ there is a variable part of the $F$-wheels forced to $T$ adjacent to the clause vertex adjacent to $a$, we have (P1) for any 1-in-3-satisfying assignment. 

Further, we observe the collection gadget consists of the variable sets $\{a_{r,j}, a_{s,j}, a_{t,j}\}$, $\{b_{r,j}, b_{s,j}, b_{t,j}\}$, $F$-wheels and a double clause gadget. The truth values of the $F$-wheels are fixed and due to Lemma~\ref{lem:doubleclause_gadget}, there are exactly three 1-in-3-satisfying assignments for the double clause gadget. The remaining variables $\{b_{r,j},b_{s,j},b_{t,j}\}$ are adjacent to a clause adjacent to a variable part of an $F$-wheel forced to $F$ and a variable $x \in \{x_{r,j},x_{s,j},x_{t,j}\}$. Due to Lemma~\ref{lem:doubleclause_gadget}, exactly one variable in $\{x_{r,j},x_{s,j},x_{t,j}\}$ can be set to $T$ and therefore, we conclude (P2) for any 1-in-3-satisfying assignment. 
\end{proof}

With the definitions of the gadgets in mind, we want to puzzle together new \textsc{Sat} instances with guaranteed properties.

\subsection{Parsimonious Reductions}
Recall that a parsimonious reduction preserves the number of satisfying assignments. In this case, we want to preserve the number of 1-in-3-satisfying assignments of \textsc{Positive 1-in-3-Sat}. 
\begin{lem}\label{lem:pars-1-in-3-at-least-two}
There is a parsimonious reduction from \textsc{Positive 1-in-3-Sat} to \textsc{Positive 1-in-3-Sat} with at least two appearances per variable. 
\end{lem}

\begin{proof}
Let $\mathcal{I}=(X,C)$ be an instance of \textsc{Positive 1-in-3-Sat}. 

We obtain an instance $\mathcal{I}'=(X',C')$ of \textsc{Positive 1-in-3-Sat} with at least two appearances per variable as follows. Let $Y \subseteq X$ denote the subset of variables appearing exactly once in $C$. Then, we obtain $C'$ from $C$ by exhaustively replacing each clause that contains a variable in $Y$ with a double clause gadget. Now, every variable appears at least twice since every variable in the double clause gadget appears at least twice. Due to Lemma \ref{lem:doubleclause_gadget} this is a parsimonious reduction.
\end{proof}

On top of the monotonicity of \textsc{Positive 1-in-3-Sat}, we want to construct a reduction so that we can section the set of clauses so that variable vertices neighbor all types of clause vertices.

\begin{lem}\label{lem:pars-3-part-1-in-3}
There is a parsimonious reduction from \textsc{Positive 1-in-3-Sat} with at least two appearances per variable to \textsc{3-Partitioned 1-in-3-Sat}. 
\end{lem}

\begin{proof}
Let $\mathcal{I}=(X,C)$ be an instance of \textsc{Positive 1-in-3-Sat} with at least two appearances per variable.  

We construct an instance $\mathcal{I}'=(X',C')$  of \textsc{3-Partitioned 1-in-3-Sat} from $\mathcal{I}$ as follows. For each variable $x_i \in X$, we replace the $q \geq 2$ appearances of $x_i$ in clauses $c_{x_i,1}, c_{x_i,2}, \ldots, c_{x_i,q} \in C$ with separate new variables $x_{i,1}, x_{i,2}, \ldots, x_{i,q}$. Let $C_\text{orig}$ denote the set of clauses obtained by this replacement operation. Then, for each $i \in \{1,2,\ldots, |X|\}$, we introduce an equality gadget $EQ(X_i)$ for $X_i = \{x_{i,1},\, x_{i,2},\, \ldots,\, x_{i,q}\}$ (see Figure~\ref{fig:equality_gadget}). Thereby, we  introduce the auxiliary variables

\[
A_i = \{a_{i,1},\, a_{i,2},\, \ldots,\, a_{i,q}\} \text{ and } B_i = \{b_{i,1},\, b_{i,2},\, \ldots,\, b_{i,q}\} \text{ for } 1\leq i \leq |X|.
\]
Note that each variable in $A_i \cup B_i$ has exactly two appearances. For each clause $c_j = \{x_r, x_s, x_t\} \in C$, we introduce a collection gadget $CO(c_j)$ (see Figure~\ref{fig:collection_gadget}). This gadget consumes one additional appearance of     
$
a_{r,j},\, b_{r,j},\, a_{s,j},\, b_{s,j},\, a_{t,j} \text{ and } b_{t,j}. 
$
By inspecting the vertices in Figures~\ref{fig:equality_gadget} and~\ref{fig:collection_gadget}, it is straightforward to verify that the constructed graph is cubic and every variable vertex in $X'$ is adjacent to clause vertices with pairwise distinct labels. It now follows that $\mathcal{I'}$ with
\begin{align*}
X' &= \bigcup_{i = 1}^{|X|}\operatorname{var}(EQ(X_i)) \cup \bigcup_{c_j \in C} \operatorname{var}(CO(c_j)) = \bigcup_{i = 1}^{|X|}(X_i \cup A_i \cup B_i) \cup \bigcup_{c_j \in C} \operatorname{var}(CO(c_j)) \text{ and}\\ 
C' &= C_\text{orig} \cup \bigcup_{i = 1}^{|X|}\operatorname{cl}(EQ(X_i)) \cup \bigcup_{c_j \in C} \operatorname{cl}(CO(c_j))
\end{align*}
is an instance of \textsc{3-Partitioned 1-in-3-Sat}. 

Next, we show that $\mathcal{I}$ is a yes-instance if and only if $\mathcal{I'}$ is a yes-instance. 

First, let $\beta\colon X \rightarrow \{T, F\}$ be a truth assignment that 1-in-3-satisfies $C$. We construct a truth assignment $\beta'\colon X' \rightarrow \{T, F\}$ that 1-in-3-satisfies $C'$. For each $x_{i,j} \in X_i$, we set $\beta'(x_{i,j}) = \beta(x_i)$. Clearly, $\beta'$ 1-in-3-satisfies each clause in $C_\text{orig}$. By Lemma~\ref{lem:collection_gadget} (P1), the truth assignment $\beta'(a_{i,j}) = F$ is forced for each $a_{i,j} \in \bigcup_{i=1}^{|X|} A_i$.  Then, all variables in $\bigcup_{i=1}^{|X|} B_i$ also have forced truth values, that is, $\beta'(b_{i,j}) \in \{T, F\}\setminus \{\beta(x_i)\}$. Now, $\beta'$ 1-in-3-satisfies each clause in each equality gadget. For each clause $c_j = \{x_r, x_s, x_t\} \in C$, we have $\beta'(b_{i,j}) = T$ for exactly two $i \in \{r,s,t\}$ since $\beta$ 1-in-3-satisfies $c_j$ and $\beta'(b_{i,j})$ has the opposite truth value of $\beta(x_i)$ for all $1 \leq i \leq |X|$. Then, by Lemma~\ref{lem:collection_gadget} (P2), there exist unique truth values for the auxiliary variables that $\beta'$ needs to assign in each collection gadget such that all contained clauses are 1-in-3-satisfied. Hence, $\beta'$ 1-in-3-satisfies each clause in $C'$.    

Second, let $\beta'\colon X' \rightarrow \{T, F\}$ be a truth assignment that 1-in-3-satisfies $C'$. We define $\beta\colon X \rightarrow \{T, F\}$ by $\beta(x_i) = \beta'(x_{i,1})$. As $\beta'$ 1-in-3-satisfies $C_\text{orig}$, it follows from Lemma~\ref{lem:equality_gadget} that $\beta$ 1-in-3-satisfies $C$.  
 
Finally, we show that the reduction preserves the number of 1-in-3-satisfying truth assignments. Let
\begin{align*}
S &= \{\beta\colon X\rightarrow \{T, F\} \mid \beta \text{ 1-in-3-satisfies } C \} \text { and}\\
S' &= \{\beta'\colon X'\rightarrow \{T, F\} \mid \beta' \text{ 1-in-3-satisfies } C' \}
\end{align*}
As the two directions above describe injective mappings $f\colon S \rightarrow S'$ and $g\colon S' \rightarrow S$, we conclude $|S| = |S'|$. We conclude the proof by remarking that the reduction is polynomial.
\end{proof}

We can generalize this result to its counting variant: The problem of determining the number of solutions of \textsc{3-Partitioned 1-in-3-Sat} is \#\textsc{P}-complete.

\begin{cor}\label{cor:hash3part}
The class \textsc{3-Partitioned 1-in-3-Sat} is \textsc{NP}-complete and its counting variant is \#\textsc{P}-complete.
\end{cor}
\begin{proof}
Lemma~\ref{lem:pars-1-in-3-at-least-two} and Lemma~\ref{lem:pars-3-part-1-in-3} constitute a polynomial time parsimonious reduction from \textsc{Positive 1-in-3-Sat} to \textsc{3-Partitioned 1-in-3-Sat}. Since \textsc{Positive 1-in-3-Sat} is \textsc{NP}-complete by Schaefer's dichotomy theorem~\cite{Schaefer1978} and its counting variant is \#\textsc{P}-complete by Hunt, Marathe, Radhakrishnan and Stearns~\cite[Corollary~3.11]{Hunt1998}, the claim follows. 
\end{proof}

As we have now considered the attributes of `positive' and `3-partitioned', we can move on towering more attributes to enrich the \textsc{Sat} hardness results.

\section{Hardness under Planarity Constraints}\label{sec:hardnessplanarity}
Let $\mathcal{I}=(X,C,P)$ be an instance of \textsc{3-Partitioned 1-in-3-Sat} and $G$ its bipartite incidence graph. Throughout the section, we fix an instance and the embedding of $G$. We have decided to present very detailed proofs and step by step explanations of the constructed gadgets. In later sections, we increasingly shorten proofs relying on gadgets.

\subsection{Gadgets}
We aim to obtain planarity by replacing crossings in the graph embedding with gadgets called \emph{crossover gadgets}.

\begin{figure}[h]
\centering
\includegraphics[width=.95\textwidth]{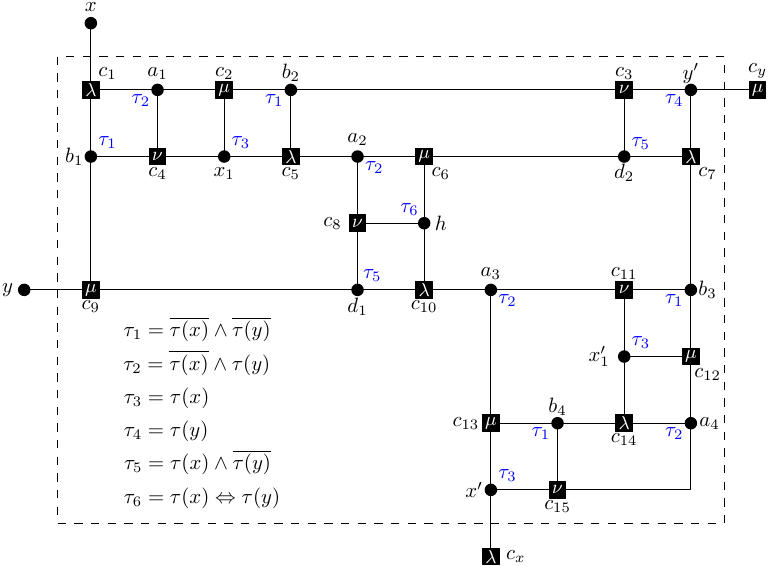}
    \caption{Parsimonious crossover gadget for the case that the crossing edges are incident to clauses of different partitions. The equality verifiers (4-cycles) enforce that subsets of vertices obtain the same truth value. The orthogonal embedding was computed using the graph editor yEd~\cite{yEd2025}.}
\label{fig:crossover_gadget_diff}
\end{figure}

In order to facilitate which variable is connected to which, we call the clause $c_x$ that contains $x$ and correspondingly with $c_y$ that contains $y$. The clauses within the gadget are enumerated as usual.

\paragraph{Crossover gadget (different partitions)} Given an orthogonal embedding  of $G$, the crossover gadget allows to replace a crossing between two edges $\{x, c_x\}$ and $\{y, c_y\}$ with $x,y\in X$ and $c_x,c_y \in C$, where $c_x$ and $c_y$ are contained in different partitions, with a planar construction (see Figure~\ref{fig:crossover_gadget_diff}).

Depending on the partition allocation of the clause vertices, we need a different crossover gadget. We differentiate between the cases whether clause vertices next to a crossing are allocated to different or all to the same partition.

\begin{sidewaysfigure}
    \centering
    \includegraphics[width=\textwidth]{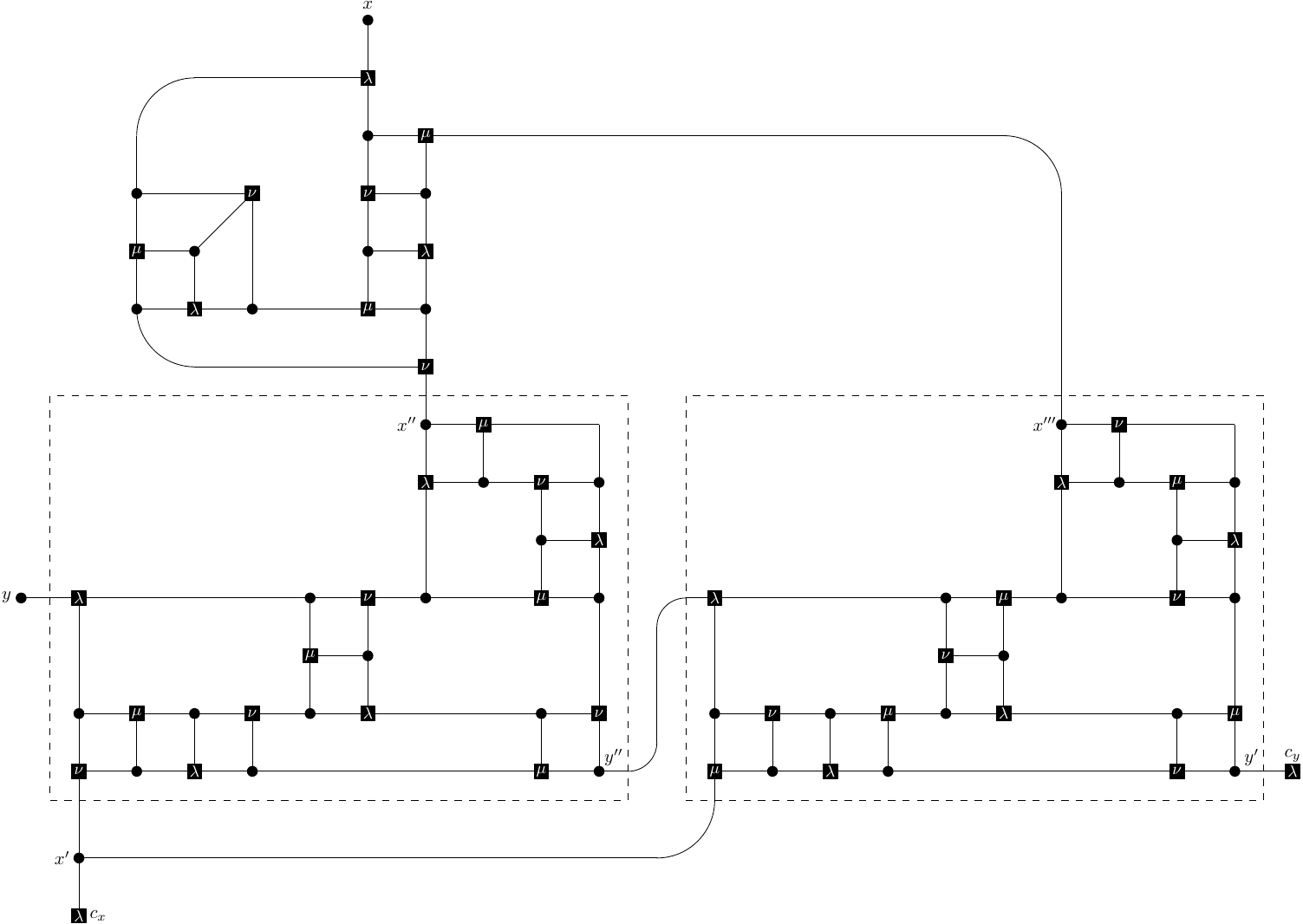}
    \caption{Parsimonious crossover gadget for the same partition}
    \label{fig:crossover_gadget_same}
\end{sidewaysfigure}

\paragraph{Crossover gadget (same partition)} Given an orthogonal embedding  of $G$, the crossover gadget allows to replace a crossing between two edges $\{x, c_x\}$ and $\{y, c_y\}$ with $x,y\in X$ and $c_x,c_y \in C$, where $c_x$ and $c_y$ are contained in the same partition, with a planar construction (see Figure~\ref{fig:crossover_gadget_same}). In order to have unique partition labels when the partition $\lambda \in \{1,2,3\}$ of $c_x$ and $c_y$ is given, we require $\mu < \nu$. The gadget contains two crossover gadgets with input clauses of different partitions encircled in dotted squares.

For each crossover gadget, there are four possible orientations determined by the direction of the horizontal variable \[d_\text{hor} \in \{x \rightarrow x', x' \leftarrow x\}\] and the direction of the vertical variable \[d_\text{ver} \in \{\uparrow_{y}^{y'}, \downarrow_{y'}^{y} \}.\] The direction indicates which side the variable originates from and which side leads to the corresponding clause. For example, $d_\text{hor} = x \rightarrow x'$ means that the variable $x$ originates on the left of the horizontal line segment at the crossing, whereas $d_\text{hor} = x' \leftarrow x$ means that $x$ originates on the right. Analogously, $d_\text{ver}$ indicates the origin of the variable $y$ with respect to the vertical line segment. 

Distinct constructions for each orientation are unnecessary, since all four versions can be obtained by mirroring the gadget along the $x$-axis and/or the $y$-axis.

\begin{lem}\label{lem:crossover_diff}
Let $CR(x,y,\lambda, \mu)$ be a crossover gadget with variables
\[
A = \{x,y\} \cup \{x',y'\} \cup \{x_1, x_1'\} \cup \{a_i, b_i \mid 1\leq i \leq 4\} \cup \{d_1, d_2, h\} 
\]
and clauses $C_A = \{c_j \mid 1\leq j \leq 15\}$ as illustrated in Figure~\ref{fig:crossover_gadget_diff}. Then, the following statements are true:
\begin{enumerate}
\item[(i)] Each truth assignment $\tau\colon \{x,y\} \rightarrow \{T,F\}$ has a unique extension $\beta\colon A \rightarrow \{T,F\}$ that 1-in-3-satisfies each clause in $CR(x,y,\lambda, \mu)$.
\item[(ii)] For each unique extension $\beta$ in (i), we have $\beta(x') = \tau(x)$ and $\beta(y') = \tau(y)$.
\end{enumerate}
\end{lem}

\begin{proof}
We prove (i) and (ii) together. Each 4-cycle is an equality verifier and thus, 
\begin{align*}
\beta(a_1) &= \beta(a_2) = \beta(a_3) =  \beta(a_4), \\
\beta(b_1) &= \beta(b_2) = \beta(b_3) =  \beta(b_4), \\
\beta(d_1) &= \beta(d_2), \\
\beta(x)   &= \beta(x_1), \\
\beta(x')  &= \beta(x'_1). 
\end{align*}
We show that the truth value of each variable in $A \setminus\{x,y\}$ is determined by the truth values $\tau(x)$ and $\tau(y)$ of $x$ and $y$. To this end, let 
\begin{align*}
\tau_1 &= \overline{\tau(x)} \wedge \overline{\tau(y)}, \\ \tau_2 &= \overline{\tau(x)} \wedge \tau(y), \\ \tau_3 &= \tau(x), \\ \tau_4 &= \tau(y), \\ \tau_5 &= \tau(x) \wedge \overline{\tau(y)}, \quad \text{and } \\ \tau_6 &= \tau(x) \Leftrightarrow \tau(y). 
\end{align*}

We show that $\beta(b_1) = \tau_1$ if $\beta$ 1-in-3-satisfies $C_A$. Due to clauses $c_1$ and $c_9$, we have $\beta(b_1) \Rightarrow \tau_1$. Assume towards a contradiction that $\tau_1 \Rightarrow \beta(b_1)$ does not hold. Then, $\beta(x) = \beta(y) = \beta(b_1) = F$. Thus, with clause $c_1$, we obtain $\beta(a_i) = T$ for $i \in \{1,2,3,4\}$. By clause $c_8$, we have $\beta(d_1) = F$. But then, clause $c_9$ is not 1-in-3-satisfied, a contradiction. Hence, also $\tau_1 \Rightarrow \beta(b_1)$ holds. This establishes $\beta(b_1) = \tau_1$ and
\[
\beta(b_1) = \beta(b_2) = \beta(b_3) =  \beta(b_4) = \tau_1.
\]
Next, we consider $a_1$. By clause $c_1$, we have $\beta(a_1) \Leftrightarrow \overline{\tau_3} \wedge \overline{\tau_1} \Leftrightarrow \overline{\tau(x)} \wedge (\tau(x) \vee \tau(y)) \Leftrightarrow \tau_2$. This establishes     
\[
\beta(a_1) = \beta(a_2) = \beta(a_3) =  \beta(a_4) = \tau_2.
\]
By clause $c_9$, we have $\beta(d_1) \Leftrightarrow \overline{\tau_4} \wedge \overline{\tau_1} \Leftrightarrow \overline{\tau(y)} \wedge (\tau(x) \vee \tau(y)) \Leftrightarrow \tau_5 $. Hence, $\beta(d_1) = \beta(d_2) = \tau_5.$ \\

By clause $c_8$, we have $\beta(h) \Leftrightarrow \overline{\tau_5} \wedge \overline{\tau_2} \Leftrightarrow 
(\overline{\tau(x)} \vee \tau(y)) \wedge (\tau(x) \vee \overline{\tau(y)}) \Leftrightarrow (\overline{\tau(x)} \wedge \overline{\tau(y)}) \vee (\tau(x) \wedge \tau(y)) \Leftrightarrow \tau_6$. Hence, $\beta(h) = \tau_6$. 

By clause $c_3$, we have 
\[
\beta(y') \Leftrightarrow \overline{\tau_5} \wedge \overline{\tau_1} \Leftrightarrow (\overline{\tau(x)} \vee \tau(y)) \wedge (\tau(x) \vee \tau(y)) \Leftrightarrow \tau_4.
\]
Hence, $\beta(y') = \tau_4 = \tau(y)$.

By clause $c_{13}$, we have 
\[
\beta(x') \Leftrightarrow \overline{\tau_1} \wedge \overline{\tau_2} \Leftrightarrow (\tau(x) \vee \tau(y)) \wedge (\tau(x) \vee \overline{\tau(y)}) \Leftrightarrow \tau_3.
\]
Hence, $\beta(x') = \tau_3 = \tau(x)$.

This establishes~(ii). Recalling that the equality verifiers enforce $\beta(x_1) = \beta(x) = \tau_3$ and $\beta(x_1') = \beta(x') = \tau_3$, it now follows that the truth values for $x$ and $y$ set by $\beta$ determine the truth values of all variables in $A$. This establishes~(i).  
\end{proof}

Likewise, we prove this for the crossover gadget with same partitions.

\begin{lem}\label{lem:crossover_same}
Let $CR(x,y,\lambda, \lambda)$ be a crossover gadget with variables
\[
A = \{x,y, x'\} \cup \{a_i, w_i \mid 1\leq i \leq 4\} \cup \operatorname{var}(CR(x',y,\nu       , \lambda)) \cup \operatorname{var}(CR(x',y'',\mu, \lambda)) 
\]
and clauses 
\[
C_A = \{c_j \mid 1\leq j \leq 9\} \cup \operatorname{cl}(CR(x',y,\nu         , \lambda)) \cup \operatorname{cl}(CR(x',y'',\mu, \lambda)) 
\]
as illustrated in Figure~\ref{fig:crossover_gadget_same}. Then, each truth assignment $\tau\colon \{x,y\} \rightarrow \{T,F\}$ has a unique extension $\beta\colon A \rightarrow \{T,F\}$ that 1-in-3-satisfies each clause in $CR(x,y,\lambda, \lambda)$.
\end{lem}

\begin{proof}
The clauses $c_7,c_8$ and $c_9$ are an $F$-wheel. Hence, 
\[
\beta(w_1) = T \text{ and } \beta(w_2) = \beta(w_3) = \beta(w_4) = F.
\] 
Next, we deduce the truth values for $\{a_1,a_2, a_3, a_4, x'\}$ in $\{\tau(x), \overline{\tau(x)}, F\}$:  
 
\begin{enumerate}[label=(\arabic*)]
\item By clause $c_1$, we have $\beta(a_1) = \overline{\tau(x)}$.
\item By the equality verifier $\{w_4,a_3,a_4\}, \{a_2,a_3,a_4\}$, we have $\beta(a_2) = \beta(w_4) = F$.
\item By clause $c_3$, we have $\beta(a_3) \Leftrightarrow \overline{\beta(a_1)} \wedge \overline{\beta(a_2)} \Leftrightarrow \tau(x)$. Hence, $\beta(a_3) = \tau(x)$.
\item By the equality verifier $\{a_1,a_2,a_3\}, \{a_4,a_2,a_3\}$, we have $\beta(a_4) = \beta(a_1) = \overline{\tau(x)}$.
\item By Lemma~\ref{lem:crossover_diff} and clause $c_2$, we have $\beta(x') \Leftrightarrow \overline{\beta(a_1)} \wedge \overline{\beta(a_2)} \Leftrightarrow \tau(x)$. Hence, $\beta(x') = \tau(x)$.
\end{enumerate}

By Lemma~\ref{lem:crossover_diff}, if $x'$ has truth value $\tau(x)$ and $y$ has truth value $\tau(y)$, then there exist unique truth values for the variables $\operatorname{var}(CR(x',y,\nu, \lambda))$ such that each clause in $\operatorname{cl}(CR(x',y,\nu, \lambda))$ is 1-in-3-satisfied. In particular, $\beta(x'') = \beta(x') = \tau(x)$ and $\beta(y'') = \beta(y) = \tau(y)$ follow. Again, by Lemma~\ref{lem:crossover_diff} unique truth values for $\operatorname{var}(CR(x',y'',\mu, \lambda))$ follow. In particular, $\beta(x'') = \beta(x') = \tau(x)$ and $\beta(y') = \beta(y'') = \tau(y)$.

Hence, we have $\beta(y') = \tau(y)$ and, by (5), $\beta(x') = \tau(x)$, so $\beta$ is an extension of $\tau$. Further, each variable now has a truth value that either depends on $\tau(x)$ and $\tau(y)$ or is in $\{T,F\}$. Recalling that $\beta(x'') = \beta(x') = \tau(x)$, it is now easy to verify that each clause of the gadget is 1-in-3-satisfied. This establishes~(i).  
\end{proof}
Therefore, we have proven the characteristics for the gadgets so that we can replace crossings with crossover gadgets.

\subsection{Parsimonious Reduction for \textsc{3-Partitioned 1-in-3-Sat}}
The crossover gadgets yield a reduction. In the following, we formalize how to replace the crossing of an associated embedding.
\begin{thm}\label{thm:planar-3-part}
There is a parsimonious polynomial time reduction from \textsc{3-Partitioned 1-in-3-Sat} to \textsc{Planar 3-Partitioned 1-in-3-Sat}.
\end{thm}

\begin{proof}
Let $\mathcal{I}=(X,C,P)$ be an instance of \textsc{3-Partitioned 1-in-3-Sat}. Let $OE_\mathcal{I}$ be an orthogonal embedding of $G$ on a $(|X| + |C|) \times (|X| + |C|)$-grid with at most two bends per edge. Such an embedding exists and can be computed in linear time using the algorithm by Biedl and Kant~\cite{Biedl1998}.

As we later replace crossings with crossover gadgets, we first refine the above grid and scale $OE_\mathcal{I}$ accordingly in order to  make space to modify it by inserting embeddings of gadgets. Clearly, there is a positive integer $s < 20$ such that the embeddings depicted in Figure~\ref{fig:crossover_gadget_diff} and Figure~\ref{fig:crossover_gadget_same}, respectively, fit on a grid of size $(2s+1) \times (2s+1)$ such that $x$, $y$, $c_x$ and $c_y$ enter the grid in the middle of their respective sides of the gadget. We obtain another orthogonal embedding $OE_{\mathcal{I},s }$ of $G(X,C)$ from $OE_\mathcal{I}$ by multiplying the coordinates of vertices, bends and crosses by $2s+1$ and scaling the line segments accordingly. 

We now obtain a new instance $\mathcal{I}'=(X', C',P')$ of \textsc{3-Partitioned 1-in-3-Sat} as follows. Let $(q,r)$ be integer coordinates of a crossing in $OE_{\mathcal{I},s}$. Let $x,y$ be the two variables crossing at $(q,r)$ with
\[
d_\text{ver} \in \{x \rightarrow x', x' \leftarrow x\} \text{ and } d_\text{hor} \in \{\uparrow_{y}^{y'}, \downarrow_{y'}^{y} \}.
\]
Here, $d_\text{ver}$ indicates whether the variable $x$ origins from the left of the horizontal line segment at the crossing, i.e.\ $d_\text{ver} = x \rightarrow x'$, or from its right side, i.e.\ $d_\text{ver} = x' \leftarrow x$. Analogously, $d_\text{hor}$ indicates the origin of the variable $y$ with respect to the vertical line segment. Let $c_x$ and $c_y$ be the clauses in $C$ such that $\{x,c_x\}$ and $\{y,c_y\}$ are edges in $G$ that cross at $(q,r)$ in $OE_{\mathcal{I},s}$. We now replace the part of the embedding that has the size of the embeddings of the crossover gadgets on a $(2s+1) \times (2s+1)$-grid with $(q,r)$ in its center with the appropriate crossover gadget. That is, if $c_x$ and $c_y$ belong to different partitions in $C = \bigcup_{i=1}^3 P_i$, we use the gadget depicted in Figure~\ref{fig:crossover_gadget_diff} and, otherwise, we use the gadget depicted in Figure~\ref{fig:crossover_gadget_same}. In both cases, the partition labels $\lambda$, $\mu$ and $\nu$ of the crossover gadget are determined by the partition labels of $c_x$ and $c_y$---in the latter case due to the additional requirement $\mu < \nu$. Further, the orientation of the crossover gadget is determined by $d_\text{ver}$ and $d_\text{hor}$. Let $X'$ and $C'$ denote the variables and clauses of the resulting instance. As each variable vertex in the crossover gadget is adjacent to three clauses with pairwise different partition labels in $\{1,2,3\}$ and the gadget preserves this property for $x$ and $y$, the clauses $C'$ decompose into three partitions of $X'$. The replacement of a crossing as alluded to above yields an orthogonal embedding $OE_{\mathcal{I}',s}$ of $G'$ with one less crossing than $OE_{\mathcal{I},s}$. Furthermore, the remaining crossings do not require scaling again, as the replacement was local to the crossing at $(q,r)$. By Lemmas~\ref{lem:crossover_diff} and~\ref{lem:crossover_same}, the reduction from $\mathcal{I}$ to $\mathcal{I'}$ is parsimonious. 

By repeating the above reduction, we obtain  a parsimonious reduction from $\mathcal{I}$ to an instance $\mathcal{I}_\text{plan}$ of \textsc{Planar 3-Partitioned 1-in-3-Sat}. Since there are less than $(|X| + |C|)^2$ crossings in $OE_\mathcal{I}$, we repeat the above reduction at most $(|X| + |C|)^2$ times. Noting that the refining of the grid and scaling was done only once, the embedding of the final orthogonal embedding---which is also planar---is on a grid of size 
\[
(2s+1)(|X| + |C|) \times (2s+1)(|X| + |C|).
\]
The theorem now follows.
\end{proof}

Similar to the reasoning in Corollary \ref{cor:hash3part}, we can immediately conclude that:

\begin{cor}
\textsc{Planar 3-Partitioned 1-in-3-Sat} is \textsc{NP}-complete and its counting variant is \#\textsc{P}-complete.
\end{cor}
Here, we conclude the section on planarity and turn our attention to a different property.

\section{Hardness under Linearity Constraints}\label{sec:hardnesslinearity}
Recall that the formula is called \emph{linear} if each pair of distinct clauses shares at most one variable. 
\subsection{Gadgets}
\begin{figure}[h]
    \centering
    \includegraphics[width=\textwidth]{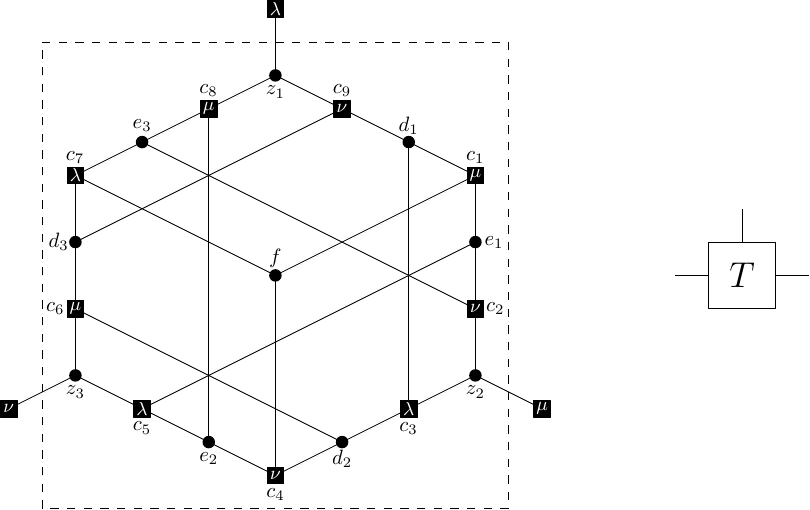}
    \caption{$T$-square and its abbreviation.}
    \label{fig:t_square}
\end{figure}

\paragraph{$T$-square} The $T$-square consists of variables $z_1, z_2, z_3$, and $f$, as well as six auxiliary variables and nine linear clauses as illustrated in Figure \ref{fig:t_square}. The variables $z_1$, $z_2$, and $z_3$ are forced to take the truth value~$T$ in any 1-in-3-satisfying truth assignment.

\begin{figure}[h]
    \centering
    \includegraphics[width=.7\textwidth]{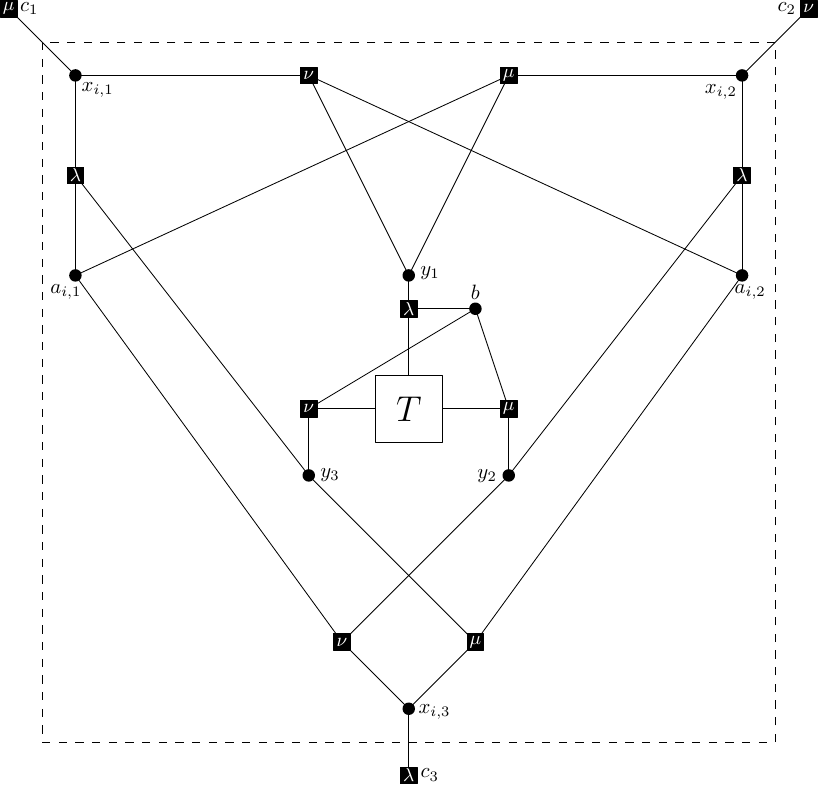}
    \caption{Linearity gadget for three partitions.}
    \label{fig:lin_3}
\end{figure}

\paragraph{Linear equality gadget} The linear equality gadget $LEQ(X_i)$ for a set of variables $X_i = \{x_{i,1}, x_{i,2}, x_{i,3}\}$ ensures the three variables to have the same truth value in any 1-in-3-satisfying truth assignment. The gadget simulates a variable vertex and contains a $T$-square together with the variables $y_1, y_2, y_3, b$ and three additional linear clauses, two auxiliary variables $a_{i,1}, a_{i,2}$, and six additional linear clauses as illustrated in Figure \ref{fig:lin_3}. The variables $y_1$, $y_2$, and $y_3$ are forced to take the truth value $F$ in any 1-in-3-satisfying truth assignment.  \\

\begin{lem}\label{lem:t_square}
Let $TS$ be a $T$-square with variables
\[ A = \{z_1,z_2,z_3\} \cup \{f\} \cup \{d_i, e_i \mid 1 \le i \le 3\} \]
and clauses $C_A = \{c_j \mid 1 \le j \le 9\}$ as illustrated in Figure \ref{fig:t_square}. Then there is exactly one truth assignment $\beta\colon A \rightarrow \{T, F\}$ that 1-in-3-satisfies each clause in $C_A$. This truth assignment satisfies $\beta(z_1) = \beta(z_2) = \beta(z_3) = T$.
\end{lem}

\begin{proof} 
First, we prove that there is no truth assignment $\beta\colon A \rightarrow \{T, F\}$ with $\beta(f) = F$ that 1-in-3-satisfies each clause in~$C_A$. Assume towards a contradiction that there exists a satisfying truth assignment with $\beta(f) = F$. To 1-in-3-satisfy clause $c_1$, either $\beta(d_1) = T$ or $\beta(e_1) = T$. We assume $\beta(d_1) = T$. Then, with clauses $c_3$ and $c_9$, we obtain $\beta(d_2) = \beta(d_3) = F$. With clause~$c_7$ we then obtain $\beta(e_3) = T$ which, by clause $c_8$, gives us $\beta(e_2) = F$. But then, clause $c_4$ is not 1-in-3-satisfied, a contradiction. Now assume $\beta(e_1) = T$. Then, with clauses $c_5$ and $c_2$, we obtain $\beta(e_2) = \beta(e_3) = F$. With clause~$c_7$ we then obtain $\beta(d_3) = T$ which, by clause $c_6$, gives us $\beta(d_2) = F$. But then, clause $c_4$ is not 1-in-3-satisfied, leading to another contradiction.  \\

As there is no satisfying truth assignment with $\beta(f) = F$, consider a truth assignment with $\beta(f) = T$. Through clauses $c_1, c_4, c_7$ this forces $\beta(d_i) = \beta(e_i) = F$ for $i \in \{1,2,3\}$, which by the remaining clauses forces $\beta(z_1) = \beta(z_2) = \beta(z_3) = T$. This truth assignment 1-in-3-satisfies every clause in $C_A$, and as every assignment is forced, it is the only one to do so. 
\end{proof}

A $T$-square can be extended to a linear equality gadget by adding only a few elements. Namely, let $LEQ(X_i)$ be a linear equality gadget built around a $T$-square $TS$ as illustrated in Figure~\ref{fig:lin_3}. Its variables are
\[ \operatorname{var}(LEQ(X_i)) = X_i \cup \{a_{i,1}, a_{i,2}\} \cup \{y_1,y_2,y_3\} \cup \{b\} \cup \operatorname{var}(TS). \]
Besides the clauses of $TS$, it contains the three clauses $\{y_1, b, z_1\}$, $\{y_2, b, z_2\}$, $\{y_3, b, z_3\}$, which together with $TS$ form the $F$-square $FS$, and six clauses that each contain one variable of $X_i$, one of $a_{i,1}, a_{i,2}$, and one of $y_1, y_2, y_3$. Due to the forced assignment of the $T$-square's variables $z_1,z_2,z_3$ to $T$, as shown by Lemma~\ref{lem:t_square}, the three clauses added by the $F$-square can be 1-in-3-satisfied if and only if $\beta(b) = \beta(y_1) = \beta(y_2) = \beta(y_3) = F$. Subsequently, there is exactly one truth assignment $\beta\colon \operatorname{var}(FS) \rightarrow \{T, F\}$ that 1-in-3-satisfies each clause in $\operatorname{cl}(FS)$. \\

\begin{lem}\label{lem:linear_3} Let $LEQ(X_i)$ be a linear equality gadget for a set of variables $X_i = \{x_{i,1}, x_{i,2}, x_{i,3}\}$ as illustrated in Figure \ref{fig:lin_3}. Then, there are exactly two truth assignments $\beta_1$ and~$\beta_2$ for $\operatorname{var}(LEQ(X_i))$ that 1-in-3-satisfy each clause in $\operatorname{cl}(LEQ(X_i))$. Further, $\beta_1$ and $\beta_2$ satisfy 
\begin{align*}
\beta_1(x_{i,1}) = \beta_1(x_{i,2}) = \beta_1(x_{i,3}) = T, && \beta_1(a_{i,1}) = \beta_1(a_{i,2}) = F, \\
\beta_2(x_{i,1}) = \beta_2(x_{i,2}) = \beta_2(x_{i,3}) = F, && \beta_2(a_{i,1}) = \beta_2(a_{i,2}) = T.
\end{align*}
and all clauses are linear.
\end{lem}

\begin{proof} The only assignment that 1-in-3-satisfies the $F$-square forces its variables $y_1,y_2,y_3$ to $F$. This forces the following to be true for any 1-in-3-satisfying truth assignment $\beta$ for $LEQ(X_i)$: 
\begin{align*}
& (\beta(a_{i,1}) \oplus \beta(x_{i,1})) \wedge (\beta(a_{i,1}) \oplus \beta(x_{i,2})) \wedge (\beta(a_{i,1}) \oplus \beta(x_{i,3})) \\ \wedge & (\beta(a_{i,2}) \oplus \beta(x_{i,1})) \wedge (\beta(a_{i,2}) \oplus \beta(x_{i,2})) \wedge (\beta(a_{i,2}) \oplus \beta(x_{i,3})).
\end{align*}
Hence, 
\[
\beta(a_{i,k}) \ne \beta(x_{i,j}) \text{ for } k \in \{1,2\} \text{ and } j \in \{1,2,3\},
\]
which implies 
\begin{align*}
& \beta(a_{i,1}) = \beta(a_{i,2}) \text{, and} \\
& \beta(x_{i,1}) = \beta(x_{i,2}) = \beta(x_{i,3}).
\end{align*}
There are only two possible truth assignments that satisfy this, $\beta_1$ and $\beta_2$. 

Furthermore, all clauses in $\operatorname{cl}(LEQ(X_i))$ are linear if and only if for every $v \in \operatorname{var}(LEQ(X_i))$ the cardinality of the union of clauses containing $v$ is equal to seven. This is due to $v$ appearing in three clauses with three variables each, where the only overlap between the clauses is $v$, reducing the number of distinct variables by two (one appearance of $v$ remains). Hence, 
$$| \{ u \in c_v \mid c_v \in \operatorname{cl}(LEQ(X_i)) \wedge v \in c_v \} | = 3 \cdot 3 - 2 =7.$$ 
Keeping in mind that the clauses $c_1, c_2$ and $c_3$ outside the gadget are distinct and cannot contain another variable of $\operatorname{var}(LEQ(X_i))$ besides $x_{i,1}, x_{i,2}$ and $x_{i,3}$, respectively, this property can be verified for all 19 variables appearing in $LEQ(X_i)$. 
\end{proof}

Analogously, we can construct a parsimonious reduction for four partitions that can be found in the Appendix.

\begin{lem}\label{lem:lin_3_p} There is a parsimonious reduction from \textsc{3-Partitioned 1-in-3-Sat} to \textsc{Linear 3-Partitioned 1-in-3-Sat}. 
\end{lem}

\begin{proof} Let $\mathcal{I} = (X,C)$ be an instance of \textsc{3-Partitioned 1-in-3-Sat}. 

We construct an instance $\mathcal{I}' = (X', C')$ of \textsc{Linear 3-Partitioned 1-in-3-Sat} from $\mathcal{I}$ as follows. For each variable $x_i \in X$, we replace the three appearances of $x_i$ with separate new variables $x_{i,1}, x_{i,2}, x_{i,3}$. Let $C_\text{orig}$ denote the set of clauses obtained by this replacement operation. Then, for each $i \in \{1,2, \ldots, |X|\}$ we introduce a linear equality gadget $LEQ(X_i)$ (see Figure~\ref{fig:lin_3}).

Because of the replacement, no variable appearing in a clause $c \in C_\text{orig}$ also appears in another clause~$c'~\in~C_\text{orig}$. As Lemma \ref{lem:linear_3} shows, all clauses in $\operatorname{cl}(LEQ(X_i))$ are linear. Hence, all clauses in $C_\text{orig} \cup \bigcup^{|X|}_{i=1} \operatorname{cl}(LEQ(X_i))$ are linear. It now follows that $\mathcal{I}'=(X',C')$ with $X' = \bigcup^{|X|}_{i=1} \operatorname{var}(LEQ(X_i))$ and $C' = C_\text{orig} \cup \bigcup^{|X|}_{i=1} \operatorname{cl}(LEQ(X_i))$ is an instance of \textsc{Linear 3-Partitioned 1-in-3-Sat}.

Let $\beta\colon X \rightarrow \{T, F\}$ be a truth assignment that 1-in-3-satisfies $C$. We construct a truth assignment $\beta'\colon X' \rightarrow \{T, F\}$ that 1-in-3-satisfies $C'$. For each $x_{i,j}$ introduced by the replacement of~$x_i$, we set $\beta'(x_{i,j}) = \beta(x_i)$. Clearly, $\beta'$ 1-in-3-satisfies each clause in $C_\text{orig}$. By Lemma \ref{lem:linear_3}, if the truth assignment $\beta'(x_{i,j})$ is given, $\beta'(a_{i,k}) = \overline{\beta'(x_{i,j})}$ is forced, and all clauses in $LEQ(X_i)$ are 1-in-3-satisfied. Hence, $\beta'$ 1-in-3-satisfies each clause in $C'$. Second, let $\beta'\colon X' \rightarrow\{T, F\}$ be a truth assignment that 1-in-3-satisfies~$C'$. We define $\beta\colon X \rightarrow \{T, F\}$ by $\beta(x_i) = \beta'(x_{i,1} )$. As $\beta'$ 1-in-3-satisfies $C_\text{orig}$, it follows from Lemma \ref{lem:linear_3} that $\beta$ 1-in-3-satisfies $C$. 

We conclude that the reduction is parsimonious since there are exactly as many assignments that 1-in-3-satisfy $C$ as there are for $C'$ and that the construction is polynomial.
\end{proof}

The concept of the construction used in the reduction presented in Lemma \ref{lem:lin_3_p} can be extended to four partitions with similar gadgets. The corresponding gadgets and proofs of their properties can be found in the Appendix. 

\begin{lem}\label{lem:lin_4_p} There is a parsimonious reduction from \textsc{3-Partitioned 1-in-3-Sat} to \textsc{Linear 4-Partitioned 1-in-3-Sat}.
\end{lem}

\begin{proof} Let $\mathcal{I} = (X,C)$ be an instance of \textsc{3-Partitioned 1-in-3-Sat}. 

We construct an instance $\mathcal{I}' = (X', C')$ of \textsc{Linear 4-Partitioned 1-in-3-Sat} from $\mathcal{I}$ as follows. First, we create a fourth partition by duplicating all clauses of one of the partitions in $\mathcal{I}$. Every variable now appears in four different partitions, and the number of satisfying truth assignments remains unchanged, as the clauses added are simply a duplication of clauses already existing in $C$. The rest of the proof is analogous to the proof of Lemma \ref{lem:lin_3_p}. 
\end{proof}

\subsection{\textsc{NP}-completeness and \#\textsc{P}-completeness for $k \ge 3$ partitions}

Having established hardness for $k=3$ and $k=4$ partitions, we now consider greater $k$ and show that hardness holds for all $k\ge3$ partitions. The proof is carried out by reducing from \textsc{Linear $k$-Partitioned 1-in-3-Sat} to \textsc{Linear $(k+2)$-Partitioned 1-in-3-Sat}.

\begin{thm}\label{thm:Linear$k$-Partitioned1-in-3-Sat}
\textsc{Linear $k$-Partitioned 1-in-3-Sat} is \textsc{NP}-complete and its counting variant is \#\textsc{P}-complete for $k \ge 3$.
\end{thm}

Consider an instance $\mathcal{I}$ of \textsc{Linear $k$-Partitioned 1-in-3-Sat} with variable set $X = \{x_1,\ldots,x_n\}$ and partitions $P_1, \ldots, P_k$. Note that $n$ must be divisible by three because otherwise, $X$ could not be divided into pairwise disjoint clauses of three variables that constitute each partition. W.l.o.g.\ we may assume that
\begin{align*}
P_1 &= \{\{x_{3j-2}, x_{3j-1}, x_{3j}\} \mid j \in \{1,2,\ldots,\frac{n}{3}\}\} \\
&= \{\{x_1, x_2, x_3\},\{x_4, x_5, x_6\}, \ldots, \{x_{n-2},x_{n-1},x_n\}\}.
\end{align*}
Let $\mathcal{I}'$ and $\mathcal{I}''$ be two copies of $\mathcal{I}$ with variable set $Y$ and $Z$, respectively. Then, for example, we have 
\begin{align*}
P'_1 &= \{\{y_{3j-2}, y_{3j-1}, y_{3j}\} \mid j \in \{1,2,\ldots,\frac{n}{3}\}\} \\
&= \{\{y_1, y_2, y_3\},\{y_4, y_5, y_6\}, \ldots, \{y_{n-2},y_{n-1},y_n\}\}, \\ 
P''_1 &= \{\{z_{3j-2}, z_{3j-1}, z_{3j}\} \mid j \in \{1,2,\ldots,\frac{n}{3}\}\} \\
&= \{\{z_1, z_2, z_3\},\{z_4, z_5, z_6\}, \ldots, \{z_{n-2},z_{n-1},z_n\}\}.
\end{align*}
Then, $\mathcal{I}$ is a yes-instance if and only if the following instance $\mathcal{I}'''$ of \textsc{Linear $(k+2)$-Partitioned 1-in-3-Sat} is a yes-instance. 

\begin{align*}
P'''_r &= P_r \cup P'_r \cup P''_r, \quad 1 \leq r \leq k, \\
P'''_{k+1} &= \{ 
\{ x_{3j-2}, y_{3j-1}, z_{3j} \}, \{ y_{3j-2}, z_{3j-1}, x_{3j} \}, \{ z_{3j-2}, x_{3j-1}, y_{3j} \} \mid j \in \{1,2,\ldots,\frac{n}{3}\} 
\} \\
&= \{\{x_1,y_2,z_3\}, \{y_1, z_2, x_3\}, \{z_1,x_2,y_3\},\ldots\}, \\
P'''_{k+2} &= \{ 
\{ x_{3j-2}, z_{3j-1}, y_{3j} \}, \{ y_{3j-2}, x_{3j-1}, z_{3j} \}, \{ z_{3j-2}, y_{3j-1}, x_{3j} \} \mid j \in \{1,2,\ldots,\frac{n}{3}\} 
\} \\
&= \{\{x_1,z_2,y_3\}, \{y_1, x_2, z_3\}, \{z_1,y_2,x_3\},\ldots\}
\end{align*}

\begin{proof}
First, let $\mathcal{I}$ be 1-in-3-satisfiable. Then, there is a truth assignment $\tau$ that sets exactly one variable to $T$ in each clause of $\mathcal{I}$, including those in $P_1$. This truth assignment forces the truth assignments $\tau'$ and $\tau''$ for $Y$ and $Z$:

If $\tau(x_{3j-2}) = T$ then, due to $P_1 \subset P_1'''$, $\tau(x_{3j-1}) = \tau(x_{3j}) = F$. The clauses in $P_{k+1}'''$ and $P_{k+2}'''$ force $\tau'(y_{3j-1}) = \tau'(y_{3j}) = \tau''(z_{3j-1}) = \tau''(z_{3j}) = F$. To 1-in-3-satisfy $P_1' \subset P_1'''$ and $P_1'' \subset P_1'''$, this forces $\tau(y_{3j-2}) = \tau(z_{3j-2}) = T$.

The same holds for the other indices $3j-1$ and $3j$. Hence, let $\tau'$ and $\tau''$ be truth assignments for~$Y$ and $Z$ defined by $\tau(x_i) = \tau'(y_i) = \tau''(z_i)$. This yields a truth assignment $\tau'''$ for $X \cup Y \cup Z$ that 1-in-3-satisfies $P'''_{k+1}$ and $P'''_{k+2}$. It also obviously satisfies $P'''_r$ for $1 \le r \le k$. \\

Now, let $\mathcal{I}'''$ be satisfiable. Then, there is a truth assignment $\tau'''$ that 1-in-3-satisfies each clause of~$\mathcal{I}'''$. Since the clauses of $\mathcal{I}$ are a subset of the clauses of $\mathcal{I}'''$, the truth assignment $\tau$ with $\tau(x_i) = \tau'''(x_i)$ also 1-in-3-satisfies $\mathcal{I}$. \\

Finally, we show that the reduction preserves the number of 1-in-3-satisfying truth assignments. Let
\begin{align*}
S &= \{\tau\colon X\rightarrow \{T, F\} \mid \tau \text{ 1-in-3-satisfies } P_1 \cup \dots \cup P_k \} \text { and}\\
S' &= \{\tau'''\colon X \cup Y \cup Z \rightarrow \{T, F\} \mid \tau''' \text{ 1-in-3-satisfies } P_1''' \cup \dots \cup P_k''' \cup P_{k+1}''' \cup P_{k+2}''' \}
\end{align*}
As the two directions above describe injective mappings $f\colon S \rightarrow S'$ and $g\colon S' \rightarrow S$, we conclude $|S| = |S'|$. 

We conclude the proof by remarking that the reduction is polynomial.

With this, we obtain \textsc{NP}- and \#\textsc{P}-completeness for all odd $k \ge 3$ and all even $k \ge 4$. Lemmas \ref{lem:lin_3_p} and \ref{lem:lin_4_p} prove \textsc{NP}- and \#\textsc{P}-completeness for $k = 3$ and $k = 4$, respectively. Hence, we can deduce \textsc{NP}- and \#\textsc{P}-completeness for all $k \ge 3$.
\end{proof}

\begin{cor}\label{cor:$k$-Partitioned 1-in-3-Sat}
The problem \textsc{$k$-Partitioned 1-in-3-Sat} is \textsc{NP}-complete and its counting variant is \#\textsc{P}-complete for $k \ge 3$.
\end{cor}

\section{Hardness under Connectivity Constraints}\label{sec:hardnessconnectivity}

In the following, we extend the \textsc{3-Partitioned 1-in-3-Sat} hardness result to \textsc{Sat} formulas with a 3-connected incidence graph. We reduce from \textsc{3-Partitioned 1-in-3-Sat} in several steps. First, we add a grid structure to the incidence graph. Then we replace the grid vertices with crossover gadgets and obtain a 3-connected \textsc{Sat} graph. Finally, we replace the crossings of the \textsc{Sat} graph to obtain planarity.\\

Let $\mathcal{I}=(X,C,P)$ be an instance of \textsc{3-Partitioned 1-in-3-Sat}. So far, we have indexed clauses according to the variables they contain. For instance, we indexed the clause $c_{x_i}$ with $x_i$ if the clause contained the variable $x_i$ and added a second index to the variable for different occurrences. From now on, we allow a different, additional notation, and index variables according to the clauses in which they are contained.\\
Let the number of variables be $n$. Without loss of generality, we fix an order of the variables in a clause $c_i = \{x_{c_i,1}, x_{c_i,2},x_{c_i,3}\}$, $1 \leq i \leq |C|$, and we also fix an arbitrary order of the clauses. We index the variables in the clause according to their containment in the clause (first index) and according to their position in the clause (second index). This enables us to assign a number, the \textit{position}, to each variable appearance as follows: 

\[
\begin{array}{cccccccccc}
(x_{c_1,1}, & x_{c_1,2}, & x_{c_1,3}), & (x_{c_2,1}, & x_{c_2,2}, & x_{c_2,3}), & \cdots & (x_{c_n,1}, & x_{c_n,2}, & x_{c_n,3}) \\
1 & 2 & 3 & 4 & 5 & 6 & \cdots & 3n-2 & 3n-1 & 3n
\end{array}
\]

Since each variable appears in exactly three clauses in $C$, each $x_{i} \in X$ corresponds to exactly three numbers between 1 and $3n$ (that is, the exact positions where $x_i$ appears). Let $\operatorname{pos}(i,1)$, $\operatorname{pos}(i,2)$ and $\operatorname{pos}(i,3)$ denote the positions of $x_i$. Since the variables occur three times and the ordering of the same variable has no impact on the formula, we may assume that $\operatorname{pos}(n,1) > \operatorname{pos}(n,2) > \operatorname{pos}(n,3)$.

We define the graph~$Q= (V^Q,E^Q)$ by
\begin{align*}
  V^Q   &= \{q_{i,j} \mid 1 \leq i \leq 3n,\ 1 \leq j \leq 3n\}, \\
  E^Q &= \bigl\{\{q_{i,j}, q_{i,j+1}\} \mid 1 \leq i \leq 3n,\ 1 \leq j < 3n\bigr\} \\
      &\quad \cup \bigl\{\{q_{i,j}, q_{i+1,j}\} \mid 1 \leq i < 3n,\ 1 \leq j \leq 3n\bigr\}.
\end{align*}
Using the above notation, we define the graph~$G_{\text{grid}}(X,C,P)=(V^{\text{grid}},E^{\text{grid}})$ by
\begin{align*}
V^{\text{grid}} &= X \cup C \cup V^Q,\\
E^{\text{grid}} &=  E^Q \cup \{\{x_i, q_{3i-2,1}\}, \{x_i, q_{3i-1,1}\}, \{x_i, q_{3i,1}\} \mid x_i \in X \}\\
& \cup \{\{c_j, q_{1,3j-2}\}, \{c_j, q_{1,3j-1}\}, \{c_j, q_{1,3j}\} \mid c_j \in C \}\\
& \cup \{\{q_{3i-2,3n}, q_{3n,\operatorname{pos}(i,1)}\}, \{q_{3i-1,3n}, q_{3n,\operatorname{pos}(i,2)}\}, \{q_{3i,3n}, q_{3n,\operatorname{pos}(i,3)}\} \mid x_i \in X  \}.
\end{align*}
\begin{figure}[h]
    \centering
 \includegraphics[width=.5\textwidth]{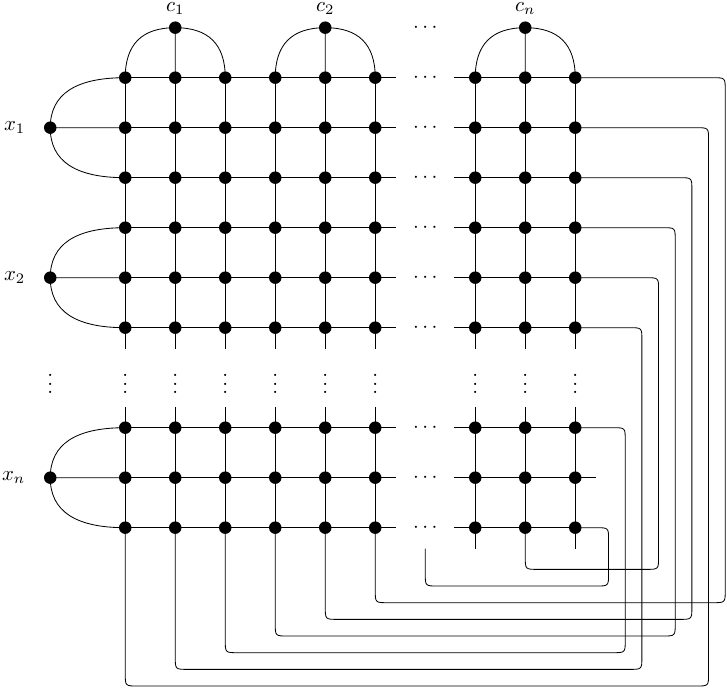}
    \caption{Illustration of $G_{\text{grid}}(\mathcal{I})$}
    \label{fig:3-conn_gadget-grid}
\end{figure}

We refer to the vertices in~$X$ as variable vertices, and to the vertices in~$C$ as clause vertices. 

We fix a drawing for $G_{\text{grid}}(\mathcal{I})$ such that without loss of generality two distinct edges cross at most once as shown in Figure~\ref{fig:3-conn_gadget-grid}. This drawing determines an orientation and thus induces the notions of leftmost and
rightmost column as well as top and bottom row of the grid. The edge set~$E^{\text{grid}}$ then consists of the vertical and horizontal edges of the grid~$Q$, edges connecting the clause vertices in~$C$ to the top row, edges connecting the variable vertices in~$X$ to the leftmost column, and edges connecting the vertices of the rightmost column to the bottom row of the grid. The latter depend on the formula instance, which determines how the loose ends on the right of the grid are joined to those below it. The figure therefore only indicates these edges schematically. Moreover, the condition $\operatorname{pos}(n,1) > \operatorname{pos}(n,2) > \operatorname{pos}(n,3)$ ensures that the corner vertex $q_{3n, 3n}$ is adjacent to a vertex $q_{3n, j}$ with $j \leq 3n-2$, which is used in the proof for 3-connectivity in the following lemma. Also note that all vertices in~$Q$ have degree four. 

\begin{lem}\label{lem:gridis3con}
The graph $G_{\text{grid}}(\mathcal{I})$ is 3-connected.
\end{lem}

\begin{proof}
It is easy to see (e.g., by using {\v{S}}pacapan's connectivity formula~\cite[Theorem~1]{Spacapan2008}) that grids are 2-connected, hence, so is~$Q$. We show 3-connectedness of \(G_{\text{grid}}(\mathcal{I})\) by arguing that the deletion of at most two vertices in $G_{\text{grid}}(\mathcal{I})$ leaves $G_{\text{grid}}(\mathcal{I})$ connected.  Observe that if \(S\subseteq V^Q \), \(|S|=2\), and \(Q\setminus S\) is not connected, then \(S\) consists of the two neighbors of a corner of
\(Q\): Choose a row \(R'\) and a column \(C'\) avoiding \(S\), then \(R'\cup C'\) is connected in
\(Q \setminus S\). Let \(v\) lie in another component. The horizontal path from
$v$ to \(C'\) and the vertical path from \(v\) to \(R'\) must each
contain one vertex of \(S\). If $v$ has an adjacent neighbor that is not in $S$, there exists a path from $v$ to $C' \cup R'$, which is a contradiction. Hence, \(v\) is a corner and the two vertices of \(S\) must be its two neighbors.

We distinguish cases according to the set $S \subseteq V^{\text{grid}}$, with $|S|\leq 2$. 
If~\(|S\cap V^Q |\leq 1\), then \(Q\setminus (S\cap V^Q )\) is connected since $Q$ is 2-connected. Every vertex of
\((X\cup C)\setminus S\) has three distinct neighbors in \(V^Q\), at
least one of which remains after deleting \(S\). Hence, \(G_{\mathrm{grid}}(\mathcal{I})\) is connected. 

If \(|S\cap V^Q| = 2\) and ~\(Q\setminus (S\cap V^Q)\) is not connected, by the observation above, \(S\) consists of the two neighbors of a corner. The corners
\(q_{1,1}\), \(q_{1,3n}\), and \(q_{3n,1}\) remain connected to the
rest of the grid via~\(x_1\), \(c_n\), and \(x_n\), respectively.
Finally, \(q_{3n,3n}\) is adjacent to
\(q_{3n,\operatorname{pos}(n,3)}\), and
\(\operatorname{pos}(n,3)\leq 3n-2\), so this vertex is not deleted.
Hence, \(G_{\mathrm{grid}}(X,C,P)\setminus S\) is connected in all cases.
\end{proof}

In the following, we replace all grid vertices in $G_{\text{grid}}(\mathcal{I})$ with a
crossover gadget (see the opposing-partition gadget in Figure~\ref{fig:crossover_gadget_diff} and the equal-partition gadget in Figure~\ref{fig:crossover_gadget_same}). Recall that
a crossover gadget contains two input vertices, which are clause vertices, and two output
vertices, which are variable vertices. Each vertex has two edges that point towards vertices in the gadget and a free edge that can be attached to a graph. In $G_{\text{grid}}(\mathcal{I})$, each grid vertex has degree four, and we call its edges
the upper, the lower, the left and the right edge, respectively. We replace each grid vertex with a
crossover gadget such that the left and the lower edge are the free edges of the two input vertices and the upper and the right edge are the free edges of the two output vertices.

\begin{lem}\label{lem:possiblecrosspart}
There is a unique choice of a gadget (equal partition or opposing partition) at each grid
vertex such that the resulting graph is the incidence graph of an instance of \textsc{3-Partitioned 1-in-3-Sat}.
\end{lem}

\begin{proof}
By construction of the parsimonious crossover gadgets for equal and opposing partitions,
the right edge of a crossover gadget must lead to a clause vertex of the same partition as
its left input vertex. Similarly, the upper edge of a crossover gadget must lead to a
clause vertex of the same partition as its lower input vertex. Hence, all left input
vertices in a row have the same partition, and all lower input vertices in a column have
the same partition.

Since the partitions of the clause vertices~$c_j \in C$, which are adjacent to the vertices of the first row, are
given by the instance $\mathcal{I}$, the partition propagates to all lower input vertices
in the same column. Since the lower edges of the gadgets in the last row are adjacent  to
the right edges of the gadgets in the rightmost column, the partition further propagates
to all left input vertices in every row. Hence, each grid vertex $q_{i,j}$ is crossed
horizontally and vertically with a determined partition, and the gadget is forced: equal
partitions yield the equal-partition gadget, opposing partitions the opposing-partition gadget.

It remains to show that the resulting graph is a valid instance of
\textsc{3-Partitioned 1-in-3-Sat}. Inside each crossover gadget, every variable already
appears in exactly three clauses from pairwise distinct partitions, except for the output
vertices. The right (upper) output vertex lacks exactly the partition of the left (lower) input vertex.
Furthermore, each variable~$x_i \in X$ is adjacent to the left input vertices of the gadgets replacing $q_{3i-2,1}$, $q_{3i-1,1}$ and $q_{3i,1}$. By the
propagation argument from above and since $\mathcal{I}$ is an instance of \textsc{3-Partitioned 1-in-3-Sat}, $q_{3i-2,1}$, $q_{3i-1,1}$ and $q_{3i,1}$ belong to distinct partitions. Similarly, each
clause vertex~$c_j \in C$ is adjacent to exactly three output vertices, namely the upper output
vertices of the gadgets replacing $q_{1,3j-2}$, $q_{1,3j-1}$ and $q_{1,3j}$, respectively, and each of
these output vertices lacks exactly the partition of~$c_j$. Finally, condition $|\{i \in \{1, \ldots, k\} \mid \exists c \in P_i \colon \{x,x'\} \subseteq c\}| < k$ holds:
two variables corresponding to variable vertices from different gadgets, or an original variable and a gadget variable, share at most one clause, namely the clause vertex incident to their connecting edge and the property is true in single gadgets by
construction.

We conclude that the resulting graph corresponds to a valid instance of
\textsc{3-Partitioned 1-in-3-Sat}.
\end{proof}

Since we aim at constructing a 3-connected instance from an instance of \textsc{3-Partitioned 1-in-3-Sat}, we prove a general replacement lemma. Inspired by Hasan, Mondal and Rahman~\cite[Lemma 1]{Hasan2022}, this lemma ensures that replacing the grid vertices with a gadget preserves 3-connectivity. In order to prove 3-connectivity, we use two equivalent definitions of $k$-connectedness. A graph $G$ with at least $k+1$ vertices is $k$-connected if it remains connected after the removal of any $k-1$ vertices. By Menger’s Theorem, this is equivalent to the existence of $k$ internally vertex-disjoint paths between any two distinct vertices of $G$.
 
\begin{lem}\label{lem:upsizeto3con}
Let $G$ be a 3-connected graph with a vertex $r$ of degree three or four, where $w_1, \ldots , w_3$ and $w_4$, respectively are the neighbors of $r$. Let $H = (V^H, E^H)$ be a 2-connected graph with a dedicated subset $D \subseteq V^H$ with $3 \leq |D| \leq 4$,
and furthermore, for each $z \in V^H \setminus D$, let there exist three internally vertex-disjoint paths from $z$ to three distinct vertices $d_1, d_2, d_3 \in D$. Let $G'=(V^{G'},E^{G'})$ be a graph obtained by replacing~$r$ with~$H$ and connecting~$w_1, \ldots , w_4$ to the vertices in~$D$ such that each vertex in $D$ obtains exactly one new edge. Then $G'$ is 3-connected.
\end{lem}
\begin{proof}
Assume for a contradiction that $G'$ is not 3-connected. Thus there are vertices $u$ and $v$ in $V^{G'}$ that are not connected by three internally vertex-disjoint paths. Since $G$ is 3-connected, at least one of them needs to be in $V^H$. If exactly one of them is contained in $V^H$, we have a contradiction since for each $z \in V^H$, there exist three internally vertex-disjoint paths from~$z$ to~$d_1, d_2, d_3 \in D$, and each vertex in $D$ is connected to $G$ which is 3-connected. Hence assume that both $u$ and $v$ are in $V^H$ and therefore, there must be two vertices $x$ and $y$ in $V^H$ so that $G' \setminus \{x,y\}$ decomposes into at least two connected components. Assume that both $u$ and $v$ are in $V^H \setminus D$ and lie in different connected components. Intuitively, the vertices in $D$ represent `exits' to $G$ and hence, if at least one of $u,v$ is in $D$, we have that $G'$ is 3-connected. Call the restriction to $H$ of respective components $H_u$ and $H_v$ (rectangles in Figure \ref{fig:pathsinreplacementlemma}).
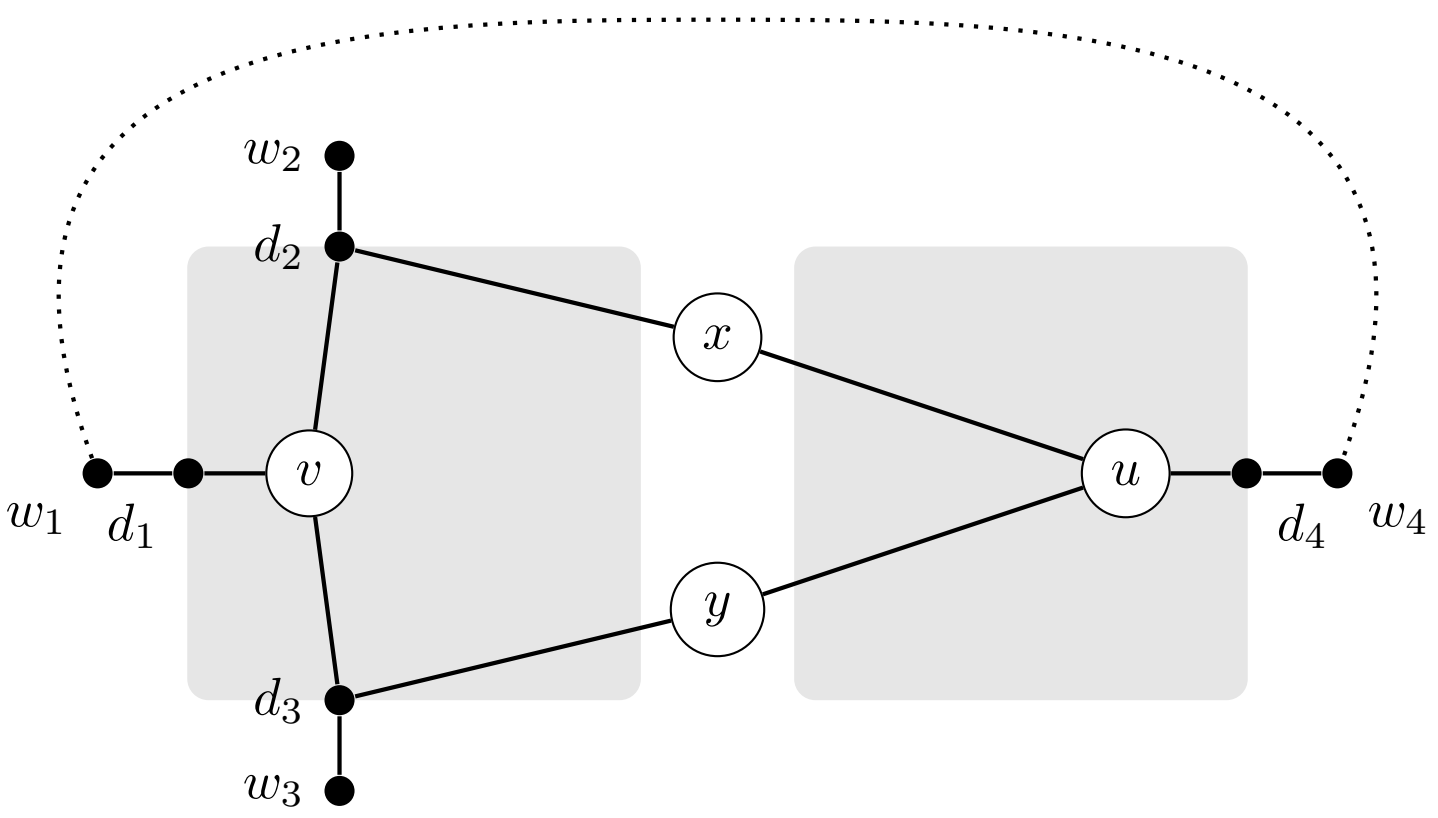
\begin{figure}[h]
    \centering
\begin{tikzpicture}[>=stealth]
\tikzset{
    comp/.style = {fill=black!10,rounded corners},
    bdot/.style = {circle,fill=black,inner sep=2pt},
    dottedpath/.style = {dotted,thick},
    outpath/.style = {thick}
}

\node[comp,minimum width=3.0cm,minimum height=3.0cm,anchor=east] (C1) at (3.5,0) {};
\node[comp,minimum width=3.0cm,minimum height=3.0cm,anchor=west] (C2) at (4.5,0) {};

\node[bdot,label=below left:$d_1$] (L) at (2-1.5,0) {};
\node[bdot,label=left:$d_2$] (T) at (2-0.5,1.5) {};
\node[bdot,label=left:$d_3$] (B) at (2-0.5,-1.5) {};
\node[bdot,label=below right:$d_4$] (R) at (6+1.5,0) {};

\node[bdot,label=below left:$w_1$] (W1) at (-0.1,0) {};
\node[bdot,label=left:$w_2$] (W2) at (1.5,2.1) {};
\node[bdot,label=left:$w_3$] (W3) at (1.5,-2.1) {};
\node[bdot,label=below right:$w_4$] (W4) at (8.1,0) {};

\node[draw,circle,fill=white] (v) at (1.3,0) {$v$};
\node[draw,circle,fill=white] (u) at (6.7,0) {$u$};
\node[draw,circle,fill=white] (x) at (4,0.9) {$x$};
\node[draw,circle,fill=white] (y) at (4,-0.9) {$y$};

\draw[dottedpath]
    (W1)
    .. controls (-1.2,3.0) and (1.8,3.0) ..
    (4,3.0)
    .. controls (6.2,3.0) and (9.2,3.0) ..
    (W4);

\draw[outpath] (v) -- (T);
\draw[outpath] (v) -- (L);
\draw[outpath] (v) -- (B);

\draw[outpath] (u) -- (x);
\draw[outpath] (x) -- (T);

\draw[outpath] (u) -- (y);
\draw[outpath] (y) -- (B);

\draw[outpath] (u) -- (R);

\draw[outpath] (L) -- (W1);
\draw[outpath] (T) -- (W2);
\draw[outpath] (B) -- (W3);
\draw[outpath] (R) -- (W4);

\end{tikzpicture}

\caption{Example of paths through $H$. The two gray shapes
depict the two connected components $H_u$ and $H_v$ and the dotted path connects $w_1$
to $w_4$ in $G$}
\label{fig:pathsinreplacementlemma}
\end{figure}

Because we assumed that for every $z \in V^H \setminus D$, there exist three internally vertex-disjoint paths from~$z$ to three distinct $d_1, d_2, d_3 \in D$ and $H_u$ and $H_v$ are separated by the two vertices $x$ and $y$, we have that each connected component in $H$ contains at least one element of $D$. Consequently, there is a path from $v$ to a $d \in D$ that avoids $x$ and $y$ and similarly, a path from $u$ to $d' \in D$ that does not intersect $x$ and $y$. Without loss of generality, assume that $d$ is connected to $w_1$ and $d'$ is connected to $w_2$. Since $G$ was 3-connected, there is a path from $w_1$ to $w_2$ in $G \setminus \{r\}$. Concatenating these paths yields the path $(v, d, w_1,\dots, w_2, d', u)$ connecting the components and avoiding the vertices~$x$ and~$y$. This contradicts the assumption that \(\{x,y\}\) separates \(u\) and \(v\). Hence, no vertex cut of size at most two exists, and therefore, \(G'\) is 3-connected.
\end{proof}

Let $G_{\text{plan}}(X,C,P)$ denote the graph obtained by replacing every crossing of the drawing with one of the two crossover gadgets. To this end, we want to ensure that a crossover gadget can always be placed such that the incoming edges are attached to the correct outer vertices. 
In particular, the resulting graph must remain bipartite with respect to variable and clause vertices, and every variable must occur in clauses from all partitions.

We have shown a similar result in Lemma \ref{lem:possiblecrosspart} for the grid vertices. Since the crossings do not lie on a grid, we orient all edges from variable vertices to clause vertices for better readability. Moreover, we label the crossings according to how they are connected by edges. Intuitively, the crossings on an edge can be thought of as pearls on a string, which line up once the string is pulled straight. 

\begin{lem}\label{lem:crosswelldefined}
The choice of crossover gadgets at the crossings is unique, and $G_{\text{plan}}(X,C,P)$ is the incidence graph of an instance of  \textsc{Planar 3-Partitioned 1-in-3-Sat}.
\end{lem}
\begin{proof}
By Lemma \ref{lem:possiblecrosspart}, we can assume that all grid vertices are replaced by crossover gadgets such that the resulting graph is the incidence graph of an instance of \textsc{3-Partitioned 1-in-3-Sat}.

First, we label the crossings inductively. A crossing receives the label  $\text{cross}_{i,1}$ if it is adjacent to $q_{i,3n}$. Each remaining crossing receives the label $\text{cross}_{i,j+1}$ if it is joined to the crossing $\text{cross}_{i,j}$ by an edge segment that contains no further crossing. Hence, all crossings receive a unique label and are thereby assigned a row and a column in an extension of the grid graph.

Next, we replace all crossings with a crossover gadget. We start with the top crossing $\text{cross}_{1,1}$. Its upper input vertex is joined to the clause~$c_n$, and its left input vertex is joined to a clause of the gadget that replaced the grid vertex $q_{3n,1}$. Since the partitions of both clauses at the input vertices are fixed, there is exactly one crossover gadget which can replace the crossing $\text{cross}_{1,1}$. Inductively, we replace all crossings $\text{cross}_{i,1}$ for $i \in \{1, \dots, 3n\}$. Further, we replace all remaining crossings $\text{cross}_{i,j}$ for $i \geq 2$ with the same type of crossover gadget used to replace $\text{cross}_{i-1,j}$. In every step, the choice is unique, since the partitions at the two input vertices determine whether the crossover gadget for equal or for opposing input partitions is required. Consequently, there is a unique way to replace all crossings with  crossover gadgets.
\end{proof}

Note that replacing a crossing with a vertex does not necessarily preserve 3-connectivity, since two previously disjoint paths may then intersect in the new vertex. We therefore show that replacing a crossing with a crossover gadget preserves the property that any two vertices outside the gadget are joined by three internally vertex-disjoint paths.

\begin{lem}\label{lem:requmetparti3con}
The graph $G_{\text{plan}}(\mathcal{I})$ is 3-connected.
\end{lem}
\begin{proof}
By Lemma \ref{lem:gridis3con}, $G_{\text{grid}}(\mathcal{I})$ is 3-connected. Let $G$ be the graph obtained from $G_{\text{grid}}(\mathcal{I})$ by replacing all grid vertices with crossover gadgets. Since the two crossover gadgets depicted in Figure~\ref{fig:crossover_gadget_diff} and Figure~\ref{fig:crossover_gadget_same} satisfy the requirements of Lemma~\ref{lem:upsizeto3con}, 
$G$ is 3-connected. However, $G$ is not necessarily planar. We show that $G$ remains  3-connected when all crossings are replaced with crossover gadgets. By Lemma \ref{lem:crosswelldefined}, the replacement of crossings is well-defined.\\

Let $u$ and $v$ be two vertices of~$G$, and consider three internally vertex-disjoint paths between them.  If none of the three paths is involved in a crossing, replacing the crossings with crossover gadgets cannot harm the property of being internally vertex-disjoint. Hence, we may assume that the paths have crossings. Moreover, we assume that exactly two paths cross in one crossing point since a drawing in which three paths cross in a common point can easily be modified such that only two paths cross in one point. 

We need to reason that if two internally vertex-disjoint paths from $u$ to $v$ cross, there are still three internally vertex-disjoint paths from $u$ to $v$ after we replaced the crossing with a crossover gadget. We do so, by arguing that there are internally vertex-disjoint paths through the crossover gadget. To this end, we fix the orientation of the gadget: let $d_1$, $d_2$, $d_3$ and $d_4$ denote the top, left, bottom and right vertex of the set~$D$ of the gadget, respectively. Since both crossover gadgets are depicted on a grid, with one vertex of $D$ on each side, this is well-defined.

When we replace a crossing of two paths $p_1$ and $p_2$, assume $p_1$ as the path going through $d_1$ and $d_3$, and $p_2$ as the path going through $d_2$ and $d_4$. The outer boundary of each crossover gadget is a unique cycle containing all vertices of~$D$. Consequently, there are internally vertex-disjoint paths connecting $d_i$ and $d_{i+1}$ for all $i \in \{1,2,3\}$. Since both $p_1$ and $p_2$ lead to $v$, we obtain two new paths: the first one follows $p_1$ from $u$ to $d_1$, continues to $d_4$, and then follows~$p_2$. The second one follows $p_2$ from $u$ to $d_2$, continues to $d_3$, and then follows~$p_1$ (see Figure \ref{fig:crossovergadgetpathsillustration}). Applying this construction inductively to all crossings yields three internally vertex-disjoint paths from $u$ to $v$ in $G_{\text{plan}}(\mathcal{I})$.

\begin{figure}[h]
    \centering
\begin{tikzpicture}[>=stealth]
\tikzset{
    comp/.style = {fill=black!10,rounded corners},
    bdot/.style = {circle,fill=black,inner sep=2pt},
    boldpath/.style = {line width=1.6pt},
    ponepath/.style = {dotted,line width=1.4pt},
    ptwopath/.style = {dash pattern=on 7pt off 4pt,line width=1.4pt},
    cycleedge/.style = {thin}
}

\node[comp,minimum width=3.0cm,minimum height=3.0cm] (G) at (0,0) {};

\node[draw,circle,fill=white] (u) at (-3.5,0) {$u$};
\node[draw,circle,fill=white] (v) at (3.5,0) {$v$};

\node[bdot,label=above right:$d_1$] (d1) at (0,1.5) {};
\node[bdot,label=above left:$d_2$] (d2) at (-1.5,0) {};
\node[bdot,label=below left:$d_3$] (d3) at (0,-1.5) {};
\node[bdot,label=above right:$d_4$] (d4) at (1.5,0) {};

\draw[cycleedge,rounded corners] (d1) -- (-1.5,1.5) -- (d2);
\draw[cycleedge,rounded corners] (d3) -- (1.5,-1.5) -- (d4);

\draw[boldpath,->,rounded corners] (d1) -- (1.5,1.5) -- (d4);
\draw[boldpath,->,rounded corners] (d2) -- (-1.5,-1.5) -- (d3);

\draw[ponepath,->] (u)
    .. controls (-3,1.6) and (-1.2,2.2) ..
    node[pos=0.5,above left] {$p_1$} (d1)
    -- (d3)
    .. controls (1.2,-2.2) and (3,-1.6) .. (v);

\draw[ptwopath,->] (u)
    -- node[midway,below] {$p_2$} (d2)
    -- (d4)
    -- (v);

\end{tikzpicture}

\caption{Two alternative paths from $u$ to $v$ through a crossover gadget (gray shape)}
\label{fig:crossovergadgetpathsillustration}
\end{figure}
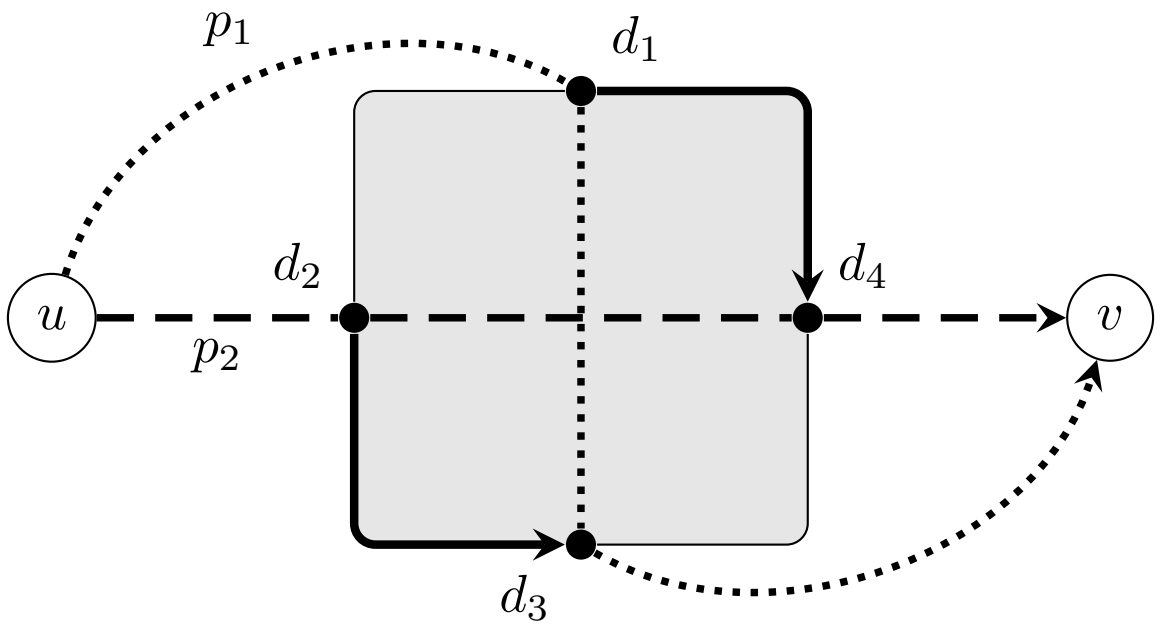
\end{proof}

Since $G_{\text{plan}}(\mathcal{I})$ is 3-connected, this yields a reduction to \textsc{Planar 3-Connected 3-Partitioned 1-in-3-Sat}.

\begin{thm}\label{thm:planr3con}
There is a parsimonious polynomial time reduction from \textsc{3-Partitioned 1-in-3-Sat} to \textsc{Planar 3-Connected 3-Partitioned 1-in-3-Sat}.
\end{thm}
\begin{proof}
Take an instance $\mathcal{I}=(X,C,P)$ of \textsc{3-Partitioned 1-in-3-Sat} and let $\mathcal{I}'$ be the instance associated with graph $G_{\text{plan}}(X,C,P)$. Due to the reduction in Theorem \ref{thm:planar-3-part}, $\mathcal{I}$ has equally many 1-in-3-satisfying assignments as $\mathcal{I}'$. With the previous Lemma \ref{lem:requmetparti3con} and since $G_{\text{plan}}(X,C,P)$ is planar by construction, we have obtained a reduction from \textsc{3-Partitioned 1-in-3-Sat} to \textsc{Planar 3-Connected 3-Partitioned 1-in-3-Sat}. The reduction runs in polynomial time, as there are $\mathcal{O}(n^2)$ grid vertices, at most $\binom{3n}{2}$ crossings and all gadgets are of constant size.
\end{proof}

Moreover, we can construct 3-connected linear formulas (at the cost of planarity) since the reduction from Theorem~\ref{thm:Linear$k$-Partitioned1-in-3-Sat} preserves 3-connectivity. The \textsc{Sat} graph of $\mathcal{I}'''$ consists of three copies of $G(\mathcal{I})$ and the $2n$ clause vertices of $P'''_{k+1} \cup P'''_{k+2}$, each of which is adjacent to one variable vertex of every copy. If $G(\mathcal{I})$ is 3-connected, deleting at most two vertices leaves every copy connected, and since each vertex is, or is adjacent to, at most two of these $2n \geq 6$ clause vertices, one of them keeps all three neighbors and connects the copies, while every other one that is not deleted keeps a neighbor.

\begin{cor}\label{cor:lin3conkpart}
\textsc{Linear 3-Connected $k$-Partitioned 1-in-3-Sat} with $k\geq 3$ is \textsc{NP}-complete and its counting variant is \#\textsc{P}-complete.
\end{cor}

Since both \textsc{Planar 3-Partitioned 1-in-3-Sat} and \textsc{Linear 3-Partitioned 1-in-3-Sat} are \textsc{NP}-complete, a natural question is whether hardness persists if planarity and linearity are required simultaneously, for instance for \textsc{Planar Linear 3-Connected 3-Partitioned 1-in-3-Sat}.
The answer is negative in a strong sense: such instances do not exist (Corollary~\ref{cor:nolinplan}), so the problem is trivial. Hence, not all combinations of structural properties are possible. \\

Recall that the \emph{girth} of a graph is the length of the shortest cycle in the graph. Hud\'{a}k and Madaras~\cite{Hudak2009} used Euler's polyhedral formula to show that each planar
graph with minimum degree at least~3 has girth at most~5. The bound on the girth also holds for planar graphs with \emph{average degree} at least 3. Since bipartite graphs have even girth and the \textsc{Sat} graph is bipartite, the following observation is a consequence of the bound for general planar graphs. 

\begin{obs}\label{obs:girth}
Every simple planar bipartite graph with average vertex degree at least 3 has girth 4. 
\end{obs}

Consequently, for \textsc{Sat} formulas, there is a tradeoff between planarity and linearity:

\begin{cor}\label{cor:nolinplan}
There is no linear instance of \textsc{Planar 1-in-3-Sat} in which  each variable appears at least three times.
\end{cor}

\begin{proof} Take an instance of \textsc{Planar 1-in-3-Sat} in which  each variable appears at least three times. By Observation \ref{obs:girth}, the length of the shortest cycle is four. But if a cycle has length four then clauses share two variables and consequently, the instance cannot be linear.
\end{proof}

This raises the question whether 3-connectivity and planarity can be combined for $k \neq 3$.

\section{\textsc{Planar $k$-Partitioned 1-in-3-Sat}\label{sec:4part} for $k>3$}

\begin{figure}[h!]
    \centering
    \includegraphics[width=1\linewidth]{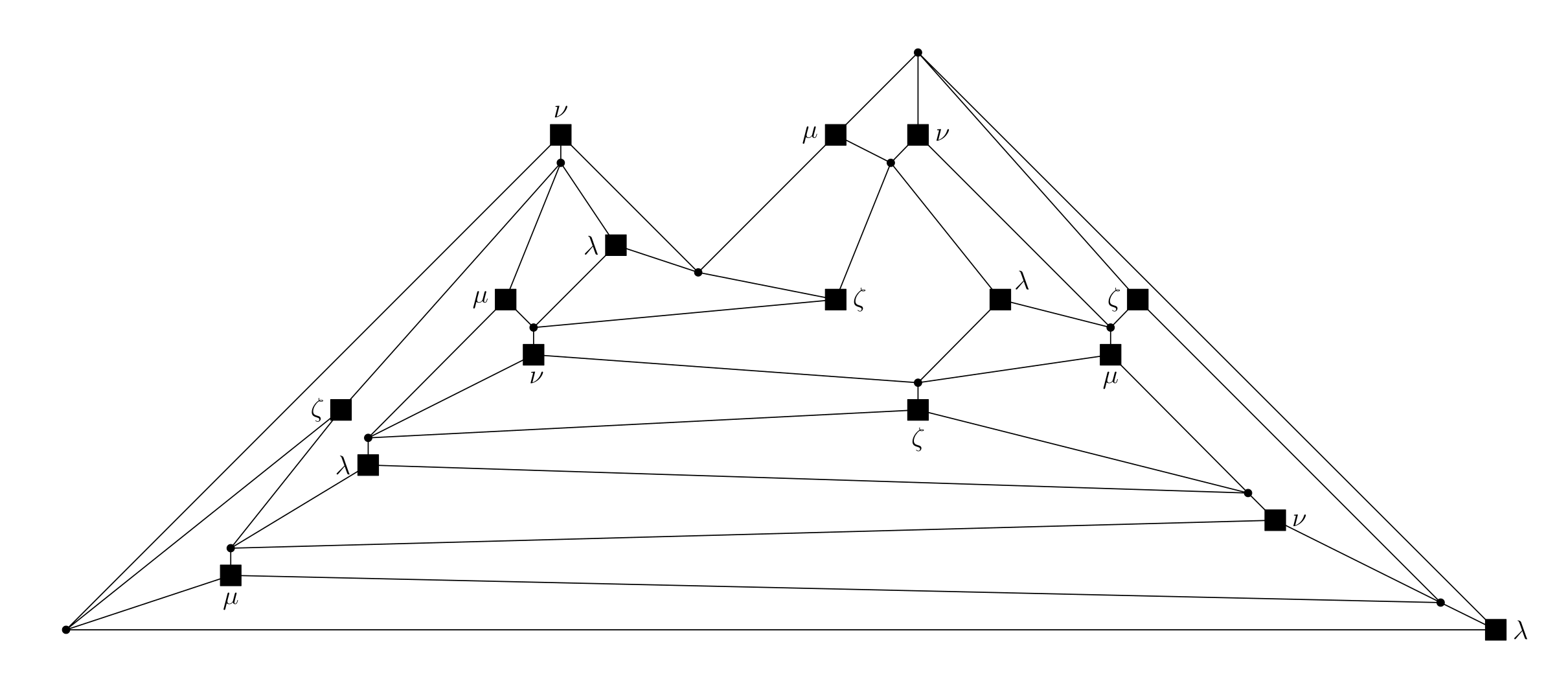}
    \caption{A no-instance of \textsc{Planar 3-Connected 4-Partitioned 1-in-3-Sat}}
    \label{fig:counterexample4planpart}
\end{figure}

We first study \textsc{Planar 3-Connected 4-Partitioned 1-in-3-Sat}. This problem has both yes- and no-instances. Figure~\ref{fig:counterexample4planpart} shows a no-instance.

Nonetheless, a substantial class of instances of \textsc{Planar 4-Partitioned 1-in-3-Sat} is always satisfiable, namely those for which the length of every face is divisible by four. For this, first we first show by a double-counting argument that connected \textsc{Sat} graphs of 4-partitioned instances are 2-connected. 

\begin{lem}\label{lem:noartpoint}
The \textsc{Sat} graph of every instance of \textsc{4-Partitioned 1-in-3-Sat} is 2-connected.
\end{lem}
\begin{proof}
Let $\mathcal{I}=(X,C,P)$ be an instance of \textsc{4-Partitioned 1-in-3-Sat} with partition $P=\langle P_1, \dots, P_4 \rangle$ and let $G$ be the corresponding (connected) \textsc{Sat} graph. Let~$v$ be a vertex of~$G$, let $K$ be a component of $G\setminus v$ and set
$r=|X \cap K|$ as the number of variable vertices in $K$. As $G$ is connected, $K$ contains a neighbor
of $v$. Moreover, every neighbor of a vertex of $K$, other
than $v$, lies in $K$. In both cases below we count the edges
between $X \cap K$ and $P_i\cap K$ in two ways.

\emph{Case 1:} $v$ is a variable vertex.
Let $\varepsilon_i\in\{0,1\}$ indicate whether the neighbor of $v$
in $P_i$ lies in $K$. Each variable of $K$ has all four neighbors in $K$, exactly one of them in~$P_i$. Each clause of $P_i\cap K$ has three
neighbors in $K$, except for the neighbor of~$v$ in $P_i$ (if it lies in~$K$), which has only two.
Hence, $r=3|P_i\cap K|-\varepsilon_i$,
so $\varepsilon_i\equiv -r \pmod 3$ for every $i$. With
$\varepsilon_i\in\{0,1\}$, we conclude that all $\varepsilon_i$ are equal, and since
$K$ contains a neighbor of $v$, they all equal $1$. Thus every
component of $G\setminus v$ contains all four neighbors of $v$ and hence, $G\setminus v$ is connected.

\emph{Case 2}: $v\in P_j$ is a clause vertex.
Let $d$ be the number of neighbors of $v$ in $K$, so that
$1\le d\le 3$. For~$i \neq j$, every variable of~$K$ has its neighbor in~$P_i$ in~$K$, and every clause of~$P_i \cap K$ has all three neighbors in~$K$. Hence,
$r=3|P_i\cap K|$ for $i \neq j$.
In $P_j$, exactly $d$ variables of $K$ have $v$ as their neighbor, which yields
$r-d=3|P_j\cap K|$.
Hence, $3\mid d$, and so $d=3$. As in Case~1, it follows that
$G\setminus v$ is connected.
\end{proof}

The following proposition shows that these instances are trivial. Its proof yields a linear-time algorithm.

\begin{pro}\label{pro:allevencyclesatisfiable}
Let $\mathcal{I}=(X,C,P)$ be an instance of \textsc{Planar 4-Partitioned 1-in-3-Sat} whose \textsc{Sat} graph has a plane embedding in which the length of every face is divisible by four. Then $\mathcal{I}$ is 1-in-3-satisfiable, and a satisfying assignment of $\mathcal{I}$ can be computed in linear time.
\end{pro}
\begin{proof}
Let~$G$ be the \textsc{Sat} graph of~$\mathcal{I}$ with such an embedding. By Lemma \ref{lem:noartpoint}, $G$ is 2-connected, so every face is bounded by a cycle. We construct a graph $C_G$ as follows.
\begin{enumerate}
    \item Let $f_1, \dots, f_d$ be the faces of length greater than four, i.e., length at least eight (including the outer face). Since $G$ is bipartite, the faces have even length and alternate between clause and variable vertices. Within each face $f_i \in \{f_1, \dots, f_d\}$, connect every two variable vertices that are consecutive on the boundary of $f_i$, i.e., that are separated by a single clause vertex on the boundary of~$f_i$, and draw these edges inside $f_i$. Since $G$ is 2-connected, the variables are distinct. Hence, within each $f_i$, we inserted a cycle $Z_i$ whose length is half the length of $f_i$, and therefore at least four.
    \item For each face of length four (consisting of two variable and two clause vertices), connect its two variable vertices with an edge. Now consider a variable vertex $v$. Since $v$ has degree four in $G$, it is incident with four faces. Let $\ell$ be the number of these faces that have length greater than four, so the remaining $4-\ell$ faces have length four. In each face of length greater than four, $v$ receives two edges of the inserted cycle. In each face of length four, $v$ receives one edge. Hence, after this step, $v$ has degree $8 + \ell$.
    \item For each inserted cycle $Z_i$, we add a new vertex $w_i$ inside $f_i$ and connect it to all vertices of $Z_i$. The vertex $w_i$ has degree $|Z_i| = \text{length}(f_i)/2$, which is even because the length of~$f_i$ is divisible by four. Each variable vertex $v$ gains one edge for each of the $\ell$ faces of length greater than four that it lies on, so $v$ now has degree $8 + 2\ell$.
    \item Delete all clause vertices together with their incident edges. Since every variable appears exactly four times in $\mathcal{I}$, each variable vertex loses exactly four edges and has thus degree $4 + 2\ell$ in the resulting graph~$C_G$. Hence, each vertex of $C_G$ has even degree.
\end{enumerate}

After Step 4, every clause vertex is enclosed in a triangle: If a clause vertex is included in three faces of length four in~$G$, this is clear. If a clause vertex lies on a face of greater length, it is enclosed in a triangle since all these faces have augmented cycles. Consequently, $C_G$ is a plane near-triangulation, possibly with parallel edges, in which every vertex has even degree, and the three variables of each clause bound a face. Such a near-triangulation admits a $3$-coloring of its vertices~\cite{TsaiWest2011}, in which the three variables of each clause receive distinct colors. Setting the variables of one color class to $T$ and all other variables to $F$ therefore yields a 1-in-3-satisfying assignment.

The graph $C_G$ can be constructed in linear time, and a 3-coloring of a triangulation with even degrees can be computed in linear time~(Diks, Kowalik and Kurowski~\cite{DiksKowalikKurowski2002}, Tsai and West~\cite{TsaiWest2011}). 
Hence, a 1-in-3-satisfying assignment for~$\mathcal{I}$ can be found in linear time.
\end{proof}

Since a 3-connected graph is in particular connected, Proposition~\ref{pro:allevencyclesatisfiable} is a corrected version of Theorem~4 in the work by Hasan, Mondal and Rahman~\cite{Hasan2022} which claims satisfiability without the condition on the face lengths. The instance in Figure~\ref{fig:counterexample4planpart} shows that this condidition cannot be dropped. \\

Without the restriction on the face lengths, the problem becomes hard. In the remainder of this section, we show that \textsc{Planar 4-Partitioned 1-in-3-Sat} is \textsc{NP}-hard. 

The method is similar to the one we used to establish planarity and linearity (see Sections \ref{sec:hardnessplanarity} and \ref{sec:hardnesslinearity}). There, we replaced a single clause or variable vertex by a gadget with the same number of connections. Here, the gadgets have a higher degree than vertices they replace. Therefore, all vertices of the instance must be replaced at the same time to yield a valid \textsc{Sat} instance. Recall that the input/output vertices of a gadget are its clause vertices of degree less than three and its variable vertices of degree less than four. To describe how one gadget connects to another, we introduce  \emph{glue vertices} in addition to the input/output vertices that are part of the gadget. These are vertices that are adjacent to vertices of the gadget without belonging to it and that are later merged with an input/output vertex of another gadget.

\subsection{Gadgets}
We provide a gadget that ensures that every variable occurs exactly four times, as well as gadgets that ensure that the formula is 4-partitioned.

\begin{figure}
    \centering    \includegraphics[width=0.5\linewidth]{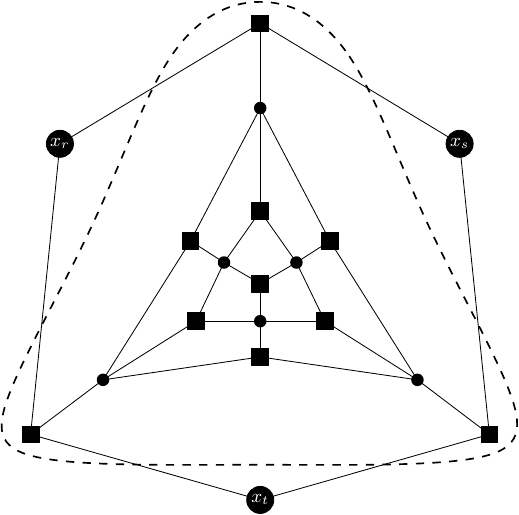}
    \caption{The 3-to-4-gadget.}
    \label{fig:4withoutpartNPgadget}
\end{figure}

\paragraph{3-to-4-gadget} The 3-to-4-gadget (Figure~\ref{fig:4withoutpartNPgadget}) consists of six variable vertices and ten clause vertices. Its glue vertices are $x_r, x_s$ and $x_t$. The gadget increases the number of variable occurrences in the corresponding formula from three to four. The resulting formula is not 4-partitioned.

\begin{figure}[ht]
\centering\includegraphics[scale=0.9]{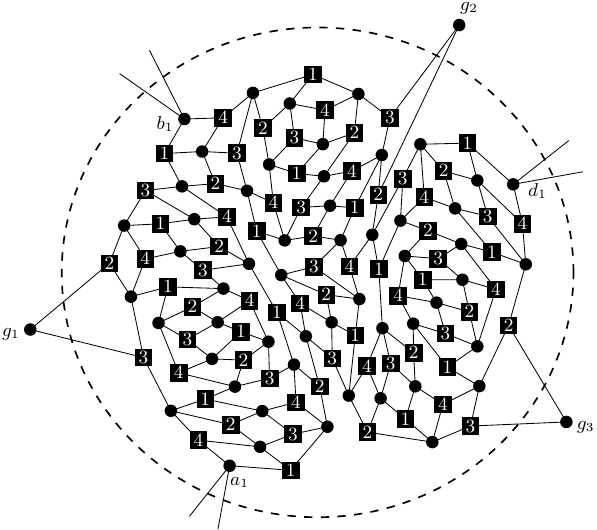}
\caption{The $C$-gadget.}
\label{fig:C}
\end{figure}

\paragraph{$C$-gadget} The $C$-gadget (Figure~\ref{fig:C}) consists of 54 variable vertices $a_1,\dots,a_8$, $b_1,b_2$,
$d_1,\dots,d_8$, $e_1,e_2$, $f$, $h$, $k_1,\dots,k_4$, $p_1,\dots,p_4$, $q_1,\dots,q_4$,
$z_1,\dots,z_{16}$, $o_1,\dots,o_4$ and 72 clause vertices. Its
output vertices are $a_1,\, b_1,\, d_1$, and the vertices $g_1$, $g_2$ and $g_3$ are its glue vertices.
This gadget simulates a clause vertex.

\begin{figure}[ht]
\centering\includegraphics[scale=1]{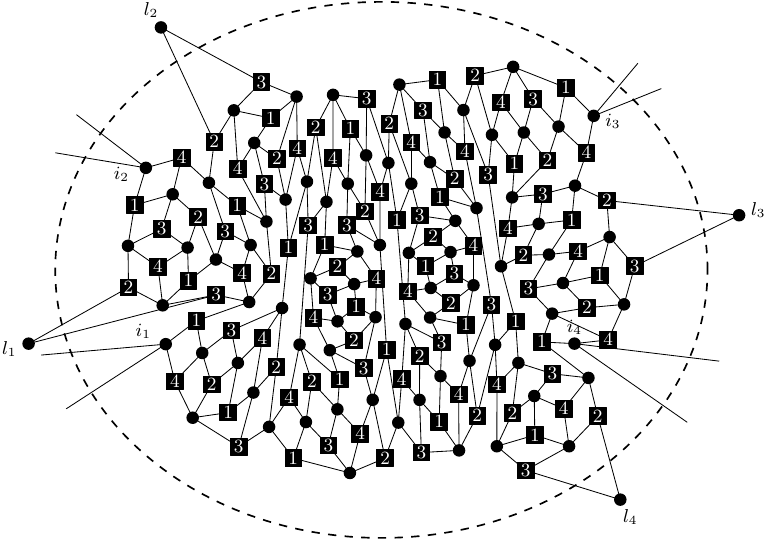}
\caption{The $X$-gadget. }
\label{fig:Xgadget}
\end{figure}

\paragraph{$X$-gadget} The $X$-gadget (Figure~\ref{fig:Xgadget}) consists of 78 variable vertices $i_1,\dots,i_{20}$, $b_1,\dots,b_{20}$,
$z_1,\dots,z_{32}$, $o_1,\dots,o_6$ and 104 clause vertices. Its
output vertices are $i_1,\, i_2,\, i_3,\, i_4$, and the vertices $l_1,\dots,l_4$ are its glue vertices.
This gadget simulates a variable vertex.\\

In Figures~\ref{fig:C} and~\ref{fig:Xgadget}, every output vertex lies on the outer face and is
immediately followed clockwise by its glue vertex. Visibly, these gadgets are larger than the previous ones, but their properties can be proved similarly as before. The full proof can be found in the Appendix.

\begin{pro}\label{pro:bringing4parttogether}
The $C$-gadget and the $X$-gadget are 4-partitioned and planar. Furthermore, a truth assignment
$\beta$ of the output vertices of such a gadget extends to a 1-in-3-satisfying assignment
of the gadget if and only if
\begin{enumerate}
\item[(C)] $\beta(u)=T$ for exactly one $u\in\{a_1,b_1,d_1\}$, for the $C$-gadget,
\item[(X)] $\beta$ gives $i_1,i_2,i_3,i_4$ one common value, for the $X$-gadget with its glue vertices as additional variables and $\beta(l_j)=\beta(i_j)$ for $j\in\{1,\dots,4\}$.
\end{enumerate}
In both cases, the extension is unique.
\end{pro}

\subsection{Parsimonious reduction for \textsc{Planar 4-Partitioned 1-in-3-Sat}}
With the 3-to-4-gadget, we increase the number of variable occurrences at the cost of the partition property. With the $X$-gadget and the $C$-gadget, we construct an instance that is 4-partitioned from an instance of \textsc{Planar 3-Connected Positive 1-in-3-Sat-E4}.\\

\begin{pro}\label{pro:NPplan3con}
There is a parsimonious reduction from \textsc{Planar 3-Connected 3-Partitioned 1-in-3-Sat} to \textsc{Planar 3-Connected Positive 1-in-3-Sat-E4}.
\end{pro}
\begin{proof}
Take an instance $\mathcal{I}$ of \textsc{Planar 3-Connected 3-Partitioned 1-in-3-Sat} with a fixed embedding. Here, every variable occurs exactly three times. We replace all clauses $c=\{x_s,x_r,x_t\}$ labeled with the partition $\mu$ with the 3-to-4-gadget and connect the gadget to the variables $x_s,x_r,x_t$. Every variable is connected to exactly one partition labeled with $\mu$ and hence, by replacing the clauses with the gadget each variable occurs exactly four times.

We observe that the gadget is 1-in-3-satisfiable if and only if the clause $c=\{x_s,x_r,x_t\}$ is 1-in-3-satisfiable. Further, if one variable in $\{x_s,x_r,x_t\}$ is set to $T$, then, the assignment uniquely extends to an assignment of the whole gadget. 3-connectivity as well as planarity are sustained. Subsequently, replacing the clauses labeled with $\mu$ yields a parsimonious reduction from \textsc{Planar 3-Connected 3-Partitioned 1-in-3-Sat} to \textsc{Planar 3-Connected Positive 1-in-3-Sat-E4}.
\end{proof}

Take an instance $\mathcal{I}=(X,C)$ of \textsc{Planar 3-Connected Positive 1-in-3-Sat-E4} and compute a plane embedding of the \textsc{Sat} graph $G$ of $\mathcal{I}$, which takes linear time by the algorithm of Hopcroft and Tarjan~\cite{HopcroftTarjan74}. We construct an instance $\mathcal{I'}$ of \textsc{Planar 4-Partitioned 1-in-3-Sat} as follows:\\

\begin{enumerate}
\item For every clause $c \in C$ take a $C$-gadget with output vertices $a_{c,1},b_{c,1},d_{c,1}$ (the copies of $a_{1},b_{1},d_{1}$) and with glue vertices $g_{c,1},g_{c,2}$ and $g_{c,3}$.
\item For every variable $x\in X$ take an $X$-gadget with output vertices
$i_{x,1},i_{x,2},i_{x,3},i_{x,4}$ (the copies of $i_1,\dots,i_4$) and with glue vertices
$l_{x,1},l_{x,2},l_{x,3}$ and $l_{x,4}$.
\item Replace every clause vertex by a $C$-gadget and every variable vertex by an $X$-gadget.
\item For every edge $\{c,x\}$ in $G$, merge a vertex $g\in\{g_{c,1},g_{c,2},g_{c,3}\}$ of the
$C$-gadget of $c$ with the glue vertex $i \in \{i_{x,1},i_{x,2},i_{x,3},i_{x,4}\}$ of the $X$-gadge that replaces the variable $x$. Then
merge the output vertex $a \in \{a_{c,1},b_{c,1},d_{c,1}\}$ that immediately precedes $g$ clockwise
with the glue vertex $l \in \{l_{x,1},l_{x,2},l_{x,3},l_{x,4}\}$ that immediately follows $i$
clockwise.
\end{enumerate}

\begin{figure}[h]
\centering
\includegraphics{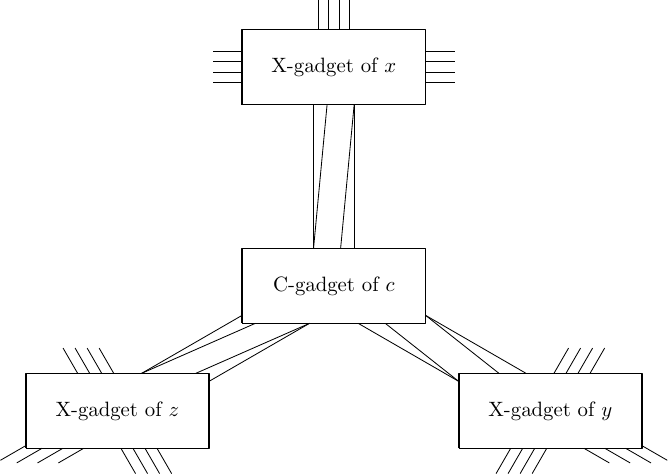}
\caption{Construction of a 4-partitioned instance.}
\label{fig:local}
\end{figure}

Figure~\ref{fig:local} shows the replacement for the clause vertex $c= \{x,y,z\}$ and the variable vertices~$x,y$ and~$z$.

\begin{thm}\label{thm:plan4partmain}
There is a parsimonious reduction from \textsc{Planar 3-Connected Positive 1-in-3-Sat-E4}
to \textsc{Planar 3-Connected 4-Partitioned 1-in-3-Sat}.
\end{thm}
\begin{proof}
Take a constructed instance $\mathcal{I}'$. With every glue vertex replaced by its output vertex, every
variable lies in exactly one clause of each partition: an output vertex lies in clauses of $P_1$ and
$P_4$ and its input vertices in $P_2$ and $P_3$, and every other variable already lies in one clause
of each partition. In Figures~\ref{fig:C} and~\ref{fig:Xgadget}, every output vertex is
immediately followed by its glue vertex on the outer face, merging the respective vertices gives a plane drawing. To see that $\mathcal{I}'$ is 3-connected take two vertices $u$ and $v$ in the \textsc{Sat} graph of $\mathcal{I}'$. All gadgets are 2-connected and there are three internally vertex-disjoint paths to output and glue vertices of the respective gadget. According to the proof of Lemma \ref{lem:upsizeto3con}, we can construct a third path that goes from $u$ to the output and glue vertices of the respective gadget and comes back to $v$. If the vertices are contained in different gadgets, then the three paths that exist in the \textsc{Sat} graph of $\mathcal{I}$ also induce three paths in $\mathcal{I}'$. By Proposition \ref{pro:bringing4parttogether}, the construction is a parsimonious reduction, and since all gadgets have constant size, the construction takes polynomial time.
\end{proof}

In addition to the hardness result, we show that \textsc{Planar Positive 1-in-3-Sat} in which every variable occurs exactly four times is fixed-parameter tractable with respect to the number of faces of lengths four in the corresponding \textsc{Sat} graph. We refer to the book by  Niedermeier~\cite{niedermeier2006invitation} for an introduction to fixed-parameter tractability. In order to obtain fixed-parameter tractability, we decompose the graph into subgraphs within which a truth assignment easily propagates to all variable vertices. We consider subgraphs that consist of faces of length four and the outer face, and call a face of length four a \emph{quadrilateral}.

\begin{figure}[h]
    \centering
    \includegraphics[width=0.5\linewidth]{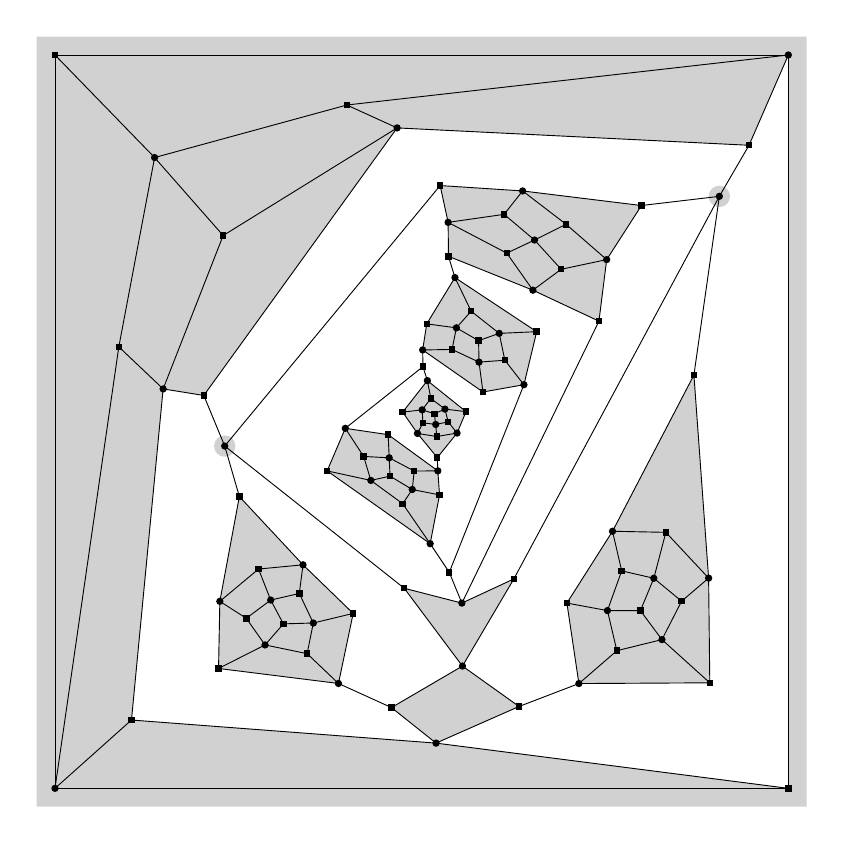}
    \caption{Quadrilateral components (highlighted in gray) of an instance of \textsc{Planar Positive 1-in-3-Sat} in which every variable occurs exactly four times.}
    \label{fig:quadcom}
\end{figure}

Let $G$ be a plane embedding of the \textsc{Sat} graph of an instance $\mathcal{I}= (X,C)$ of \textsc{Planar Positive 1-in-3-Sat}. We decompose $G$ into \emph{quadrilateral components} $K_1, \dots, K_s$. 
We say that two quadrilaterals are adjacent if they share an edge. A
quadrilateral component is a subgraph that is a maximal collection of quadrilaterals that are connected via edge adjacency, or a single variable that is only contained in faces that are not a quadrilateral. In particular, a quadrilateral that shares no edge with another quadrilateral forms a component on its own. An example with many quadrilateral components is shown in Figure~\ref{fig:quadcom}. The quadrilateral components of an embedding are unique.

\begin{lem}
\label{lem:propagation}
Let $K$ be a quadrilateral component. Then there are at most three assignments
of the variables of $K$ that can be extended to an assignment that
1-in-3-satisfies all clauses of $K$, and every 1-in-3-satisfying assignment of
$\mathcal{I}$ restricts to one of them. If the truth value of a variable of $K$
is fixed, at most two of these assignments remain, and at most one if the
variable is set to $T$. These at most three assignments can be computed in
time linear in the number of variables of $K$.

\end{lem}

\begin{proof}
If two clauses share two variables, then an assignment of the two variables uniquely propagates to the other variables of both clauses. Since every quadrilateral consists of two clauses sharing two variables, truth values propagate from face to face. Hence, if $f$ is a quadrilateral and there is a sequence of quadrilaterals from $f$ to a quadrilateral~$f'$, in which consecutive quadrilaterals are adjacent, and the variables of~$f'$ are assigned a truth value, then all variables included in $f$ also have fixed truth values by propagation. 
There are at most three possibilities of assigning truth values to the at most three variables of a clause and therefore, there are at most three possible assignments for all variables of $K$. If $K$ consists of a single variable, it has two assignments. Otherwise, every variable $v$ of $K$ is one of the two shared variables of some
quadrilateral of $K$. In any of the possible assignments, at most one of these two
variables is set to $T$, and their values determine the assignment of the whole
subformula. Hence, setting $v$ to $T$ leaves at most one possible assignment
of the subformula, and setting $v$ to $F$ leaves at most two. Every 1-in-3-satisfying assignment of $\mathcal{I}$ in particular 1-in-3-satisfies all clauses of $K$. Hence its restriction to the variables of $K$ can be extended to such an assignment and is one of the at most three
assignments.
Since the number of clauses is linear in the number of variables, all of them can be computed in linear time.
\end{proof}

By Euler's formula, quadrilaterals are considerably more frequent in \textsc{4-Partitioned 1-in-3-Sat} than in \textsc{3-Partitioned 1-in-3-Sat} since a higher degree of the variable vertices forces a smaller avarage face length.

\begin{thm}\label{thm:planptime}
Let $\mathcal{I} =(X,C)$ be an instance of \textsc{Planar Positive 1-in-3-Sat} in which every
variable occurs exactly four times, let~$n$ be its number of variables, and fix an embedding of its plane \textsc{Sat} graph~$G$. Let $s$ be
the number of quadrilateral components of $G$ in this embedding. Then a 1-in-3-satisfying assignment of~ $\mathcal{I}$ can be constructed, or shown not to exist, in time~$\mathcal{O}(2^s n)$. In particular, the problem is fixed-parameter tractable with respect to $s$.
\end{thm}
\begin{proof}
Construct the quadrilateral components $K_1, \dots, K_s$ in $\mathcal{O}(n)$ time. By Lemma~\ref{lem:propagation}, each component has at most three assignments
that extend to a 1-in-3-satisfying assignment of its clauses. These include
the restriction of every 1-in-3-satisfying assignment of $\mathcal{I}$, and all of them
can be computed in linear time.

We use the backtracking procedure given in Algorithm~\ref{alg:backtracking} to compute a 1-in-3-satisfying assignment for~$\mathcal{I}$ from the set of partial assignments of the components. We iterate over the quadrilateral components. When propagation stops, we choose a new component $K_j$ such that $K_j$ shares
a variable with $K$, a clause of $K_j$ contains a variable of $K$, or $K_j$
contains, or has a clause containing, an unassigned variable of a clause in
which one variable is set to $F$ and the other two are unassigned~($\star$).
Since $G$ is connected, such a component always exists. Components that have
at most one assignment consistent with the assignment of $K$ are chosen first.

\begin{algorithm}[t]
\caption{Backtracking over component assignments}
\label{alg:backtracking}
\begin{algorithmic}[1]
\Require Components $K_1, \dots, K_s$ of $G$, each with its at most three assignments from Lemma~\ref{lem:propagation}
\Ensure A 1-in-3-satisfying assignment of $F$, or ``unsatisfiable''
\State $K \gets \{K_1\}$
\State Pick an initial assignment of $K_1$
\State Propagate the assignment of $K$ and add all components that thereby receive an assignment to $K$
\While{there is a conflict or there are unassigned variables}
    \If{there is a conflict}
        \State \textsc{Backtrack} \Comment{return ``unsatisfiable'' if not possible}
    \Else
        \State Pick a neighboring component $K_j$ without an assignment ($\star$)
        \If{$K_j$ has an assignment consistent with $K$}
            \State Assign $K_j$ such an assignment
        \Else
            \State \textsc{Backtrack} \Comment{return ``unsatisfiable'' if not possible}
        \EndIf
    \EndIf
    \State Propagate the assignment of $K$ and add all components that thereby receive an assignment to $K$
\EndWhile
\State \Return assignment of $K$
\end{algorithmic}
\end{algorithm}

The algorithm is exhaustive. Whenever a conflict occurs or no consistent assignment exists, \textsc{Backtrack} returns to the most recent choice with untried options and continues with the next one. Hence, every combination of component assignments is either explored or discarded because it extends a partial assignment that already contains a conflict. Consequently, the algorithm returns ``unsatisfiable'' only if no consistent combination of component assignments exists.

The running time depends on the depth of the search tree and the branching factor. In every node of the search tree, the variables of at least one quadrilateral component are assigned, so the search tree has depth at most $s$. Fixing the truth value of a variable in a quadrilateral component $K_j$ leaves at
most two assignments of $K_j$, and at most one if the variable is set to $T$. By ($\star$), we either choose a new component $K_j$ that shares a variable with $K$, or one of whose clauses contains a variable of $K$, or that contains, or has a clause containing, an unassigned variable of a
clause in which one variable is set to $F$ and the other
two are unassigned. In the first case, $K_j$ contains an assigned variable, which leaves at most
two branches. In the second case, a clause $\{a,b,p\}$ of $K_j$ with $a,b$ in
$K_j$ contains an assigned variable $p$. Then $a$ and $b$ are both $F$ if $p$
is set to $T$, and exactly one of them is set to $T$ otherwise. Since $a$ and $b$ determine
the assignment of $K_j$, this also leaves at most two branches.
In the third case, $K_j$ contains~$y$ or a clause $\{a,b,y\}$ with $a,b$ in
$K_j$, for a clause $\{x,y,z\}$ with $x$ set to $F$. In the latter case, $y$ is set to $T$
if and only if $a$ and $b$ are $F$, so at most one assignment of $K_j$ sets
$y$ to $T$ and at most two set it to $F$. The same holds for $z$ and its
component. If $K_j$ also contains $z$ or a clause containing $z$, at most one branch sets
$y$ to $T$ and at most one sets $z$ to $T$. The variable vertices $y$ and $z$ can also lie in different quadrilateral components. In this case, at most one assignment of $K_j$ sets $y$ to $T$, then $z$ is set to $F$,
which leaves at most two assignments for the component of $z$. At most two
assignments of $K_j$ set $y$ to set to $F$, then $z$ is set to $T$, which leaves at most one.
Hence, both components together contribute a factor of at most four. Consequently, the search tree has at
most $3 \cdot 2^{s-1}$ leaves. We conclude that the algorithm terminates and has a worst-case running time of $\mathcal{O}(3\cdot 2^{s-1} n)=\mathcal{O}(2^{s} n)$. 
\end{proof}

Note that instances with many quadrilateral components are comparatively rare and that instances may have a high treewidth but a low number of quadrilateral components. 

Overall, we see that the number of such instances of \textsc{$k$-Partitioned 1-in-3-Sat} decreases as~$k$ grows. This is reflected in the following impossibility result based on a counting argument.

\begin{thm}\label{thm:doublestardied}
All \textsc{Planar 5-Partitioned 1-in-3-Sat} instances are unsatisfiable.
\end{thm}
\begin{proof}
Take an instance $\mathcal{I}=(X,C,P) \in \textsc{Planar 5-Partitioned 1-in-3-Sat}$ and let $G=(V,E)$ be its \textsc{Sat} graph. First, observe that every edge joins a clause vertex and a variable vertex, so $|E| = 5|X| = 3|C|$, hence
$|C| = \frac{5}{3}|X|$. By Euler's formula,
$$
|F| = 2 - |V| + |E| = 2 - \frac{8}{3}|X| + 5|X| = 2 + \frac{7}{3}|X| .
$$
Since $G$ is bipartite, every face has even length. Summing the
lengths of all faces counts every edge twice, so
$$
\sum_{k\geq 2} 2k\,|F_{2k}| = 2|E| = 10|X|
\qquad\text{and}\qquad
\sum_{k\geq 2} |F_{2k}| = |F| = 2 + \frac{7}{3}|X| .
$$
This yields $$
\sum_{k\geq 3}(k-2)\,|F_{2k}| \;=\; \frac{|X|}{3}-4 .
$$

\ \\
Assume for contradiction that the instance is satisfiable, and let $X^1$ and
$X^0$ be the sets of variables set to $T$ and $F$, respectively. Every clause
contains exactly one variable set to $T$, so $5|X^1|$ equals the number of clauses,
which is $\frac{5}{3}n$. Hence $|X^0| = \frac{2}{3}n$.

For $t\in X^1$ and $w\in X^0$, let $a(t,w)$ be the number of clauses
containing both $t$ and $w$. Each of the five clauses containing $w$ contains
exactly one variable set to $T$, so $\sum_{t\in X^1} a(t,w) = 5$ is odd. Hence the
number $O$ of pairs $(t,w)$ with $a(t,w)$ odd satisfies
$O \geq |X^0| = \frac{2}{3}n$.

Fix $t\in X^1$ and let $c_1,\dots,c_5$ be the clauses that contain $t$ labeled clockwise around $t$ according to the embedding. The face containing the vertex between $c_i$ and
$c_{i+1}$ at $t$ contains incident vertices $\dots, y_i, c_i, t, c_{i+1}, z_{i+1},
\dots$. For each $i$, let $y_i$ be the vertex just before $c_i$, and $z_{i+1}$ the vertex just after $c_{i+1}$, on the boundary of the face in the corner between $c_i$ and $c_{i+1}$ at $t$. Then $y_i$ and $z_i$ are the two variables of $c_i$ other than $t$, so $c_i = \{t, y_i, z_i\}$. Thus $a(t,w)$ is the number of
occurrences of $w$ in the five pairs $(y_i, z_{i+1})$. If the face is a
quadrilateral, then $y_i = z_{i+1}$, and this pair does not affect the parity
of $a(t,w)$. Hence every face containing $t$ of length at least six
makes at most two values $a(t,w)$ odd. So $O$ is at most twice the number of
corners at variables set to $T$ lying in such faces.

Any two consecutive variables on the boundary of a face are distinct and lie
in a common clause, so they are not both set to $T$. Hence a face of length $2k$
has at most $\lfloor k/2\rfloor \leq k-2$ corners at variables set to $T$, for
$k \geq 3$. Together with $
\sum_{k\geq 3}(k-2)\,|F_{2k}| \;=\; \frac{|X|}{3}-4$, this gives the contradiction:
$$
\frac{2}{3}n \;\leq\; O \;\leq\; 2\sum_{k\geq 3}\left\lfloor\frac{k}{2}\right\rfloor |F_{2k}|
\;\leq\; 2\sum_{k\geq 3}(k-2)\,|F_{2k}| \;=\; \frac{2}{3}n - 8,
$$
\end{proof}

The theorem shows that planarity, connectedness, and 1-in-3-satisfiability are not compatible in the case of $k=5$. Since every 3-connected graph is, in particular, connected, and a 5-partitioned clause set implies a variable occurrence of five, Theorem~\ref{thm:doublestardied} implies that there are no yes-instances of \textsc{Planar 3-Connected 5-Partitioned 1-in-3-Sat}.

This illustrates the complexity jump from \textsc{Planar 3-Connected 4-Partitioned 1-in-3-Sat} to \textsc{Planar 3-Connected 5-Partitioned 1-in-3-Sat}: the latter can be decided in constant time. Moreover, the complexity of \textsc{Planar 3-Connected $k$-Partitioned 1-in-3-Sat} is trivial for $k>5$:
\begin{obs}\label{obs:noplank>5}
For $k>5$, there are no instances of \textsc{Planar $k$-Partitioned 1-in-3-Sat}.
\end{obs}
\begin{proof}
The \textsc{Sat} graph of an instance of \textsc{3-Connected $k$-Partitioned 1-in-3-Sat} has an average degree of~$\frac{6k}{k+3}$, which is at least 4 for $k\geq 6$. However, every \textsc{Sat} graph is bipartite, and by Euler's formula, the average degree of a planar bipartite graph is strictly less than four. Consequently, there are neither yes- nor no-instances of \textsc{Planar 3-Connected $k$-Partitioned 1-in-3-Sat} for $k>5$.
\end{proof}

\section{A Link to \textsc{$k$-Dimensional Matching} and \textsc{Exact Cover by $k$-Sets}}\label{sec:matchingandexactcover}

The main body of our results relies on proving hardness for \textsc{$k$-Partitioned 1-in-3-Sat} variants. These results have a natural connection to \textsc{Exact Cover by $k$-Sets} and \textsc{$k$-Dimensional Matching} problems.

\problemStatement{$k$-Dimensional Matching ($k$DM-E$3$)}{
Non-empty and pairwise disjoint sets $U_1, U_2, \ldots, U_k$ and a collection of $k$-tuples
\[
T_U \subseteq U_1 \times U_2 \times \cdots \times U_k
\]
such that each element in $U := \bigcup_{i=1}^k U_i$ appears in exactly three $k$-tuples in~$T_U$, and no $u \in U_i$ and $u' \in U_j$ with $i \neq j$ occur in exactly the same three $k$-tuples.
}{
Is there a subset $M \subseteq T_U$ with $|M| = |U_1|$ such that no two
distinct tuples in $M$ agree in any coordinate?
}

We call such a subset $M \subseteq T_U$ with $|M| = |U_1|$ such that no two
distinct tuples in $M$ agree in any coordinate a \textit{perfect matching}.
The problem of whether there is a perfect matching is the perfect matching problem on cubic $k$-uniform strongly $k$-colourable hypergraphs $(U_1, \dots, U_k, T_U)$ (see Hazan, Safra and Schwartz~\cite{hazan2003complexity}), with the additional requirement that no two elements of different classes lie in exactly the same three tuples. Since this requirement only restricts the instances, all hardness results below also hold for the unrestricted problem. In the following, we will use the graph representation of both the matching problems and the satisfiability problems, and we distinguish the two graphs by subscript, writing $G_{\textsc{Sat}}$ and $G_{\textsc{Mat}}$.\\

Let $G_{\textsc{Mat}}(U_1, \dots, U_k, T_U)$ be the bipartite incidence graph with vertices $\bigcup_{i=1}^k U_i \cup T_U$ and edges
\[
\{\{u, t\} \mid u \in U\text{, } t \in T_U, \text{ and } u \in t\}.
\]
Analogously to the \textsc{Sat} graph, we say that an instance $\mathcal{I}_{\textsc{Mat}} = (U_1, \dots, U_k, T_U)$ of \textsc{$k$DM-E3} is \textit{planar} if $G_{\textsc{Mat}}(U_1, \dots, U_k, T_U)$ is planar. An instance $\mathcal{I}_{\textsc{Mat}}$ of \textsc{$k$DM} is \textit{linear} if no two tuples have more than
one element in common, i.e., $d_H(t, t') \geq k-1$ for each pair of distinct $k$-tuples $t,t' \in T_U$, where $d_H$ is the Hamming distance, counting the number of positions in which the two tuples differ. \\

Further, for an instance $\mathcal{I} = (X,C,P)$ of \textsc{$k$-Partitioned 1-in-3-Sat}, define $\varphi\colon X \rightarrow \cart_{i=1}^k P_i$ by $\varphi(x) = (c_{x,1},c_{x,2},\ldots,c_{x,k})$, where $c_{x,i}$ is the unique clause of~$P_i$ that contains~$x$; this clause exists and is unique because $P_i$ is a partition of~$X$. That is, $\varphi$ maps a variable $x \in X$ to the $k$-tuple representing the clauses in partitions $P_1$, $P_2$, \ldots, $P_k$ that contain $x$. Moreover, $\varphi$ is injective: no two distinct variables $x$ and $x'$ share a clause in every partition, so $\varphi(x)$ and $\varphi(x')$ differ in at least one coordinate. Hence, given a tuple in the image of $\varphi$, there is only one variable that is contained in all clauses in the tuple.

Further, if we \emph{associate} an instance $\mathcal{I}_{\textsc{Mat}}=(U_1, \dots, U_k,T_U)$ of \textsc{$k$DM-E3} with a given \textsc{$k$-Partitioned 1-in-3-Sat} instance $\mathcal{I}_{\textsc{Sat}}$ as defined below, the result indeed satisfies the requirements of \textsc{$k$DM-E3} and is therefore a valid instance. By definition, $\varphi(x) \in P_1 \times \cdots \times P_k = U_1 \times \cdots \times U_k$ for each $x \in X$. A clause $c \in P_i$ is the $i$-th coordinate of $\varphi(x)$ if and only if $x \in c$. Since $c$ contains exactly three variables and $\varphi$ is injective, $c$ appears in exactly three distinct $k$-tuples. Furthermore, note that if two clauses $c \in P_i$ and $c' \in P_j$ with $i \neq j$ occurred in exactly the same three tuples, then $\{\varphi(x) \mid x \in c\} = \{\varphi(x) \mid x \in c'\}$, and the injectivity of $\varphi$ would imply $c = c'$, contradicting the disjointness of $P_i$ and $P_j$.

\begin{defi}
Let $k \geq 2$. Let $\mathcal{I}_{\textsc{Sat}} = (X,C,P)$ be an instance of \textsc{$k$-Partitioned 1-in-3-Sat}. We define the instance $\mathcal{I}_{\textsc{Mat}}=(U_1, \dots ,U_k,T_U)$ of \textsc{$k$DM-E3} \emph{associated with} $\mathcal{I}_{\textsc{Sat}}$ by $U_i=P_i$ for all $i$ and $T_U = \{\varphi(x) \mid x \in X\}$. Conversely, let $\mathcal{I}_{\textsc{Mat}}'=(U_1, \dots , U_k, T_U)$ be an instance of \textsc{$k$DM-E3}. We define \textit{the instance $\mathcal{I'}_{\textsc{Sat}} = (X, C, P)$ of} \textsc{$k$-Partitioned 1-in-3-Sat} \textit{associated with} $\mathcal{I}_{\textsc{Mat}}'$ as follows:
\begin{align*}
X &= \{x_t \mid t \in T_U\},\\
P_i &= \{\{x_t \mid t=(u_1,\ldots, u_k) \in T_U \text{ and } u_i = u\} \mid u \in U_i\} \quad \text{for each } i \in \{1, \ldots, k\},\\
P &= \langle P_1, P_2, \ldots, P_k \rangle, \text{ and}\\
C &= \bigcup_{i = 1}^k P_i.
\end{align*}
\end{defi}

We obtain that the \textsc{$k$-Partitioned 1-in-3-Sat} instance $\mathcal{I'}_{\textsc{Sat}}$ associated with an instance $\mathcal{I}_{\textsc{Mat}}'=(U_1, \dots ,U_k,T_U)$ of \textsc{$k$DM-E3} is also indeed an instance of \textsc{$k$-Partitioned 1-in-3-Sat}: For each $u \in U_i$, by definition, there are exactly three tuples $t=(u_1, \dots, u_k) \in T_U$ with $u_i= u$, hence, each clause in $C$ contains exactly three variables. Since no two tuples are equal, we have $|\{i \in \{1, \ldots, k\} \mid \exists c \in P_i \colon \{x,x'\} \subseteq c\}| < k$ for~$\mathcal{I'}_{\textsc{Sat}}$. Further, each~$P_i$ is a partition of $X$ as, for each $t=(u_1,\ldots, u_k) \in T_U$, $u_i = u$ must be set to $T$ for exactly one $u \in U_i$, as $T_U$ is formed over $U_1 \times U_2\times \dots \times U_k$. Consequently, each variable $x_t\in X$ belongs to exactly one clause in~$P_i$. The partitions $P_1, \ldots, P_k$ are pairwise disjoint: for $u \in U_i$ and $u' \in U_j$ with $i \neq j$, we have $\{x_t \mid t=(u_1,\ldots, u_k) \in T_U \text{ and } u_i = u\} = \{x_t \mid t=(u_1,\ldots, u_k) \in T_U \text{ and } u_j = u'\}$ if and only if $u$ and $u'$ occur in exactly the same three tuples, which the definition of \textsc{$k$DM-E3} excludes. It now follows that~$\mathcal{I'}_{\textsc{Sat}}$ is an instance of \textsc{$k$-Partitioned 1-in-3-Sat}, and we have established a method to translate back and forth between instances of \textsc{$k$DM-E3} and \textsc{$k$-Partitioned 1-in-3-Sat}. Both translations can be computed in time linear in the size of the input, so they serve as polynomial-time reductions in both directions.

Note that for an instance of \textsc{$k$DM-E3} called $\mathcal{I}_{\textsc{Mat}}=(U_1, \dots, U_k, T_U)$ and its associated instance of \textsc{$k$-Partitioned 1-in-3-Sat} called $\mathcal{I}_{\textsc{Sat}} = (X, C, P)$ and further, a \textsc{$k$DM-E3} instance $\mathcal{I}_{\textsc{Mat}}'$ associated with $\mathcal{I}_{\textsc{Sat}}$, we have that $\mathcal{I}_{\textsc{Mat}}'$ can be obtained from $\mathcal{I}_{\textsc{Mat}}$ by renaming the elements in $U$.

We can investigate the relationship between satisfying assignments and perfect matchings of associated \textsc{1-in-3-Sat} and matching instances.

\begin{pro}\label{pro:one-to-one-sat-mat}
Let $\mathcal{I}_{\textsc{Sat}}= (X, C, P)$ be an instance of \textsc{$k$-Partitioned 1-in-3-Sat}. There is a one-to-one correspondence between the truth assignments for $X$ that 1-in-3-satisfy each clause in~$C$ and the perfect matchings of the instance $\mathcal{I}_{\textsc{Mat}} = (U_1, \dots, U_k, T_U)$ of \textsc{$k$DM-E3} associated with~$\mathcal{I}_{\textsc{Sat}}$.
\end{pro}

\begin{proof}
Let $\beta\colon X \rightarrow \{T,F\}$ be a truth assignment. We define
\[
M_\beta = \{\varphi(x) \mid x \in X \text{ and } \beta(x) = T\}.
\]
Recall that a clause $c \in P_i$ is the $i$-th coordinate of $\varphi(x)$ if and only if $x \in c$. Hence, the tuples in $M_\beta$ that contain~$c$ are exactly the tuples $\varphi(x)$ with $x \in c$ and $\beta(x) = T$. If $\beta$ 1-in-3-satisfies $C$, then exactly one variable in each clause $c \in C$ is set to $T$, so each element $c \in U$ appears in exactly one tuple of $M_\beta$, and $M_\beta$ is a perfect matching. Note that the perfect matchings are exactly the solutions asked for in \textsc{$k$DM-E3}: a set $M \subseteq T_U$ covers each element of~$U$ exactly once if and only if $|M| = |U_1|$ and no two distinct tuples in~$M$ agree in any coordinate, where the backward direction uses that $|U_i| = |T_U|/3 = |U_1|$ for all~$i$. Since $|\{i \in \{1, \ldots, k\} \mid \exists c \in P_i \colon \{x,x'\} \subseteq c\}| < k$, $\varphi$ is injective and, thus, $M_\beta \neq M_{\beta'}$ for all truth assignments $\beta \neq \beta'$ that 1-in-3-satisfy~$C$.

Conversely, let $M$ be a perfect matching of $\mathcal{I}_{\textsc{Mat}}$. Since $T_U = \{\varphi(x) \mid x \in X\}$, we can write $M = \{\varphi(x) \mid x \in X'\}$ for $X' = \varphi^{-1}(M) \subseteq X$. Let $\gamma\colon X \rightarrow \{T,F\}$ with $\gamma(x) = T$ if and only if $x \in X'$. For each clause $c \in C$, the tuples in $M$ that contain~$c$ are exactly the tuples $\varphi(x)$ with $x \in c \cap X'$. As $M$ is a perfect matching, there is exactly one such tuple, and as $\varphi$ is injective, $c$ contains exactly one variable $x$ with $\gamma(x) = T$. Thus, $\gamma$ 1-in-3-satisfies~$C$, and $M_\gamma = M$ by construction. Hence,
\[
\Psi\colon \{\beta\colon X\rightarrow \{T, F\} \mid \beta \text{ 1-in-3-satisfies } C \} \rightarrow \{M \subseteq T_U \mid M \text{ is a perfect matching of } \mathcal{I}_{\textsc{Mat}}\}
\]
with $\Psi(\beta) = M_\beta$ is a bijective mapping between the truth assignments for $X$ that 1-in-3-satisfy $C$ and the perfect matchings in the instance $\mathcal{I}_{\textsc{Mat}}$ of \textsc{$k$DM-E3}.
\end{proof}

The mapping that sends each clause $c \in C$ to the element $c \in U$ and each variable $x \in X$ to the $k$-tuple $\varphi(x) \in T_U$ is a bijection between the vertex sets of $G_{\textsc{Sat}}(X, C, P)$ and $G_{\textsc{Mat}}(U_1, \dots, U_k, T_U)$, since $U = C$ and $\varphi$ is injective with image~$T_U$. It preserves edges, since $x \in c$ holds if and only if $c$ is a coordinate of $\varphi(x)$. Treating the converse direction analogously, we obtain that $G_{\textsc{Sat}}(X, C, P)$ and $G_{\textsc{Mat}}(U_1, \dots, U_k, T_U)$ are isomorphic for $\mathcal{I}_{\textsc{Sat}} = (X, C, P) \in \textsc{$k$-Partitioned 1-in-3-Sat}$ and $\mathcal{I}_{\textsc{Mat}}=(U_1, \dots, U_k, T_U)$, the associated instance of \textsc{$k$DM-E3}. It follows that planarity and linearity are preserved.\\

This correspondence has motivated our paper. Since our hardness results for
\textsc{$k$-Partitioned 1-in-3-Sat} hold even under additional structural restrictions,
we can use the correspondence to transfer them directly from \textsc{Sat} to \textsc{$k$-Dimensional Matching}. 

\begin{cor}[Theorem~\ref{thm:planr3con}]\label{cor:3dm}
\textsc{Planar 3-Connected 3DM-E3} is \textsc{NP}-complete, and its counting variant is $\#$P-complete.
\end{cor}
This answers the question `Does planar 3-connected 1-in-3-SAT remain \textsc{NP}-hard when every variable appears in 3 clauses?' by Hasan, Mondal and Rahman~\cite{Hasan2022}. Since our reductions are parsimonious, the bijections between solutions
are computable in polynomial time, and \textsc{Positive 1-in-3-Sat} is
ASP-complete~\cite[Lemma~2.1]{Bosboom2020}, this also shows that
\textsc{Planar 3DM} is ASP-complete and that its counting variant is
\#\textsc{P}-complete, which answers the open question of Bosboom,
Demaine, Hesterberg, Kimball and Kopinsky~\cite{Bosboom2020}. It also reproves that counting the solutions of \textsc{Cubic Planar X3C} is \#\textsc{P}-complete, which was asked by Moore and Robson~\cite{Moore2001} and shown by Cai and Fan~\cite{Cai2023}.

\begin{cor}[Theorem~\ref{cor:lin3conkpart}]\label{cor:linkdm}
For every $k\geq 3$, \textsc{Linear 3-Connected $k$DM-E3} is \textsc{NP}-complete, and its counting variant is $\#$P-complete.
\end{cor}

\begin{cor}[Theorem~\ref{thm:plan4partmain}]\label{cor:4dm}
\textsc{Planar 3-Connected 4DM-E3} is \textsc{NP}-complete, and its counting variant is $\#$P-complete.
\end{cor}

\begin{cor}[Theorem~\ref{thm:planptime}]\label{cor:fpt4dm}
\textsc{Planar 3-Connected 4DM-E3} is fixed-parameter tractable when parameterized by the number of quadrilateral components.
\end{cor}

\begin{cor}[Corollary~\ref{cor:nolinplan}]\label{cor:nolin}
For $k \geq 3$, there are no instances of \textsc{Planar Linear $k$DM-E3}.
\end{cor}

\begin{cor}[Theorem~\ref{thm:doublestardied}]\label{cor:no5}
There are no yes-instances of \textsc{Planar 5DM-E3}.
\end{cor}

\begin{cor}[Observation~\ref{obs:noplank>5}]\label{cor:nok}
For $k>5$, there are no instances of \textsc{Planar $k$DM-E3}.
\end{cor}

In order to avoid redundancies, we have not listed immediate consequences such as \textsc{NP}-completeness of a class containing a class that is \textsc{NP}-complete. Similarly, for proving triviality or fixed-parameter tractability, we provided an algorithm that vacuously also works for restricted inputs.

\problemStatement{Exact Cover-E3 By $k$-Sets (X$k$C-E3)}{
A finite set $X$ and a collection $\mathcal{S}$ of $k$-element subsets
of~$X$ such that every element of $X$ occurs in exactly three members
of~$\mathcal{S}$.
}{
Is there a subcollection $\mathcal{S}' \subseteq \mathcal{S}$ such that every
element of $X$ occurs in exactly one member of~$\mathcal{S}'$?
}
For a finite set $A$ and a family $\mathcal{S}$ of subsets of $A$, the
\emph{incidence graph of an exact cover} $G(A,\mathcal{S})$ has vertex set $A \cup \mathcal{S}$
and an edge $\{u,S\}$ whenever $u \in S \in \mathcal{S}$.

\begin{thm}\label{thm:planexactcover} There is a parsimonious reduction from \textsc{$k$-Partitioned 1-in-3-Sat} to
\textsc{Exact Cover by $k$-Sets} that maintains planarity, linearity and 3-connectedness.
\end{thm}
Van Bevern, Bredereck, Chen, Froese, Niedermeier and Woeginger~\cite{bevern} asked whether \textsc{Planar Exact Cover by $4$-Sets} is \textsc{NP}-hard. This was recently answered by the MIT Hardness group (Abel, Demaine, Diomidova, Li and Zhou)
~\cite{MITHardnessGroup2026}: \textsc{Planar Exact Cover by $k$-Sets} is \textsc{NP}-hard for $k \in \{3,4\}$ and polynomial-time solvable otherwise. We present an alternative proof by reducing from \textsc{$k$-Partitioned 1-in-3-Sat}.

\begin{proof}
Let $\mathcal{I} = (X, C, P)$ with $P = \langle P_1, \ldots, P_k\rangle$ be
an instance of \textsc{$k$-Partitioned 1-in-3-Sat}. For $x \in X$ and
$i \in \{1, \ldots, k\}$, let $c_{i,x}$ denote the unique clause of~$P_i$
containing~$x$. For every clause $c \in C$, introduce a distinct new
element~$u_c$, and define
\[
  A := \{u_c : c \in C\}, \qquad
  S_x := \{u_c : c \in C,\ x \in c\} \quad \text{for all } x \in X, \qquad
  \mathcal{S} := \{S_x : x \in X\}.
\]
This takes linear time. As the elements~$u_c$ are distinct,
\begin{equation}\label{eq:incidence}
  u_c \in S_x \iff x \in c \qquad \text{for all } x \in X \text{ and } c \in C.
\end{equation}

Every clause containing~$x$ lies in some~$P_i$ and thus equals~$c_{i,x}$,
and $c_{1,x}, \ldots, c_{k,x}$ are pairwise distinct since
$P_1, \ldots, P_k$ are pairwise disjoint. Hence,
$S_x = \{u_{c_{1,x}}, \ldots, u_{c_{k,x}}\}$ has exactly $k$ elements.
The map $\sigma\colon X \to \mathcal{S}$, $x \mapsto S_x$, is a bijection:
it is surjective by definition, and $S_x = S_{x'}$ with $x \neq x'$ would
give, for every~$i$, $u_{c_{i,x}} \in S_{x'}$ and hence
$\{x, x'\} \subseteq c_{i,x} \in P_i$ by~\eqref{eq:incidence},
contradicting $|\{i \in \{1, \ldots, k\} \mid \exists c \in P_i \colon \{x,x'\} \subseteq c\}| < k$. By~\eqref{eq:incidence}, the members
of~$\mathcal{S}$ containing~$u_c$ are the sets~$S_x$ with $x \in c$, which
are exactly three distinct sets as $\sigma$ is injective. Thus,
$(A,\mathcal{S})$ is an instance of \textsc{X$k$C-E3}.

For $\beta\colon X \to \{0,1\}$, let $\Phi(\beta) := \{S_x : \beta(x) = 1\}$.
By~\eqref{eq:incidence} and the injectivity of~$\sigma$, for every $c \in C$,
the map~$\sigma$ restricts to a bijection from $\{x \in c : \beta(x) = 1\}$
onto $\{S \in \Phi(\beta) : u_c \in S\}$. Hence, $\beta$ 1-in-3-satisfies~$C$
if and only if $\Phi(\beta)$ is an exact cover of~$A$. Since $\sigma$ is a
bijection, so is $\Phi\colon \{0,1\}^X \to 2^{\mathcal{S}}$; thus, $\Phi$
restricts to a bijection between the 1-in-3-satisfying assignments
of~$\mathcal{I}$ and the exact covers of $(A,\mathcal{S})$, and the reduction
is parsimonious.

Finally, by~\eqref{eq:incidence}, the bijection $x \mapsto S_x$,
$c \mapsto u_c$ is an isomorphism from $G_{\textsc{Sat}}(X,C)$ onto
$G(A,\mathcal{S})$. Hence, planarity, linearity and 3-connectedness are maintained.
\end{proof}

This yields the following corollaries: 

\begin{cor}[Theorem~\ref{thm:planr3con}]
\textsc{Planar 3-Connected Exact Cover-E3 by 3-Sets} is \textsc{NP}-complete, and its counting variant is \#\textsc{P}-complete.
\end{cor}

\begin{cor}[Theorem~\ref{cor:lin3conkpart}]
For every $k\geq 3$, \textsc{Linear 3-Connected Exact Cover-E3 by $k$-Sets} is \textsc{NP}-complete, and its counting variant is \#\textsc{P}-complete.
\end{cor}

Equivalently, by Theorem~\ref{thm:planexactcover}, \textsc{XSat} is \textsc{NP}-complete for monotone linear formulas with exactly three literals per clause and exactly $k$ occurrences per variable for every $k\geq 3$, which confirms the case of clause size three of the \textsc{XSat} part of a conjecture by Porschen, Schmidt, Speckenmeyer and Wotzlaw~\cite{Porschen2014}.

\begin{cor}[Theorem~\ref{thm:plan4partmain}]
\textsc{Planar 3-Connected Exact Cover-E3 by 4-Sets} is \textsc{NP}-complete, and its counting variant is \#\textsc{P}-complete.
\end{cor}

\section{Conclusion}
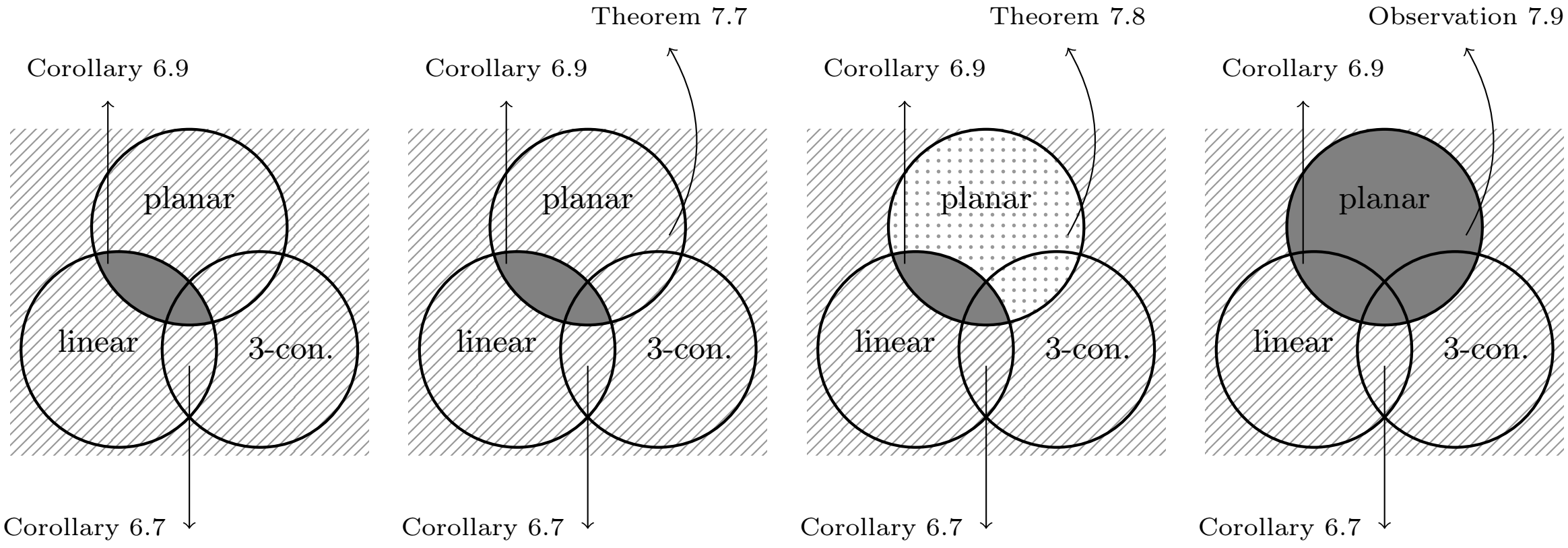
\begin{figure}[h]
\centering
\newcommand{\vennarrow}[5]{%
  \if\relax\detokenize{#5}\relax\else
    \draw[->, #3] (#1) to (#2);
    \node[#4] at (#2) {\scriptsize #5};
  \fi
}

\newcommand{\vennwithsixarrows}[8]{%
\begin{subfigure}[t]{0.22\textwidth}
\centering
\begin{tikzpicture}[scale=0.8, every node/.style={font=\small}]
  \def\r{1.2}
  \coordinate (A) at (90:1.0);
  \coordinate (B) at (210:1.0);
  \coordinate (C) at (330:1.0);

  \begin{scope}
    \clip (-2.2,-1.8) rectangle (2.2,2.2);
    \fill[pattern=north east lines, pattern color=black!40] (-2.2,-1.8) rectangle (2.2,2.2);
  \end{scope}

  #1

  \draw[thick] (A) circle (\r);
  \draw[thick] (B) circle (\r);
  \draw[thick] (C) circle (\r);
  \node[above] at ($(A)+(0,0)$) {\small planar};
  \node[below left] at ($(B)+(0.4,0.4)$) {\small linear};
  \node[below right] at ($(C)+(-0.3,0.3)$) {\small 3-con.};

  \coordinate (R1) at (1,0.9);      
  \coordinate (R2) at (-1.8,-1);    
  \coordinate (R3) at (1.95,-0.5);  
  \coordinate (R4) at (-1,0.55);    
  \coordinate (R5) at (0.6,0.1);    
  \coordinate (R6) at (0,-0.7);     
  \coordinate (R7) at (0,-1.4);

  \coordinate (T1) at ($(R1)+(0,2.3)$);
  \coordinate (T2) at ($(R2)+(0,-3.0)$);
  \coordinate (T3) at ($(R3)+(0,-3.0)$);
  \coordinate (T4) at ($(R4)+(0,2.0)$);
  \coordinate (T5) at ($(R5)+(0,-3.0)$);
  \coordinate (T6) at ($(R6)+(0,-2.0)$);

  \vennarrow{R1}{T1}{bend right}{above=3pt}{#2}
  \vennarrow{R2}{T2}{}{below=3pt, xshift=-6pt}{#3}
  \vennarrow{R3}{T3}{}{below=3pt, xshift=6pt}{#4}
  \vennarrow{R4}{T4}{}{above=2pt}{#5}
  \vennarrow{R6}{T6}{}{left=3pt}{#7}

  \path (-2.2,-3.5) rectangle (2.2,4.0);
\end{tikzpicture}
\caption{#8}
\end{subfigure}%
}

\vennwithsixarrows{
  \begin{scope}
    \clip (A) circle (\r);
    \fill[gray] (B) circle (\r);
  \end{scope}
}{}{}{}{Corollary \ref{cor:nolinplan}}{Theorem \ref{thm:planr3con}}{Corollary \ref{cor:lin3conkpart}}{3-partitioned}
\hspace{0.5em}
\vennwithsixarrows{
  \begin{scope}
    \clip (A) circle (\r);
    \fill[white] (A) circle (\r);
    \fill[pattern=north east lines, pattern color=black!40] (A) circle (\r);
  \end{scope}%
  \begin{scope}
    \clip (B) circle (\r);
    \fill[white] (C) circle (\r);
    \fill[pattern=north east lines, pattern color=black!40] (C) circle (\r);
  \end{scope}
  \begin{scope}
    \clip (A) circle (\r);
    \fill[gray] (B) circle (\r);
  \end{scope}%
}{Theorem \ref{thm:planptime}}{}{}{Corollary \ref{cor:nolinplan}}{}{Corollary \ref{cor:lin3conkpart}}{4-partitioned}
\hspace{0.5em}
\vennwithsixarrows{
  \begin{scope}
    \clip (A) circle (\r);
    \fill[white] (-2.2,-2.2) rectangle (2.2,2.2);
  \end{scope}%
  \begin{scope}
    \fill[pattern=dots, pattern color=black!40] (A) circle (\r);
  \end{scope}
  \begin{scope}
    \clip (B) circle (\r);
    \fill[pattern=north east lines, pattern color=black!40] (C) circle (\r);
  \end{scope}
  \begin{scope}
    \clip (A) circle (\r);
    \fill[gray] (B) circle (\r);
  \end{scope}
}{Theorem \ref{thm:doublestardied}}{}{}{Corollary \ref{cor:nolinplan}}{}{Corollary \ref{cor:lin3conkpart}}{5-partitioned}
\hspace{0.5em}
\vennwithsixarrows{
  \fill[gray] (A) circle (\r);
}{Observation \ref{obs:noplank>5}}{}{}{Corollary \ref{cor:nolinplan}}{}{Corollary \ref{cor:lin3conkpart}}{$k$-partitione, $k>5$}
\caption{Overview of the complexity of \textsc{$k$-Partitioned 1-in-3-Sat} for all combinations of planarity, linearity and 3-connectivity (`3-con.'). Hatched regions indicate \textsc{NP}-hardness, gray regions indicate that no instances exist, and dotted regions indicate that the problem is trivial in the sense that there are only no-instances. Arrows point to the result that establishes the complexity of the respective region.}\label{fig:Pvendiagram}
\end{figure}

In this paper, we established a complete complexity classification of \textsc{$k$-Partitioned 1-in-3-Sat}, for every $k$, with respect to all combinations of the properties `planar', `linear' and `3-connected'. From this, we derived the corresponding results for \textsc{$k$DM} and \textsc{X$k$C}. Note that in \textsc{$k$DM} we assumed that no $u \in U_i$ and $u' \in U_j$ with $i \neq j$ occur in exactly the same three $k$-tuples and in \textsc{$k$-Partitioned}, we required that $|\{i \in \{1, \ldots, k\} \mid \exists c \in P_i \colon \{x,x'\} \subseteq c\}| < k$. Figure \ref{fig:Pvendiagram} gives an overview of the results in the language of \textsc{$k$-Partitioned 1-in-3-Sat}. We also showed that the counting versions of the \textsc{NP}-complete variants are \#\textsc{P}-complete.

We started by exhibiting the idea of inserting graph gadgets into a \textsc{Sat} graph as it was done by Moore and Robson~\cite{Moore2001}.
These gadgets are designed so that the resulting formula has the desired properties.
We first applied this approach to \textsc{3-Partitioned 1-in-3-Sat}.
To obtain planarity, we replaced each edge crossing with a crossover gadget. Similarly, we inserted linearity gadgets to make the formula linear. Motivated by the question of Hasan, Mondal and Rahman~\cite{Hasan2022} whether \textsc{Planar 3-Connected 1-in-3-Sat} remains \textsc{NP}-hard when every variable appears in exactly three clauses, we focused on how to extend our results to the additional property of 3-connectivity. To this end, we added a grid structure to the graph, which again allows us to insert gadgets, and proved that the resulting graph is 3-connected. We then extended our investigation to $k > 3$. For \textsc{4-Partitioned 1-in-3-Sat}, we abstracted from the idea of a plug-in system of gadgets and  constructed gadgets that simulate a clause or a variable vertex. Although \textsc{Planar 3-Connected 4-Partitioned 1-in-3-Sat} is
\textsc{NP}-complete, we showed that it is fixed-parameter tractable when
parameterized by the number of quadrilateral components. Finally, we transferred our results to \textsc{$k$DM} and \textsc{X$k$C}. In particular, we gave an alternative answer to the question of van Bevern, Bredereck, Chen, Froese, Niedermeier and Woeginger~\cite{bevern} on the complexity of \textsc{X4C} restricted to instances in which every element occurs in exactly three sets, which was recently also resolved by the MIT Hardness Group~\cite{MITHardnessGroup2026}

We view these results as motivation for further study of restricted satisfiability problems: the $k$-partitioned variant of \textsc{Sat} provides a complexity-theoretic framework that reveals structure in related problems such as \textsc{$k$DM} and \textsc{X$k$C}. We leave the following open problem for future research.

\begin{qst}
For which $r \geq 4$ is \textsc{Cubic 3-Connected 3DM}, restricted to instances whose incidence graph has girth~$2r$, \textsc{NP}-complete? For which $r$ is its counting version \#\textsc{P}-complete? 
\end{qst}

\section*{AI use disclosure}
\begin{itemize}
    \item Sections 1--5 and \ref{sec:matchingandexactcover} were developed without the use of LLMs. LLMs were used in the remaining parts of this work as follows.
    \item Content:
    \begin{itemize}
        \item The main inspiration of Lemma \ref{lem:noartpoint} arose from conversation with Claude Opus 5.5.
        \item The gadgets in Figures \ref{fig:4withoutpartNPgadget}, \ref{fig:C} and \ref{fig:Xgadget} as well as the graph examples in Figure \ref{fig:quadcom} were found by a computer search using code generated with Claude Code (Opus 5.5). The graph in Figure \ref{fig:counterexample4planpart} was found by code generated by Codex (GPT-6.1 Sol).
        \item The derivations in the Appendix in the proof of Lemma \ref{lem:derive} were computed by Claude Code (Opus 5.5).
        \item Theorem \ref{thm:doublestardied} benefited from substantial corrections from Claude Opus 5.5.
    \end{itemize}
    \item Style and notation:
    \begin{itemize}
        \item The \LaTeX{} code for Figures~\ref{fig:pathsinreplacementlemma} and \ref{fig:Pvendiagram} was generated by GPT-6.1 Sol.
        \item The final manuscript was proofread by Claude Opus 5.5 for mathematical errors, spelling, and readability.
    \end{itemize}
    All AI-generated content, code output and suggestions were reviewed and verified by the authors, who take full responsibility for the content of this paper.
\end{itemize}

\section*{\phantom{References}} 
\addcontentsline{toc}{section}{References} 
\bibliographystyle{alpha}
\bibliography{mylit}

\section*{Appendix}
\subsection*{Appendix for Section \ref{sec:hardnesslinearity}}
\begin{figure}[h]
    \centering
    \includegraphics[width=\textwidth]{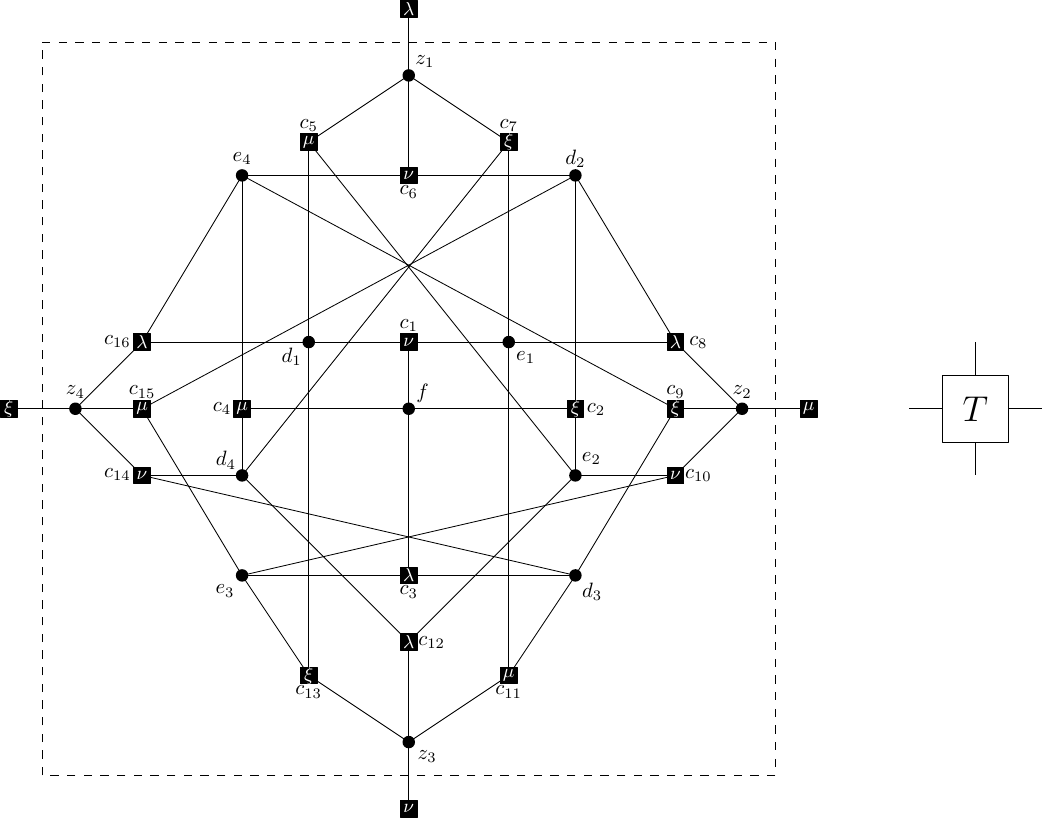}
    \caption{$T$-cross and its abbreviation.}
    \label{fig:t_cross}
\end{figure}

\addcontentsline{toc}{section}{Appendix}
\paragraph{$T$-cross} The $T$-cross consists of variables $z_1, z_2, z_3, z_4$, and $f$ as well as eight auxiliary variables and~16 linear clauses as illustrated in Figure \ref{fig:t_cross}. There is exactly one truth assignment $\beta$ that 1-in-3-satisfies the clauses of the $T$-cross. This also satisfies $\beta(z_1) = \beta(z_2) = \beta(z_3) = \beta(z_4) = T$. 

\begin{figure}[h]
    \centering
    \includegraphics[width=.75\textwidth]{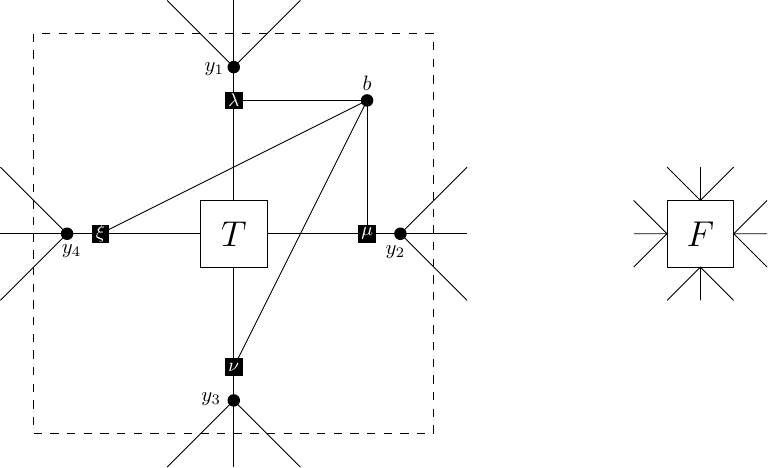}
    \caption{$F$-cross and its abbreviation.}
    \label{fig:f_cross}
\end{figure}

\paragraph{$F$-cross} The $F$-cross consists of a $T$-cross, the variables $y_1, y_2, y_3, y_4, b$, and four additional linear clauses as illustrated in Figure \ref{fig:f_cross}. There is only one truth assignment $\beta$ that 1-in-3-satisfies every clause of the $F$-cross. This also satisfies $\beta(y_1) = \beta(y_2) = \beta(y_3) = \beta(y_4) = F$.

\begin{figure}[h]
    \centering
    \includegraphics[width=.75\textwidth]{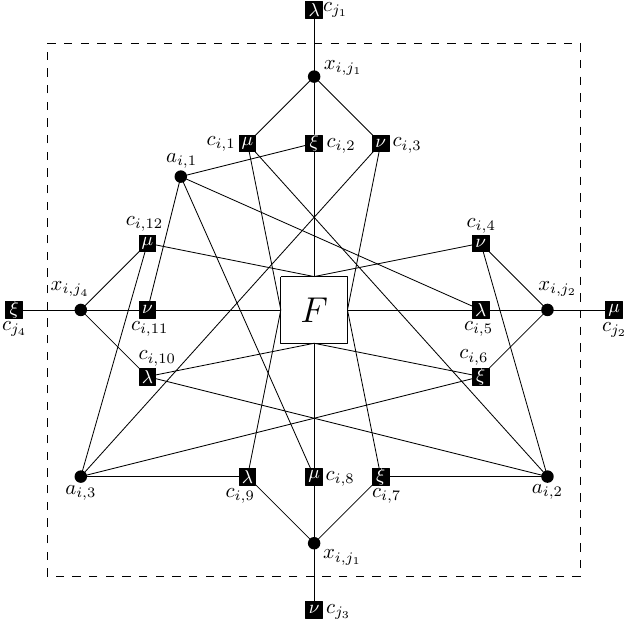}
    \caption{Linearity gadget for four partitions.}
    \label{fig:lin_4}
\end{figure}

\paragraph{Linear equality gadget} The linear equality gadget $LEQ(x_i)$ allows to replace one variable $x_i$ appearing in clauses $c_{j_1}, c_{j_2}, c_{j_3}, c_{j_4}$ with four variables $x_{i,j_1}, x_{i,j_2}, x_{i,j_3}, x_{i,j_4}$ to make the clauses $c_{j_1}, c_{j_2}, c_{j_3}, c_{j_4}$ linear. It contains an $F$-cross, three auxiliary variables, and twelve additional linear clauses as illustrated in Figure \ref{fig:lin_4}. A truth assignment $\beta$ 1-in-3-satisfies the clauses of the gadget if and only if $\beta(x_{i,j_1}) = \beta(x_{i,j_2}) = \beta(x_{i,j_3}) = \beta(x_{i,j_4})$. \\


The following lemmata are analogous to the lemmata for the gadgets for three partitions and also hold for four partitions. Their proofs are analogous to the proofs provided there as well.
\begin{lem}\label{lem:t_cross}
Let $TC$ be a $T$-cross with variables
\[ A = \{z_1,z_2,z_3,z_4\} \cup \{f\} \cup \{d_i, e_i \mid 1 \le i \le 4\} \]
and clauses $C_A = \{c_j \mid 1 \le j \le 16\}$ as illustrated in Figure \ref{fig:t_cross}. Then, there is exactly one truth assignment $\beta\colon A \rightarrow \{T, F\}$ that 1-in-3-satisfies each clause in $C_A$. This truth assignment satisfies $\beta(z_1) = \beta(z_2) = \beta(z_3) = \beta(z_4) = T$.
\end{lem}

\begin{lem}\label{lem:f_cross} Let $FC$ be an $F$-cross with variables
\[ A = \{y_1,y_2,y_3,y_4\} \cup \{b\} \cup \operatorname{var}(TC) \] 
as illustrated in Figure~\ref{fig:f_cross}. Then, there is exactly one truth assignment $\beta\colon A \rightarrow \{T, F\}$ that 1-in-3-satisfies each clause in $\operatorname{cl}(FC)$. This truth assignment satisfies $\beta(y_1) = \beta(y_2) = \beta(y_3) = \beta(y_4) = F$.
\end{lem}

\begin{lem}\label{lem:linear_4} Let $LEQ(x_i)$ be a linear equality gadget for variable $x_i$, where $x_i$ is contained in the clauses $c_{j_1}, c_{j_2}, c_{j_3}, c_{j_4}$. It replaces $x_i$ with variables 
\[ A = \{x_{i,j_1}, x_{i,j_2}, x_{i,j_3}, x_{i,j_4}\} \cup \{a_{i,1}, a_{i,2}, a_{i,3}\} \cup \operatorname{var}(FC_i) \]
and clauses $C_A = \{c_{i,j} \mid 1 \le j \le 12\} \cup \operatorname{cl}(FC_i)$ as illustrated in Figure \ref{fig:lin_4}. Then, there are exactly two truth assignments $\beta_1$ and~$\beta_2$ for $\operatorname{var}(LEQ(x_i))$ that 1-in-3-satisfy each clause in $\operatorname{cl}(LEQ(x_i))$. Further,~$\beta_1$ and $\beta_2$ satisfy 
\begin{align*}
\beta_1(x_{i,j_1}) = \beta_1(x_{i,j_2}) = \beta_1(x_{i,j_3}) = \beta_1(x_{i,j_4}) = T, && \beta_1(a_{i,1}) = \beta_1(a_{i,2}) = \beta_1(a_{i,3}) = F, \\
\beta_2(x_{i,j_1}) = \beta_2(x_{i,j_2}) = \beta_2(x_{i,j_3}) = \beta_2(x_{i,j_4}) = F, && \beta_2(a_{i,1}) = \beta_2(a_{i,2}) = \beta_2(a_{i,3}) = T.
\end{align*}
\end{lem}

\begin{lem}\label{lem:linear_4_lin} Let $LEQ(x_i)$ be a linear equality gadget for variable $x_i$. All clauses in $\operatorname{cl}(LEQ(x_i))$ are linear.
\end{lem}

\subsection*{Extended proofs for Section \ref{sec:4part}}

In Tables~\ref{tab:C} and~\ref{tab:X}, we include glue vertices.

\begin{center}\footnotesize\setlength{\tabcolsep}{3pt}
\begin{tabular}{@{}llll@{}}
\toprule
$P_1$ & $P_2$ & $P_3$ & $P_4$\\
\midrule
$\{a_{2},d_{6},k_{1}\},\{a_{4},d_{4},k_{4}\}$ & $\{a_{2},d_{3},k_{1}\},\{a_{4},d_{5},k_{4}\}$ & $\{a_{2},d_{3},k_{2}\},\{a_{3},d_{6},k_{1}\}$ & $\{a_{2},d_{2},k_{2}\},\{a_{3},d_{3},k_{1}\}$\\
$\{a_{5},d_{5},k_{3}\},\{a_{6},b_{1},f\}$ & $\{a_{5},d_{4},k_{3}\},\{a_{6},b_{2},f\}$ & $\{a_{4},b_{2},f\},\{a_{5},d_{5},k_{4}\}$ & $\{a_{4},b_{1},f\},\{a_{5},d_{4},k_{4}\}$\\
$\{b_{2},d_{3},e_{2}\},\{e_{1},h,k_{2}\}$ & $\{d_{2},e_{1},g_{2}\},\{e_{2},h,k_{2}\}$ & $\{d_{4},e_{1},g_{2}\},\{e_{2},h,k_{3}\}$ & $\{b_{2},d_{5},e_{2}\},\{e_{1},h,k_{3}\}$\\
\midrule
$\{a_{1},p_{2},z_{2}\},\{a_{3},p_{4},z_{1}\}$ & $\{a_{3},p_{2},z_{1}\},\{a_{7},p_{4},z_{4}\}$ & $\{a_{6},p_{3},z_{4}\},\{a_{7},p_{4},z_{5}\}$ & $\{a_{1},p_{1},z_{2}\},\{a_{6},p_{4},z_{4}\}$\\
$\{a_{7},p_{3},z_{4}\},\{a_{8},p_{1},z_{6}\}$ & $\{a_{8},p_{1},z_{2}\},\{g_{1},p_{3},z_{3}\}$ & $\{a_{8},p_{2},z_{2}\},\{g_{1},p_{1},z_{3}\}$ & $\{a_{7},p_{3},z_{3}\},\{a_{8},p_{2},z_{1}\}$\\
$\{z_{3},z_{5},o_{1}\},\{z_{7},z_{8},o_{2}\}$ & $\{z_{5},z_{8},o_{1}\},\{z_{6},z_{7},o_{2}\}$ & $\{z_{1},z_{6},o_{2}\},\{z_{7},z_{8},o_{1}\}$ & $\{z_{5},z_{8},o_{2}\},\{z_{6},z_{7},o_{1}\}$\\
\midrule
$\{d_{1},q_{2},z_{10}\},\{d_{2},q_{4},z_{9}\}$ & $\{d_{6},q_{3},z_{12}\},\{d_{7},q_{4},z_{13}\}$ & $\{d_{2},q_{2},z_{9}\},\{d_{7},q_{4},z_{12}\}$ & $\{d_{1},q_{1},z_{10}\},\{d_{6},q_{4},z_{12}\}$\\
$\{d_{7},q_{3},z_{12}\},\{d_{8},q_{1},z_{14}\}$ & $\{d_{8},q_{2},z_{10}\},\{g_{3},q_{1},z_{11}\}$ & $\{d_{8},q_{1},z_{10}\},\{g_{3},q_{3},z_{11}\}$ & $\{d_{7},q_{3},z_{11}\},\{d_{8},q_{2},z_{9}\}$\\
$\{z_{11},z_{13},o_{3}\},\{z_{15},z_{16},o_{4}\}$ & $\{z_{9},z_{14},o_{4}\},\{z_{15},z_{16},o_{3}\}$ & $\{z_{13},z_{16},o_{3}\},\{z_{14},z_{15},o_{4}\}$ & $\{z_{13},z_{16},o_{4}\},\{z_{14},z_{15},o_{3}\}$\\
\bottomrule
\end{tabular}
\captionof{table}{The clauses of the $C$-gadget including the glue vertices.}\label{tab:C}
\end{center}

\begin{center}\footnotesize\setlength{\tabcolsep}{3pt}
\begin{tabular}{@{}llll@{}}
\toprule
$P_1$ & $P_2$ & $P_3$ & $P_4$\\
\midrule
$\{i_{1},b_{2},z_{2}\},\{i_{2},b_{5},z_{3}\}$ & $\{i_{5},b_{5},z_{6}\},\{i_{7},b_{2},z_{4}\}$ & $\{i_{5},b_{5},z_{3}\},\{i_{7},b_{2},z_{5}\}$ & $\{i_{1},b_{2},z_{4}\},\{i_{2},b_{5},z_{7}\}$\\
$\{i_{5},b_{4},z_{6}\},\{i_{6},b_{3},z_{5}\}$ & $\{i_{8},b_{3},z_{1}\},\{i_{9},b_{1},z_{5}\}$ & $\{i_{8},b_{3},z_{8}\},\{i_{9},b_{1},z_{4}\}$ & $\{i_{5},b_{4},z_{3}\},\{i_{6},b_{3},z_{1}\}$\\
$\{i_{7},b_{1},z_{4}\},\{i_{8},b_{6},z_{1}\}$ & $\{l_{1},b_{4},z_{3}\},\{l_{2},b_{6},z_{7}\}$ & $\{l_{1},b_{4},z_{2}\},\{l_{2},b_{6},z_{1}\}$ & $\{i_{7},b_{1},z_{5}\},\{i_{8},b_{6},z_{8}\}$\\
$\{z_{7},z_{8},o_{1}\}$ & $\{z_{2},z_{8},o_{1}\}$ & $\{z_{6},z_{7},o_{1}\}$ & $\{z_{2},z_{6},o_{1}\}$\\
\midrule
$\{i_{9},b_{8},z_{10}\},\{i_{11},b_{9},z_{12}\}$ & $\{i_{6},b_{9},z_{11}\},\{i_{11},b_{10},z_{12}\}$ & $\{i_{6},b_{7},z_{11}\},\{i_{10},b_{9},z_{12}\}$ & $\{i_{9},b_{7},z_{10}\},\{i_{10},b_{10},z_{12}\}$\\
$\{i_{12},b_{7},z_{14}\},\{i_{14},b_{10},z_{9}\}$ & $\{i_{12},b_{7},z_{10}\},\{i_{14},b_{8},z_{9}\}$ & $\{i_{11},b_{10},z_{13}\},\{i_{12},b_{8},z_{10}\}$ & $\{i_{11},b_{9},z_{11}\},\{i_{12},b_{8},z_{9}\}$\\
$\{z_{11},z_{13},o_{2}\},\{z_{15},z_{16},o_{3}\}$ & $\{z_{13},z_{16},o_{2}\},\{z_{14},z_{15},o_{3}\}$ & $\{z_{9},z_{14},o_{3}\},\{z_{15},z_{16},o_{2}\}$ & $\{z_{13},z_{16},o_{3}\},\{z_{14},z_{15},o_{2}\}$\\
\midrule
$\{i_{10},b_{14},z_{17}\},\{i_{13},b_{12},z_{18}\}$ & $\{i_{10},b_{12},z_{17}\},\{i_{15},b_{14},z_{20}\}$ & $\{i_{14},b_{13},z_{20}\},\{i_{15},b_{14},z_{21}\}$ & $\{i_{13},b_{11},z_{18}\},\{i_{14},b_{14},z_{20}\}$\\
$\{i_{15},b_{13},z_{20}\},\{i_{16},b_{11},z_{22}\}$ & $\{i_{16},b_{11},z_{18}\},\{i_{17},b_{13},z_{19}\}$ & $\{i_{16},b_{12},z_{18}\},\{i_{17},b_{11},z_{19}\}$ & $\{i_{15},b_{13},z_{19}\},\{i_{16},b_{12},z_{17}\}$\\
$\{z_{19},z_{21},o_{4}\},\{z_{23},z_{24},o_{5}\}$ & $\{z_{21},z_{24},o_{4}\},\{z_{22},z_{23},o_{5}\}$ & $\{z_{17},z_{22},o_{5}\},\{z_{23},z_{24},o_{4}\}$ & $\{z_{21},z_{24},o_{5}\},\{z_{22},z_{23},o_{4}\}$\\
\midrule
$\{i_{3},b_{19},z_{27}\},\{i_{4},b_{17},z_{29}\}$ & $\{i_{13},b_{18},z_{27}\},\{i_{18},b_{19},z_{30}\}$ & $\{i_{13},b_{18},z_{26}\},\{i_{18},b_{19},z_{27}\}$ & $\{i_{3},b_{19},z_{31}\},\{i_{4},b_{17},z_{25}\}$\\
$\{i_{17},b_{16},z_{26}\},\{i_{18},b_{18},z_{30}\}$ & $\{i_{19},b_{16},z_{28}\},\{i_{20},b_{17},z_{25}\}$ & $\{i_{19},b_{16},z_{29}\},\{i_{20},b_{17},z_{32}\}$ & $\{i_{17},b_{16},z_{28}\},\{i_{18},b_{18},z_{27}\}$\\
$\{i_{19},b_{15},z_{28}\},\{i_{20},b_{20},z_{25}\}$ & $\{l_{3},b_{20},z_{31}\},\{l_{4},b_{15},z_{29}\}$ & $\{l_{3},b_{20},z_{25}\},\{l_{4},b_{15},z_{28}\}$ & $\{i_{19},b_{15},z_{29}\},\{i_{20},b_{20},z_{32}\}$\\
$\{z_{31},z_{32},o_{6}\}$ & $\{z_{26},z_{32},o_{6}\}$ & $\{z_{30},z_{31},o_{6}\}$ & $\{z_{26},z_{30},o_{6}\}$\\
\bottomrule
\end{tabular}
\captionof{table}{The clauses of the $X$-gadget including the glue vertices.}\label{tab:X}
\end{center}

The values in a 1-in-3-satisfying assignment are derived with the following rules. Rule (R1) is proven in Lemma~\ref{lem:propagation}. Rules (R2)--(R4) follow immediately from 1-in-3-satisfiability. Assume that the truth assignment $\beta$ satisfies the clause $\{x,y,z\}$.
\begin{enumerate}
\item[(R1)] If $\beta$ also satisfies $\{x',y',z'\}$, $\beta(x)=\beta(x')$ and
$\beta(y)=\beta(y')$, then $\beta(z)=\beta(z')$ (see Lemma \ref{lem:propagation}).
\item[(R2)] If $\beta(x)=\beta(y)$, then $\beta(x)=\beta(y)=F$ and $\beta(z)=T$.
\item[(R3)] If $\beta(x)=T$, then $\beta(y)=\beta(z)=F$.
\item[(R4)] If $\beta(x)=F$, then $\beta(y)=\neg\beta(z)$.
\end{enumerate}

These rules can be used to straightforwardly derive the following properties of the $X$-gadget and the $C$-gadget.
\begin{lem}\label{lem:derive}
\begin{enumerate}
\item[(a)] Let $\beta$ satisfy all clauses of the $C$-gadget, with its glue vertices $g_1,g_2,g_3$ as
additional variables. Then $\beta$ has one value on the variables with the same letter,
$\beta(g_1)=\beta(a_1)$, $\beta(g_2)=\beta(b_1)$, $\beta(g_3)=\beta(d_1)$, and
$\beta(p_j)=\neg\beta(a_m)$, $\beta(q_j)=\neg\beta(d_m)$, $\beta(z_j)=F$ and $\beta(o_j)=T$ for all
$j,m$.
\item[(b)] Let $\beta$ be a 1-in-3-satisfying assignment of the $X$-gadget with its glue vertices as additional variables and $\beta(l_j)=\beta(i_j)$ for $j\in\{1,\dots,4\}$, and let $\nu=\beta(i_1)$. Then
$\beta(i_j)=\nu$, $\beta(b_j)=\neg\nu$, $\beta(z_j)=F$ and $\beta(o_j)=T$ for all $j$.
\end{enumerate}
\end{lem}

\begin{proof}
Consider the two clauses $\{a_{2},k_{1},d_{6}\},\{a_{2},k_{1},d_{3}\}$. If $\beta(d_3)=T$, we have that $\beta(k_1)=F$ and $\beta(a_2)=F$ because the three variables are contained in the clause $\{a_{2},k_{1},d_{3}\}$. But if $\beta(k_1)=F$ and $\beta(a_2)=F$, then $\beta(d_6)=T$ because of the clause $\{a_{2},k_{1},d_{6}\}$. Similarly, if $\beta(d_3)=F$, we have that either $\beta(k_1)$ or $\beta(a_2)$ equal $T$ because the three variables are contained in the clause $\{a_{2},k_{1},d_{3}\}$. But then $\beta(d_6)=F$ because of the clause $\{a_{2},k_{1},d_{6}\}$. In any case, we have that $\beta(d_3)=\beta(d_6)$. The attentive reader might see a connection to the logic of the equality verifier in Section \ref{sec:hardness3part}.

In the same style, we deduce parities for a 1-in-3-satisfying assignment. We present them in an abbreviated fashion for both gadgets:
\paragraph{Derivation for the $C$-gadget}  {\footnotesize\raggedright
$\{a_{2},k_{1},d_{6}\},\{a_{2},k_{1},d_{3}\}\Rightarrow \beta(d_{6})=\beta(d_{3})$, $\{a_{2},d_{6},k_{1}\},\{a_{2},d_{3},k_{2}\}\Rightarrow \beta(k_{1})=\beta(k_{2})$, $\{a_{2},k_{1},d_{6}\},\{a_{2},k_{2},d_{2}\}\Rightarrow \beta(d_{6})=\beta(d_{2})$, $\{d_{6},k_{1},a_{2}\},\{d_{6},k_{1},a_{3}\}\Rightarrow \beta(a_{2})=\beta(a_{3})$, $\{e_{1},g_{2},d_{2}\},\{e_{1},g_{2},d_{4}\}\Rightarrow \beta(d_{2})=\beta(d_{4})$, $\{b_{2},e_{2},d_{3}\},\{b_{2},e_{2},d_{5}\}\Rightarrow \beta(d_{3})=\beta(d_{5})$, $\{b_{1},f,a_{4}\},\{b_{1},f,a_{6}\}\Rightarrow \beta(a_{4})=\beta(a_{6})$, $\{a_{4},f,b_{1}\},\{a_{6},f,b_{2}\}\Rightarrow \beta(b_{1})=\beta(b_{2})$, $\{k_{2},h,e_{1}\},\{k_{2},h,e_{2}\}\Rightarrow \beta(e_{1})=\beta(e_{2})$, $\{d_{2},e_{1},g_{2}\},\{d_{3},e_{2},b_{2}\}\Rightarrow \beta(g_{2})=\beta(b_{2})$, $\{h,e_{1},k_{2}\},\{h,e_{1},k_{3}\}\Rightarrow \beta(k_{2})=\beta(k_{3})$, $\{d_{6},k_{1},a_{2}\},\{d_{4},k_{3},a_{5}\}\Rightarrow \beta(a_{2})=\beta(a_{5})$, $\{a_{2},d_{6},k_{1}\},\{a_{5},d_{4},k_{4}\}\Rightarrow \beta(k_{1})=\beta(k_{4})$, $\{d_{6},k_{1},a_{2}\},\{d_{4},k_{4},a_{4}\}\Rightarrow \beta(a_{2})=\beta(a_{4})$, $\{p_{1},z_{2},a_{8}\},\{p_{1},z_{2},a_{1}\}\Rightarrow \beta(a_{8})=\beta(a_{1})$, $\{p_{1},a_{8},z_{2}\},\{p_{1},a_{8},z_{6}\}\Rightarrow \beta(z_{2})=\beta(z_{6})$, $\{a_{8},z_{2},p_{1}\},\{a_{8},z_{2},p_{2}\}\Rightarrow \beta(p_{1})=\beta(p_{2})$, $\{p_{1},a_{8},z_{2}\},\{p_{2},a_{8},z_{1}\}\Rightarrow \beta(z_{2})=\beta(z_{1})$, $\{z_{1},z_{6},o_{2}\}\colon \beta(z_{1})=\beta(z_{6})\Rightarrow \beta(z_{1})=F,\ \beta(o_{2})=T$, $\{z_{2},a_{8},p_{2}\}\colon \beta(z_{2})=F\Rightarrow \beta(a_{8})=\neg \beta(p_{2})$, $\{o_{2},z_{5},z_{8}\}\colon \beta(o_{2})=T\Rightarrow \beta(z_{5})=F,\ \beta(z_{8})=F$, $\{z_{8},z_{7},o_{1}\}\colon \beta(z_{8})=F\Rightarrow \beta(o_{1})=\neg \beta(z_{7})$, $\{o_{2},z_{6},z_{7}\}\colon \beta(o_{2})=T\Rightarrow \beta(z_{7})=F$, $\{z_{1},a_{3},p_{4}\}\colon \beta(z_{1})=F\Rightarrow \beta(p_{4})=\neg \beta(a_{3})$, $\{z_{1},a_{3},p_{2}\}\colon \beta(z_{1})=F\Rightarrow \beta(a_{3})=\neg \beta(p_{2})$, $\{o_{1},z_{3},z_{5}\}\colon \beta(o_{1})=T\Rightarrow \beta(z_{3})=F$, $\{z_{5},a_{7},p_{4}\}\colon \beta(z_{5})=F\Rightarrow \beta(a_{7})=\neg \beta(p_{4})$, $\{z_{3},g_{1},p_{1}\}\colon \beta(z_{3})=F\Rightarrow \beta(p_{1})=\neg \beta(g_{1})$, $\{z_{3},g_{1},p_{3}\}\colon \beta(z_{3})=F\Rightarrow \beta(g_{1})=\neg \beta(p_{3})$, $\{p_{1},a_{8},z_{2}\},\{p_{4},a_{6},z_{4}\}\Rightarrow \beta(z_{2})=\beta(z_{4})$, $\{q_{1},z_{10},d_{8}\},\{q_{1},z_{10},d_{1}\}\Rightarrow \beta(d_{8})=\beta(d_{1})$, $\{q_{1},d_{8},z_{10}\},\{q_{1},d_{8},z_{14}\}\Rightarrow \beta(z_{10})=\beta(z_{14})$, $\{d_{8},z_{10},q_{1}\},\{d_{8},z_{10},q_{2}\}\Rightarrow \beta(q_{1})=\beta(q_{2})$, $\{q_{1},d_{8},z_{10}\},\{q_{2},d_{8},z_{9}\}\Rightarrow \beta(z_{10})=\beta(z_{9})$, $\{z_{9},z_{14},o_{4}\}\colon \beta(z_{9})=\beta(z_{14})\Rightarrow \beta(z_{9})=F,\ \beta(o_{4})=T$, $\{z_{10},d_{8},q_{2}\}\colon \beta(z_{10})=F\Rightarrow \beta(d_{8})=\neg \beta(q_{2})$, $\{o_{4},z_{13},z_{16}\}\colon \beta(o_{4})=T\Rightarrow \beta(z_{13})=F,\ \beta(z_{16})=F$, $\{z_{16},z_{15},o_{3}\}\colon \beta(z_{16})=F\Rightarrow \beta(o_{3})=\neg \beta(z_{15})$, $\{o_{4},z_{14},z_{15}\}\colon \beta(o_{4})=T\Rightarrow \beta(z_{15})=F$, $\{z_{9},d_{2},q_{4}\}\colon \beta(z_{9})=F\Rightarrow \beta(q_{4})=\neg \beta(d_{2})$, $\{z_{9},d_{2},q_{2}\}\colon \beta(z_{9})=F\Rightarrow \beta(d_{2})=\neg \beta(q_{2})$, $\{o_{3},z_{11},z_{13}\}\colon \beta(o_{3})=T\Rightarrow \beta(z_{11})=F$, $\{z_{13},d_{7},q_{4}\}\colon \beta(z_{13})=F\Rightarrow \beta(d_{7})=\neg \beta(q_{4})$, $\{z_{11},g_{3},q_{1}\}\colon \beta(z_{11})=F\Rightarrow \beta(q_{1})=\neg \beta(g_{3})$, $\{z_{11},g_{3},q_{3}\}\colon \beta(z_{11})=F\Rightarrow \beta(g_{3})=\neg \beta(q_{3})$, $\{q_{1},d_{8},z_{10}\},\{q_{4},d_{6},z_{12}\}\Rightarrow \beta(z_{10})=\beta(z_{12})$.\par}

\paragraph{Derivation for the $X$-gadget}  {\footnotesize\raggedright
$\{b_{1},i_{7},z_{4}\},\{b_{1},i_{7},z_{5}\}\Rightarrow \beta(z_{4})=\beta(z_{5})$, $\{b_{1},z_{4},i_{7}\},\{b_{1},z_{5},i_{9}\}\Rightarrow \beta(i_{7})=\beta(i_{9})$, $\{z_{4},i_{7},b_{1}\},\{z_{5},i_{7},b_{2}\}\Rightarrow \beta(b_{1})=\beta(b_{2})$, $\{b_{1},z_{4},i_{7}\},\{b_{2},z_{4},i_{1}\}\Rightarrow \beta(i_{7})=\beta(i_{1})$, $\{b_{1},i_{7},z_{4}\},\{b_{2},i_{1},z_{2}\}\Rightarrow \beta(z_{4})=\beta(z_{2})$, $\{z_{4},i_{7},b_{1}\},\{z_{2},i_{1},b_{4}\}\Rightarrow \beta(b_{1})=\beta(b_{4})$, $\{b_{1},i_{7},z_{4}\},\{b_{4},i_{1},z_{3}\}\Rightarrow \beta(z_{4})=\beta(z_{3})$, $\{b_{1},z_{4},i_{7}\},\{b_{4},z_{3},i_{5}\}\Rightarrow \beta(i_{7})=\beta(i_{5})$, $\{b_{1},i_{7},z_{4}\},\{b_{4},i_{5},z_{6}\}\Rightarrow \beta(z_{4})=\beta(z_{6})$, $\{z_{2},z_{6},o_{1}\}\colon \beta(z_{2})=\beta(z_{6})\Rightarrow \beta(z_{2})=F,\ \beta(o_{1})=T$, $\{o_{1},z_{2},z_{8}\}\colon \beta(o_{1})=T\Rightarrow \beta(z_{8})=F$, $\{z_{3},i_{5},b_{4}\}\colon \beta(z_{3})=F\Rightarrow \beta(b_{4})=\neg \beta(i_{5})$, $\{z_{6},i_{5},b_{5}\}\colon \beta(z_{6})=F\Rightarrow \beta(i_{5})=\neg \beta(b_{5})$, $\{z_{3},i_{2},b_{5}\}\colon \beta(z_{3})=F\Rightarrow \beta(i_{2})=\neg \beta(b_{5})$, $\{o_{1},z_{6},z_{7}\}\colon \beta(o_{1})=T\Rightarrow \beta(z_{7})=F$, $\{z_{7},i_{2},b_{6}\}\colon \beta(z_{7})=F\Rightarrow \beta(i_{2})=\neg \beta(b_{6})$, $\{z_{8},i_{8},b_{6}\}\colon \beta(z_{8})=F\Rightarrow \beta(b_{6})=\neg \beta(i_{8})$, $\{z_{8},i_{8},b_{3}\}\colon \beta(z_{8})=F\Rightarrow \beta(b_{3})=\neg \beta(i_{8})$, $\{z_{5},i_{6},b_{3}\}\colon \beta(z_{5})=F\Rightarrow \beta(b_{3})=\neg \beta(i_{6})$, $\{b_{1},i_{7},z_{4}\},\{b_{3},i_{8},z_{1}\}\Rightarrow \beta(z_{4})=\beta(z_{1})$, $\{b_{7},z_{10},i_{12}\},\{b_{7},z_{10},i_{9}\}\Rightarrow \beta(i_{12})=\beta(i_{9})$, $\{b_{7},i_{12},z_{10}\},\{b_{7},i_{6},z_{11}\}\Rightarrow \beta(z_{10})=\beta(z_{11})$, $\{b_{7},i_{12},z_{10}\},\{b_{7},i_{12},z_{14}\}\Rightarrow \beta(z_{10})=\beta(z_{14})$, $\{i_{12},z_{10},b_{7}\},\{i_{12},z_{10},b_{8}\}\Rightarrow \beta(b_{7})=\beta(b_{8})$, $\{b_{7},i_{12},z_{10}\},\{b_{8},i_{12},z_{9}\}\Rightarrow \beta(z_{10})=\beta(z_{9})$, $\{z_{9},z_{14},o_{3}\}\colon \beta(z_{9})=\beta(z_{14})\Rightarrow \beta(z_{9})=F,\ \beta(o_{3})=T$, $\{z_{10},i_{12},b_{8}\}\colon \beta(z_{10})=F\Rightarrow \beta(i_{12})=\neg \beta(b_{8})$, $\{z_{11},i_{6},b_{9}\}\colon \beta(z_{11})=F\Rightarrow \beta(i_{6})=\neg \beta(b_{9})$, $\{z_{11},i_{11},b_{9}\}\colon \beta(z_{11})=F\Rightarrow \beta(b_{9})=\neg \beta(i_{11})$, $\{z_{11},z_{13},o_{2}\}\colon \beta(z_{11})=F\Rightarrow \beta(z_{13})=\neg \beta(o_{2})$, $\{o_{3},z_{13},z_{16}\}\colon \beta(o_{3})=T\Rightarrow \beta(z_{13})=F,\ \beta(z_{16})=F$, $\{o_{2},z_{15},z_{16}\}\colon \beta(o_{2})=T\Rightarrow \beta(z_{15})=F$, $\{z_{9},i_{14},b_{10}\}\colon \beta(z_{9})=F\Rightarrow \beta(b_{10})=\neg \beta(i_{14})$, $\{z_{9},i_{14},b_{8}\}\colon \beta(z_{9})=F\Rightarrow \beta(i_{14})=\neg \beta(b_{8})$, $\{b_{1},i_{7},z_{4}\},\{b_{9},i_{11},z_{12}\}\Rightarrow \beta(z_{4})=\beta(z_{12})$, $\{z_{12},i_{10},b_{10}\}\colon \beta(z_{12})=F\Rightarrow \beta(i_{10})=\neg \beta(b_{10})$, $\{b_{11},z_{18},i_{16}\},\{b_{11},z_{18},i_{13}\}\Rightarrow \beta(i_{16})=\beta(i_{13})$, $\{b_{11},i_{16},z_{18}\},\{b_{11},i_{16},z_{22}\}\Rightarrow \beta(z_{18})=\beta(z_{22})$, $\{i_{16},z_{18},b_{11}\},\{i_{16},z_{18},b_{12}\}\Rightarrow \beta(b_{11})=\beta(b_{12})$, $\{b_{11},i_{16},z_{18}\},\{b_{12},i_{16},z_{17}\}\Rightarrow \beta(z_{18})=\beta(z_{17})$, $\{z_{17},z_{22},o_{5}\}\colon \beta(z_{17})=\beta(z_{22})\Rightarrow \beta(z_{17})=F,\ \beta(o_{5})=T$, $\{z_{18},i_{16},b_{12}\}\colon \beta(z_{18})=F\Rightarrow \beta(i_{16})=\neg \beta(b_{12})$, $\{o_{5},z_{21},z_{24}\}\colon \beta(o_{5})=T\Rightarrow \beta(z_{21})=F,\ \beta(z_{24})=F$, $\{z_{24},z_{23},o_{4}\}\colon \beta(z_{24})=F\Rightarrow \beta(o_{4})=\neg \beta(z_{23})$, $\{o_{5},z_{22},z_{23}\}\colon \beta(o_{5})=T\Rightarrow \beta(z_{23})=F$, $\{z_{17},i_{10},b_{14}\}\colon \beta(z_{17})=F\Rightarrow \beta(b_{14})=\neg \beta(i_{10})$, $\{z_{17},i_{10},b_{12}\}\colon \beta(z_{17})=F\Rightarrow \beta(i_{10})=\neg \beta(b_{12})$, $\{o_{4},z_{19},z_{21}\}\colon \beta(o_{4})=T\Rightarrow \beta(z_{19})=F$, $\{z_{21},i_{15},b_{14}\}\colon \beta(z_{21})=F\Rightarrow \beta(i_{15})=\neg \beta(b_{14})$, $\{z_{19},i_{17},b_{11}\}\colon \beta(z_{19})=F\Rightarrow \beta(b_{11})=\neg \beta(i_{17})$, $\{z_{19},i_{17},b_{13}\}\colon \beta(z_{19})=F\Rightarrow \beta(i_{17})=\neg \beta(b_{13})$, $\{b_{1},i_{7},z_{4}\},\{b_{14},i_{14},z_{20}\}\Rightarrow \beta(z_{4})=\beta(z_{20})$, $\{b_{15},i_{19},z_{28}\},\{b_{15},i_{19},z_{29}\}\Rightarrow \beta(z_{28})=\beta(z_{29})$, $\{b_{15},z_{28},i_{19}\},\{b_{15},z_{29},i_{4}\}\Rightarrow \beta(i_{19})=\beta(i_{4})$, $\{z_{28},i_{19},b_{15}\},\{z_{29},i_{19},b_{16}\}\Rightarrow \beta(b_{15})=\beta(b_{16})$, $\{b_{15},z_{28},i_{19}\},\{b_{16},z_{28},i_{17}\}\Rightarrow \beta(i_{19})=\beta(i_{17})$, $\{b_{15},i_{19},z_{28}\},\{b_{16},i_{17},z_{26}\}\Rightarrow \beta(z_{28})=\beta(z_{26})$, $\{z_{28},i_{19},b_{15}\},\{z_{26},i_{13},b_{18}\}\Rightarrow \beta(b_{15})=\beta(b_{18})$, $\{b_{15},i_{19},z_{28}\},\{b_{18},i_{13},z_{27}\}\Rightarrow \beta(z_{28})=\beta(z_{27})$, $\{z_{28},i_{19},b_{15}\},\{z_{29},i_{4},b_{17}\}\Rightarrow \beta(b_{15})=\beta(b_{17})$, $\{b_{15},i_{19},z_{28}\},\{b_{17},i_{4},z_{25}\}\Rightarrow \beta(z_{28})=\beta(z_{25})$, $\{b_{15},z_{28},i_{19}\},\{b_{17},z_{25},i_{20}\}\Rightarrow \beta(i_{19})=\beta(i_{20})$, $\{b_{15},i_{19},z_{28}\},\{b_{17},i_{20},z_{32}\}\Rightarrow \beta(z_{28})=\beta(z_{32})$, $\{z_{26},z_{32},o_{6}\}\colon \beta(z_{26})=\beta(z_{32})\Rightarrow \beta(z_{26})=F,\ \beta(o_{6})=T$, $\{z_{27},i_{18},b_{18}\}\colon \beta(z_{27})=F\Rightarrow \beta(b_{18})=\neg \beta(i_{18})$, $\{z_{27},i_{18},b_{19}\}\colon \beta(z_{27})=F\Rightarrow \beta(i_{18})=\neg \beta(b_{19})$, $\{z_{27},i_{3},b_{19}\}\colon \beta(z_{27})=F\Rightarrow \beta(i_{3})=\neg \beta(b_{19})$, $\{o_{6},z_{30},z_{31}\}\colon \beta(o_{6})=T\Rightarrow \beta(z_{30})=F,\ \beta(z_{31})=F$, $\{z_{31},i_{3},b_{20}\}\colon \beta(z_{31})=F\Rightarrow \beta(i_{3})=\neg \beta(b_{20})$, $\{z_{32},i_{20},b_{20}\}\colon \beta(z_{32})=F\Rightarrow \beta(b_{20})=\neg \beta(i_{20})$.\par}
\end{proof}

This yields the following proposition:

\begin{proof}[Proof of Proposition~\ref{pro:bringing4parttogether}] By construction, both gadgets are planar and 4-partitioned.
\emph{$C$-gadget.} Let $\beta$ be a 1-in-3-satisfying assignment. By Lemma~\ref{lem:derive}(a),
applied with $\beta(g_1)=\beta(a_1)$, $\beta(g_2)=\beta(b_1)$, $\beta(g_3)=\beta(d_1)$, $\beta$ has
one value on each letter, written $\beta(a)$ and so on, and $\beta(p)=\neg\beta(a)$,
$\beta(q)=\neg\beta(d)$, $\beta(z)=F$, $\beta(o)=T$. So the clauses of the forms $\{a,p,z\}$,
$\{d,q,z\}$ and $\{z,z,o\}$ are satisfied, where a letter stands for any of its variables. The others
are of the forms $\{a,d,k\}$, $\{a,b,f\}$, $\{b,d,e\}$, $\{e,h,k\}$, where $g_1,g_2,g_3$ count as
$a,b,d$; with $T=1$ and $F=0$ they say
\[
\beta(a)+\beta(d)+\beta(k)=\beta(a)+\beta(b)+\beta(f)=\beta(b)+\beta(d)+\beta(e)
=\beta(e)+\beta(h)+\beta(k)=1.
\]
By the first three equations, at most one of the values $\beta(a),\beta(b),\beta(d)$ is $T$; if none
of them were, then $\beta(k)=\beta(f)=\beta(e)=T$, contradicting the fourth. So $\beta(u)=T$ for
exactly one $u\in\{a_1,b_1,d_1\}$, and $u$ determines all values: of the letters
$a,b,d,e,f,h,k$, exactly $a,e$ have the value $T$ if $u=a_1$, exactly $b,k$ if $u=b_1$, and exactly
$d,f,h$ if $u=d_1$. Conversely, each of these three choices satisfies all clauses.

\emph{$X$-gadget.} By Lemma~\ref{lem:derive}(b), a 1-in-3-satisfying assignment gives all four output
vertices one value $\nu$, which is determined by $\nu$. Conversely, for both values of
$\nu$, the values $\beta(i_j)=\nu$, $\beta(b_j)=\neg\nu$, $\beta(z_j)=F$, $\beta(o_j)=T$, with $\beta(l_j)=\nu$, satisfy all
clauses, since they are of the forms $\{i,b,z\}$ and $\{z,z,o\}$, where $i$ stands for any of $i_1,\dots,i_{20}$ and $l_1,\dots,l_4$.
\end{proof}
\end{document}